\documentclass[ba,preprint]{imsart}

\RequirePackage{amsthm,amsmath,amsfonts,amssymb}
\RequirePackage[authoryear]{natbib}
\RequirePackage{hyperref}
\RequirePackage{graphicx}
\RequirePackage{subcaption}
\RequirePackage[dvipsnames]{xcolor}

\pubyear{2026}
\volume{TBA}
\issue{TBA}
\firstpage{1}
\lastpage{1}

\startlocaldefs
\theoremstyle{plain}

\newtheorem{theorem}{Theorem}[section]
\newtheorem{lemma}[theorem]{Lemma}
\newtheorem{proposition}[theorem]{Proposition}
\theoremstyle{definition}
\newtheorem{remark}{Remark}
\newtheorem{definition}[theorem]{Definition}
\newtheorem{example}{Example}

\theoremstyle{remark}

\endlocaldefs

\begin{document}

\begin{frontmatter}
\title{Conformalized Bayesian Inference, with Applications to Random Partition Models}
\runtitle{Conformalized Bayesian Inference}

\begin{aug}
\author[A]{\fnms{Nicola}~\snm{Bariletto}},
\author[A]{\fnms{Nhat}~\snm{Ho}}
\and
\author[A]{\fnms{Alessandro}~\snm{Rinaldo}}
\address[A]{Department of Statistics and Data Science,
The University of Texas at Austin.}

\end{aug}

\begin{abstract}
Bayesian posterior distributions naturally represent parameter uncertainty informed by data. However, when the parameter space is complex, as in many nonparametric settings where it is infinite-dimensional or combinatorially large, standard summaries such as posterior means, credible intervals, or simple notions of multimodality are often unavailable, hindering interpretable posterior uncertainty quantification. We introduce Conformalized Bayesian Inference (CBI), a broadly applicable and computationally efficient framework for posterior inference on nonstandard parameter spaces. CBI yields a point estimate, a credible region with assumption-free posterior coverage guarantees, and a principled analysis of posterior multimodality, requiring only Monte Carlo samples from the posterior and a notion of discrepancy between parameters. The method builds a pseudo-density score for each parameter value, yielding a maximum-a-posteriori-like point estimate and a credible region derived from conformal prediction principles. The key conceptual step underlying this construction is the reinterpretation of posterior inference as prediction on the parameter space, with interesting connections to classical Bayesian point estimation and credible set constructions. A final density-based clustering step identifies representative posterior modes. We investigate a number of theoretical and methodological properties of CBI and demonstrate its practicality, scalability, and versatility in simulated and real data clustering applications with random partition models. 
\end{abstract}



\end{frontmatter}

\section{Introduction}

Bayesian statistical methods are a standard tool in probabilistic modeling, owing to their natural ability to represent uncertainty. The data-generating process is formalized as a statistical model whose parameters are inferred from the data, with uncertainty expressed through a prior and updated to a posterior given the data. In particular, given a parameter space $\Theta$, it suffices to specify a prior on $\Theta$ and compute---often approximately, e.g., via Monte Carlo---the posterior, providing a coherent basis for point estimation (a single value summarizing the posterior) and uncertainty quantification (characterizing regions of $\Theta$ carrying substantial posterior mass). This is particularly powerful in nonparametric settings, where flexibility comes from letting $\Theta$ be complex and often infinite-dimensional. However, such parameter spaces often lack a conventional (e.g., vector or ordering) structure needed for interpretable summaries like a posterior mean/mode or a credible interval. This is a pervasive challenge: consider $\Theta$ consisting of data partitions, covariance matrices, graphs, or mixing measures. Here, although Monte Carlo methods (often Markov Chain Monte Carlo, or MCMC) approximate the posterior well, the complexity of the sampled objects hinders interpretable inference, raising fundamental questions such as: what is a suitable point estimate? How does one define a credible region of reasonable size with prescribed posterior coverage? Is the posterior concentrated in one region of $\Theta$, or spread across several distinct ones? Without a familiar Euclidean-like structure on $\Theta$, no general answers exist.

These challenges have motivated solutions ranging from theoretically grounded to ad hoc. For point estimation, a common decision-theoretic approach minimizes the posterior expected loss under a chosen loss function on $\Theta\times\Theta$, typically via Monte Carlo approximation. Credible regions and multimodality are instead often addressed through problem-specific heuristics; for instance, when $\theta$ encodes a partition of the data, several approaches exist~\citep{wade2018clustering, balocchi2025understanding}. Another strategy examines low-dimensional aspects of the parameter~\citep{woody2021model,bolfarine2025lower}, offering preliminary insight but risking a loss of nuance. A general, principled framework addressing these questions simultaneously, coherently, and efficiently is therefore desirable.

Here, we make a step in this direction by introducing a broadly applicable methodology grounded in the theories of conformal prediction~\citep{vovk2005algorithmic,shafer2008tutorial,angelopoulos2023conformal} and density-based clustering~\citep{rodriguez2014clustering,ester1996density,kriegel2011density}, which we term \emph{Conformalized Bayesian Inference (CBI)}. {\color{black}Both conformal prediction and density-based clustering are well-established fields with roots in statistics and machine learning. In particular, conformal prediction addresses the following problem: given exchangeable observations $Y_1,\ldots,Y_T$ on some space $\mathcal Y$, how can one construct a region $\mathcal Y_0\subseteq \mathcal Y$ satisfying a coverage guarantee of the form $\mathbb P(Y_{T+1}\in\mathcal Y_0)\geq 1-\alpha$, where $\alpha\in(0,1)$ is user-specified? The standard setting corresponds to $Y_1,\ldots,Y_T$ representing observed data and $\mathcal Y_0$ a set-valued prediction enjoying guaranteed coverage under the unknown data-generating mechanism. Conformal prediction achieves this by exploiting the relative ranking of points $Y\in\mathcal Y$ according to a conformity score computed from the observed sample $Y_1,\ldots,Y_T$. While we defer a detailed discussion of this construction, we note that, for our purposes, the utility of this framework arises from viewing $Y_1,\ldots,Y_T$ as posterior samples and the generating mechanism as induced by the posterior distribution, so that the resulting region $\mathcal Y_0$ admits the interpretation of a posterior credible region.}

{\color{black}To set the stage for our discussion, denote by $\Pi$ the target posterior distribution on $\Theta$. We remark that, while it is customary to explicitly flag the dependence of the posterior on the data $X_1,\ldots,X_n$ that gives rise to it, e.g., writing $\Pi(\cdot\mid X_1,\ldots,X_n)$, here we depart from this convention to avoid overloading our notation. In particular, because our treatment will completely abstract from observed data and will be instead concerned with the posterior distribution as a generic measure on $\Theta$ that we are interested in summarizing, the lighter notation $\Pi$ suffices for our purposes. With this in mind, assume that} we have at our disposal $T \in \mathbb{N}$ independent and identically distributed (iid) Monte Carlo samples $\boldsymbol{\theta} = \{\theta_1, \dots, \theta_T\}$, each marginally distributed according to $\Pi$. The iid assumption may be either true exactly by the nature of the adopted Monte Carlo scheme, or can be thought to hold approximately in the case of MCMC samples, as long as the sampler chain has been properly thinned (see {Proposition~\ref{pro:mixing} in Appendix~\ref{sec:proofs}}); a more detailed discussion on these points is postponed to Section~\ref{sec:credible_region}. We then split the samples into two subsets: a training set $\boldsymbol{\theta}^1 = \{\theta_1, \dots, \theta_S\}$ and a calibration set $\boldsymbol{\theta}^2 = \{\theta_{S+1}, \dots, \theta_T\}$. The key idea is to compute, for each calibration sample $\theta \in \boldsymbol{\theta}^2$, a pseudo-density score 
\(
\theta\mapsto s(\theta; \boldsymbol{\theta}^1) \in \mathbb{R},
\)
which is estimated using the training set $\boldsymbol{\theta}^1$ and is designed to capture a notion of ``likelihood'' or ``density'' of $\theta$ under the posterior distribution. This will be achieved by defining $s(\cdot; \boldsymbol{\theta}^1)$ as a kernel density estimator (KDE) based on a discrepancy $\mathcal D:\Theta\times\Theta\to\mathbb R_+$ between parameters{\color{black}, that is
\begin{equation*}
    s(\theta; \boldsymbol{\theta}^1) := \frac{1}{S}\sum_{t=1}^S  e^{-\gamma \mathcal D(\theta, \theta_t)}
\end{equation*}
for some $\gamma>0$}.

Once the scoring rule is available, the CBI framework proceeds through the following three main steps:

\begin{itemize}
    \item \emph{Step 1: point estimation.} The calibration scores can be used to define an intuitive point estimator. Specifically, we select
    \[
    \theta_\star \in \arg\max_{\theta \in \boldsymbol{\theta}^2} s(\theta; \boldsymbol{\theta}^1).
    \]
    Because $s(\cdot; \boldsymbol{\theta}^1)$ is a posterior pseudo-density function, $\theta_\star$ may be interpreted as a pseudo-maximum-a-posteriori (pseudo-MAP) estimator.

    \item \emph{Step 2: credible region.} By exchangeability of the calibration scores, standard conformal prediction arguments yield a credible region $\mathcal C_{1-\alpha}(\boldsymbol{\theta}^2)\subseteq\Theta$, based on the quantiles of $\{ s(\theta; \boldsymbol{\theta}^1) : \theta \in \boldsymbol{\theta}^2 \}$, having a prescribed $(1 - \alpha) \times 100\%$ coverage under $\Pi$ (for $\alpha \in (0,1)$, in a precise statistical sense to be defined later). The key conceptual step leading to this construction is the reinterpretation of inference based on posterior Monte Carlo samples as prediction on the parameter space according to the posterior distribution, and to leverage conformal prediction tools accordingly. Two appealing features of this procedure are that (i) the attached coverage guarantee holds exactly for any finite Monte Carlo sample size $T$ and is independent of any feature of the target posterior $\Pi$, accommodating parameter spaces of arbitrary dimension and with almost no structure, and (ii) for a new sample $\theta$, it provides a simple way to define a standardized measure of typicality $\hat p(\theta;\boldsymbol{\theta}^2)\in[0,1]$ under $\Pi$---informally, how close $\theta$ lies to the bulk of the posterior distribution relative to the calibration samples---which may also be formally interpreted as a $p$-value under an intuitive posterior distributional null hypothesis. In particular, this construction allows one to test the posterior plausibility of scientifically meaningful hypotheses about $\theta$. For instance, if $\Theta$ is the space of data partitions arising from a mixture model and each observation represents a hospital patient, one may test whether the posterior distribution supports the hypothesis that all patients belong to a single homogeneous group: it is enough to check whether the trivial partition $\theta$ clustering all patients together lies in the credible region of pre-specified posterior coverage, or equivalently if the conformal $p$-value $\hat p(\theta;\boldsymbol{\theta}^2)$ is high enough.

    \item \emph{Step 3: multimodality analysis.} Finally, the interpretation of $\theta\mapsto s(\theta; \boldsymbol{\theta}^1)$ as a posterior pseudo-density function allows one to explore the possibly multimodal structure of the posterior distribution through density-based clustering techniques. {\color{black}Concretely, by equipping $\Theta$ with the discrepancy $\mathcal D$ and having the pseudo-density score available, one can identify multiple high-density regions which are far apart from each other (in the sense of $\mathcal D$); this reveals whether the posterior distribution places substantial mass around several, well-separated parameter values, which may be interpreted as corresponding to distinct modes of the posterior.}
    \end{itemize}

Throughout the paper, we focus on developing simple, robust, and efficient methods to compute the calibration scores and to carry out the multimodality analysis. As will become evident, the advantages of our methodology include a high degree of intuitiveness and flexibility (for example, in the choice of the metric $\mathcal D$, or in the ability to restrict point estimation and uncertainty quantification to regions of $\Theta$ that the user considers plausible or of interest), as well as the potential for efficient implementation thanks to the parallelizability of score and distance computations.

While CBI can be described in full generality without reference to any particular statistical model, it is useful to ground the discussion in a concrete setting involving a parameter space $\Theta$ of interest. Therefore, before presenting a general treatment of CBI, in Section~\ref{sec:RPM} we first introduce a fundamental class of models, namely random partition models, on which the strengths of our methodology can be clearly demonstrated. One advantage of working with random partition models is the existence of recent work addressing point estimation and uncertainty quantification based on posterior Monte Carlo samples over the space of data partitions~\citep{wade2018clustering,wade2023bayesian, balocchi2025understanding}, providing a natural conceptual benchmark for our discussion. Moreover, since clustering is an inherently difficult problem, Bayesian posteriors over data partitions often display substantial uncertainty, frequently in the form of multimodality, which our framework is specifically designed to handle. After this, in Sections~\ref{sec:why_CP}, \ref{sec:point_estimation}, \ref{sec:credible_region}, \ref{sec:multimodality} and \ref{sec:hyperparameter} we describe CBI in detail and illustrate it with simulated data analyzed with random partition models; we instead refer the reader to {Appendices~\ref{sec:mixmeasure_exp} and~\ref{sec:covariance_exp}} for additional experiments applying CBI to models where the parameter space consists of discrete mixing measures endowed with the Wasserstein-1 distance and covariance matrices endowed with the operator norm distance. Section~\ref{sec:real_data} showcases two real-world applications of CBI to Bayesian clustering of (i) the classical Galaxy velocities dataset \citep{roeder1990density}, and (ii) event-related potential (ERP) waveform functional data~\citep{song2025repulsive,kappenman2021erp}. Section~\ref{sec:discussion} concludes the article with a discussion of the current limitations and research directions related to CBI. An accompanying Python library, \texttt{cbi\_partitions}, is hosted on GitHub at \url{https://github.com/nbariletto/cbi_partitions_repo}, with its use illustrated in a step-by-step tutorial at \url{https://nbariletto.github.io/cbi_partitions_repo/}.

\section{Motivation: random partition models}\label{sec:RPM}

Given a dataset $X_1, \ldots, X_n$, a fundamental unsupervised statistical learning problem consists in partitioning it into disjoint clusters representing heterogeneous subpopulations. Denoting by $\Theta$ the space of partitions of the data, the Bayesian framework tackles the clustering problem by specifying a prior $\pi(\theta)$ over $\Theta$ and a likelihood $p(X_1, \ldots, X_n \mid \theta)$ for the observations given their partitioning according to $\theta \in \Theta$. The resulting posterior distribution
\begin{equation*}
    \Pi(\theta) \propto p(X_1, \ldots, X_n \mid \theta)\, \pi(\theta)
\end{equation*}
allows for coherent probabilistic inference on the clustering structure of the observed data. In practice, inference on clustering has been classically formulated at the latent level of a hierarchical density mixture model, where heterogeneity among subpopulations is interpreted as observations being generated from different components of a mixture model~\citep{fruhwirth2019handbook,lo1984class,escobar1995bayesian,lijoi2005hierarchical,barrios2013modeling,lijoi2020pitman}, or, more recently, through approaches based on tailored losses \citep{rigon2023generalized} or density-based ideas \citep{buch2024bayesian,bariletto2026scalable}. We refer the reader to \citet{wade2023bayesian}, \citet{grazian2023review}, and the review section of \citet{balocchi2025understanding} for comprehensive and up-to-date overviews of the field. {\color{black}Importantly, our methodology is fully general and applicable to any random partition model that induces a posterior $\Pi$ over partitions of the $n$ data points, motivating an upcoming methodological discussion that is largely agnostic to the Bayesian model generating the posterior samples of interest.}

The combinatorially large size of $\Theta$ typically prevents exact posterior evaluation, and approximate computational methods, most often in the form of MCMC sampling, are adopted instead. For instance, if the random partition model is specified within a hierarchical mixture framework, standard samplers~\citep{escobar1995bayesian,escobar1998computing,maceachern1998estimating} will, at each iteration $t = 1, \ldots, T$, draw a sample $\theta_t \sim \Pi$ (in addition to sampling the other mixture parameters). These samples are usually stored as vectors of cluster assignments $C_t = (c_{t1}, \ldots, c_{tn})$, encoding the partition (equivalence) relation $\sim_{\theta_t}$ via $X_i \sim_{\theta_t} X_j$ if and only if $c_{ti} = c_{tj}$, for all $i, j = 1, \ldots, n$. Assuming appropriate thinning of the chain and a warm start $\theta_1 \sim \Pi$, the resulting samples $\theta_1, \ldots, \theta_T$ will serve as the basis of our CBI procedure.

A common choice on the space of partitions $\Theta$ is the Variation-of-Information (VI) metric \citep{meila2007comparing}, denoted by $\mathcal D_{VI}$ and defined as follows. Take any two partitions $\theta, \theta' \in \Theta$ with associated cluster assignment vectors $C, C'$ and respective numbers of clusters $K$ and $K'$. Let $\{\bar c_1, \ldots, \bar c_K\}$ and $\{\bar c'_1, \ldots, \bar c'_{K'}\}$ denote the sets of unique cluster labels in each vector, and define
$n_{j,*} := |\{ i = 1, \ldots, n : c_i = \bar c_j \}|$, 
$n_{*,k}  := |\{ i = 1, \ldots, n : c'_i = \bar c'_k \}|$, and $n_{j,k} := |\{ i = 1, \ldots, n : c_i = \bar c_j, \, c'_i = \bar c'_k \}|$, for all $j = 1, \ldots, K$ and $k = 1, \ldots, K'$. Then,
\begin{align}\label{eq:VI}
    \mathcal D_{VI}(\theta, \theta') 
    &:= \\
    -\, \sum_{j=1}^K &\frac{n_{j,*}}{n} \log_2\left( \frac{n_{j,*}}{n} \right)
    \,-\, \sum_{k=1}^{K'} \frac{n_{*,k}}{n} \log_2\left( \frac{n_{*,k}}{n} \right)
    \,-\, 2 \sum_{j=1}^K \sum_{k=1}^{K'} \frac{n_{j,k}}{n} 
    \log_2\left( \frac{n \cdot n_{j,k}}{n_{j,*} \cdot n_{*,k}} \right).\nonumber
\end{align}

\citet{meila2007comparing} showed that $\mathcal D_{VI}$ is a proper metric on $\Theta$, and that it enjoys several appealing theoretical properties, such as its information-theoretic interpretation in terms of self-entropy and cross-entropy between partitions, while also being efficiently computable. {\color{black}Several other discrepancy measures have been proposed, such as the Binder loss~\citep{binder1978bayesian,dahl2006model,lau2007clustering}, the adjusted Rand index~\citep{fritsch2009clustering}, the normalized VI~\citep{vinh2010information,rastelli2018optimal}, and more~\citep{quintana2003clustering,dahl2022search}, which are all well suited for our purpose of creating a pseudo-KDE score. As $\mathcal D_{VI}$ has been recently exploited both methodologically \citep{wade2018clustering,balocchi2025understanding} and computationally (for instance in the popular R packages \texttt{mcclust} and \texttt{salso}), we mostly restrict our attention to it in the rest of the paper. Nevertheless, the accompanying Python library \texttt{cbi\_partitions} offers a choice between both the VI and Binder losses, the latter being another well-established option in Bayesian clustering applications.}

The large size of $\Theta$ and its discrete nature make it an ideal example of a setting where point estimation and posterior uncertainty quantification are highly non-trivial, even when Monte Carlo samples from $\Pi$ are available. {\color{black}The popularity of clustering and the success of Bayesian approaches in this domain have attracted recent interest in the topic, resulting in a growing, though still relatively modest, body of literature.} A seminal contribution is due to \citet{wade2018clustering}, who popularized the use of the VI metric in Bayesian clustering and proposed to address point estimation via posterior expected VI minimization, while constructing (approximations of) VI credible balls with prescribed posterior mass for uncertainty quantification. Computational and theoretical extensions of this framework, often but not exclusively focused on point estimation, include \citet{rastelli2018optimal, dahl2022search, nguyen2024summarizing, buch2024bayesian}, among others. An insightful perspective was recently introduced by \citet{balocchi2025understanding}, who emphasized the importance of accounting for multimodality in the posterior distribution over $\Theta$ and proposed summarizing $\Pi$ with the closest $L$-atom mixture of point masses in the Wasserstein-over-VI metric, yielding a fixed number $L\in\mathbb N$ of representative partitions corresponding to the distinct posterior modes.

\section{Why conformal prediction?}\label{sec:why_CP}
Before delving into the details of CBI, a few conceptual remarks are in order. First, as already hinted at in the introduction, our use of conformal prediction ideas operates entirely at the level of the posterior distribution and samples drawn from it, so all guarantees and hypothesis testing statements should be interpreted accordingly. For instance, ``coverage'' always means ``posterior probability content,'' rather than anything tied to sampling variability in the data. Likewise, the quantity $\hat p(\theta;\boldsymbol{\theta}^2)$ defined in Equation~\eqref{eq:conformal_pval} is a $p$-value under a specific null hypothesis on the joint posterior distribution of $\theta$ and $\boldsymbol{\theta}^2$, in the standard sense that a $p$-value $\tilde p$ under a probability law $\mathbb P$ is a $[0,1]$-valued random variable with super-uniform distribution, i.e., satisfying $\mathbb P[\tilde p\leq \lambda]\leq \lambda$ for all $\lambda\in[0,1]$; see \cite{lehmann2005testing}. That is, $\hat p(\theta;\boldsymbol{\theta}^2)$ should not be read as carrying frequentist-style evidence about a true data-generating mechanism; rather, it characterizes features of the posterior distribution that are not immediately apparent from the collected posterior samples because of the complex structure of $\Theta$, in the same spirit as one estimates properties of an unknown distribution from observed data. Our methodology thus summarizes a posterior that is already given and treated as an appropriate parameter model, while it does not speak to the frequentist properties of the underlying Bayesian model (e.g., sampling variability or consistency).

Despite the conformal framing, CBI connects closely with classical Bayesian ideas. The conformal credible set can be related to highest posterior density regions by interpreting $s(\cdot;\boldsymbol{\theta}^1)$ as a pseudo-KDE that approaches a kind of posterior density as the number of samples grows; similarly, the point estimator $\theta_\star$ (Section~\ref{sec:point_estimation}) is closely connected to both expected loss minimizers and MAP estimators. We nonetheless frame the procedure through conformal prediction because this preserves the full generality of the framework on nonstandard parameter spaces $\Theta$, where interpreting $s(\cdot;\boldsymbol{\theta}^1)$ as a genuine KDE may not be feasible or meaningful; treating it instead as a metric-based conformity score keeps it well-defined, useful, and comes with precise finite-sample coverage guarantees for the resulting credible region. The same generality is valuable for the multimodality analysis of Section~\ref{sec:multimodality}, where density may be understood in a purely topological or spatial sense rather than a probabilistic one. The conformal viewpoint also unifies our pseudo-KDE construction with earlier metric-ball-based credible regions for data partitions \citep{wade2018clustering}, enabling a principled comparison between the two once both are seen as instances of the same general framework (see Subsection~\ref{sub:balls} and Example~\ref{ex:credible_reg} below). More broadly, this generality may stimulate future work on alternative conformity scores for other nonstandard spaces---including scores with no natural connection to posterior density or simple asymptotic interpretations, but for which the automatic finite-sample guarantees coming from conformal prediction remain valid. Finally, the conformal construction yields the quantity $\hat p(\theta;\boldsymbol{\theta}^2)$, interpretable either as a $p$-value for a specific null or, heuristically, as a standardized measure of posterior typicality for $\theta$ relative to $\boldsymbol{\theta}^2$. This is a practically useful heuristic for assessing posterior support for a hypothesized value that arises naturally only from the conformal perspective, motivating our emphasis on this framework alongside its many connections to classical Bayesian concepts.

\section{Conformalized Bayesian Inference: preliminaries and point estimation}\label{sec:point_estimation}

Recall the problem we aim to solve: given iid posterior Monte Carlo samples $\boldsymbol{\theta}$, split into training $\boldsymbol{\theta}^1$ and calibration $\boldsymbol{\theta}^2$ subsets, how can we perform point estimation and uncertainty quantification in practice? Following the pipeline from the introduction, the first step is to define a scoring function $\theta \mapsto s(\theta; \boldsymbol{\theta}^1)$ capturing a notion of posterior pseudo-density~\citep{ferraty2006nonparametric}. We now propose a simple, robust solution.

We begin by choosing a metric $\mathcal D$ on $\Theta$, though for practical purposes $\mathcal D$ need only be a meaningful discrepancy measure. In many nonparametric problems, even when $\Theta$ has nonstandard structure, a distance can typically still be defined: the VI metric for partitions, the operator norm for covariance matrices, edit or spectral distances for graphs, Wasserstein distances for mixing measures, and $\mathcal L^p$ distances for densities, among others. Given $\mathcal D$, we define the following scoring rule.
\begin{definition}\label{def:score}
    Let $\theta, \theta' \in \Theta$. The \emph{$\mathcal D$-KDE score} $s(\theta; \boldsymbol{\theta}^1)$ with hyperparameter $\gamma > 0$ is given by
    \begin{equation*}
        s(\theta; \boldsymbol{\theta}^1) := \frac{1}{S}\sum_{t=1}^S \mathcal K(\theta, \theta_t), 
        \qquad 
        \mathcal K(\theta, \theta') := e^{-\gamma \mathcal D(\theta, \theta')}.
    \end{equation*}
\end{definition}

Intuitively, $s(\cdot; \boldsymbol{\theta}^1)$ acts as a KDE score, replacing the Euclidean distance of traditional settings with $\mathcal D$ to accommodate the nonstandard $\Theta$. Its interpretation as a probabilistic density, however, should not be taken literally in general, since $\Theta$ may not admit a reference measure with respect to which $s(\cdot;\boldsymbol{\theta}^1)$ is a valid density. For partitions, where $\Theta$ is large but finite, the counting measure is natural and $s(\cdot;\boldsymbol{\theta}^1)$ can be viewed as the unnormalized density of a $\mathcal D_{VI}$-smoothed version of the empirical posterior $S^{-1}\sum_{t=1}^S\delta_{\theta_t}$. The two limiting regimes of $\gamma$ illustrate this: as $\gamma\to0$, the score converges to $1$ for all $\theta$; as $\gamma\to\infty$, it converges to the relative frequency of $\theta$ in the training set. Thus $s(\cdot;\boldsymbol{\theta}^1)$ uses $\mathcal D$ to encode posterior representativeness beyond raw Monte Carlo frequency, enabling a ``borrowing of typicality'' across nearby samples whereby even a value $\theta\not\in\boldsymbol{\theta}^1$ can receive a relatively high score if it lies close to enough $\theta'\in\boldsymbol{\theta}^1$. We remark that, while other scoring rules are possible, $\mathcal D$-KDE offers a simple, robust baseline, as it relies on a clearly defined, meaningful distance to reduce pseudo-density definition on the complex $\Theta$ to an efficiently implementable KDE computation. In fact, evaluating $\mathcal K(\theta, \theta_t)$ parallelizes easily across samples, and computation can be further sped up by using only a random subset of training samples per calibration score.

At this point, our proposal for a point estimate naturally is
\(
\theta_\star \in \arg\max_{\theta \in \boldsymbol{\theta}^2} s(\theta; \boldsymbol{\theta}^1).
\)
Since $s(\cdot; \boldsymbol{\theta}^1)$ acts as a posterior pseudo-KDE, $\theta_\star$ can be interpreted as a kind of MAP estimator---albeit with due caution, as the posterior may be atomless or lack a proper density, making true MAP estimation ill-defined. When $\Theta$ is discrete (as for partitions), this interpretation is formally justified in the limit $\gamma\to\infty$, where $s(\theta;\boldsymbol{\theta}^1)$ approaches the empirical frequency of $\theta$ in $\boldsymbol{\theta}^1$ and $\theta_\star$ approaches a true empirical MAP estimator; more generally, $\theta_\star$ may be viewed as a $\mathcal D$-smoothed MAP estimator converging to the exact one as $\gamma\to\infty$.

Alternatively, $\theta_\star$ can be seen as a posterior expected loss minimizer restricted to the calibration set, with the loss $\mathcal D$ transformed via $D \mapsto \exp\{-\gamma D\}$ into a pseudo-density score.
$s(\theta;\boldsymbol{\theta}^1)=S^{-1}\sum_{t=1}^S \mathcal K(\theta, \theta_t)\,\approx\,\mathbb E_{\theta'\sim \Pi}\big[\mathcal K(\theta, \theta')\big]$
for large $S$, by the law of large numbers (for iid draws) or the ergodic theorem (with MCMC samples), so $\theta_\star$ is approximately optimal for minimizing the expected loss $-\mathcal K(\theta,\theta')$. We emphasize that, although efficient exhaustive minimization algorithms exist for random partition models \citep{wade2018clustering, dahl2022search} and could be adapted extend our maximization beyond the calibration set, here we restrict ourselves to $\boldsymbol{\theta}^2$ for three reasons: (i) the calibration scores must be computed anyway (and can be parallelized) for the later conformal uncertainty quantification step, making this more convenient than exhaustive search when point estimation is not the primary or only goal; (ii) calibration samples, being posterior draws, are already likely to achieve high scores, making exhaustive search largely redundant; and (iii) the search can be flexibly customized to incorporate user-specified restrictions, e.g., limiting it to partitions with a scientifically interpretable number of clusters.

\begin{example}\label{ex:point_est}

    \begin{figure}
        \centering
        \caption{Gaussian mixture simulated data, analyzed using a PY Gaussian density mixture model.} 
        \begin{subfigure}{0.3\linewidth}
            \centering
            \includegraphics[width=\linewidth]{figures/multimodal_exp_figs/true_partition_multimod}
            \caption{True data clustering.}
            \label{fig:true_partition}
        \end{subfigure}
        \hfill
        \begin{subfigure}{0.38\linewidth}
            \centering
            \includegraphics[width=\linewidth]{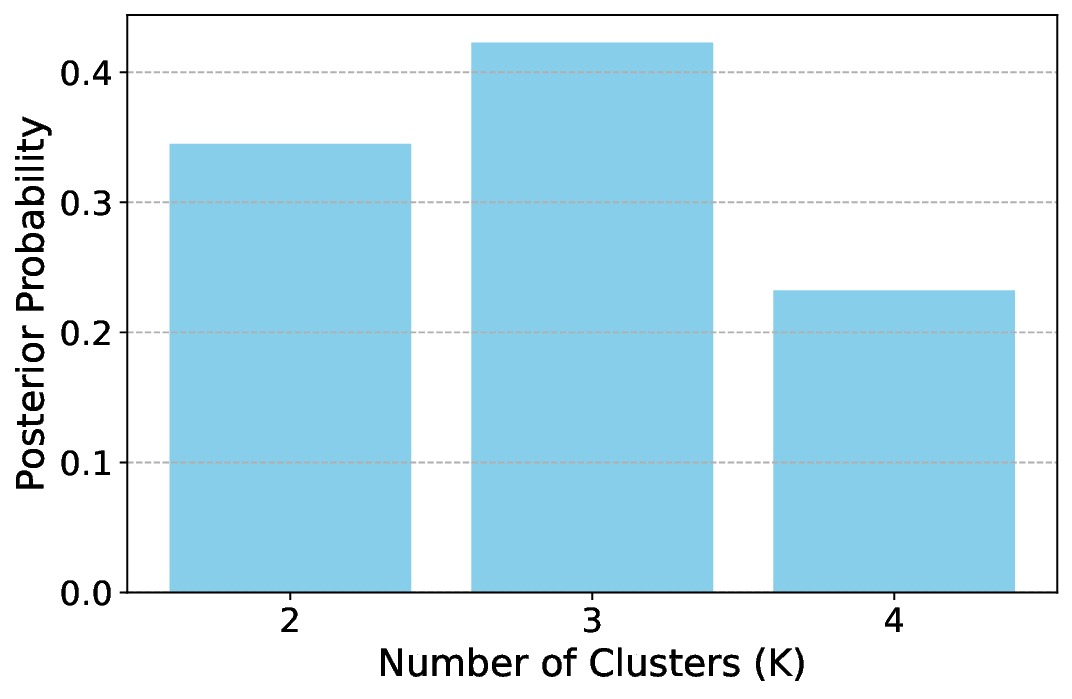}
            \caption{Posterior distribution of $K$.}
            \label{fig:post_K_multimod}
        \end{subfigure}
        \hfill
        \begin{subfigure}{0.3\linewidth}
            \centering
            \includegraphics[width=\linewidth]{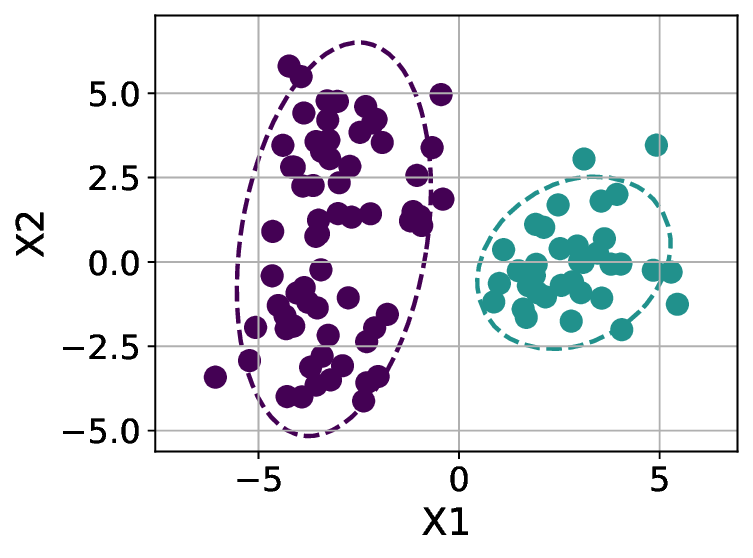}
            \caption{Point estimate.}
            \label{fig:top_partition_kde_vi_multimod}
        \end{subfigure}
    \end{figure}
    We draw 100 samples from a two-dimensional isotropic Gaussian mixture model with three equally weighted components, each with identity covariance matrix scaled by 1.5 and with respective mean vectors $(-3, -3)$, $(-3, 3)$, and $(3, 0)$. The simulation proceeds by first sampling component assignments and then generating the data according to the corresponding Gaussian component. Figure~\ref{fig:true_partition} visualizes the resulting ``true'' partition of the data based on these component assignments. The data are then analyzed by fitting a Gaussian mixture model with a Pitman-Yor process prior on the mixing measure~\citep{pitman1997two,perman1992size,deblasi2013gibbs}, using a concentration parameter of 0.03 and a discount parameter of 0.01. A standard marginal Gibbs sampler is applied via the Python package \texttt{pyrichlet} \citep{selva2025pyrichlet}, and after burn-in and thinning, 5,000 training partitions and 1,000 calibration partitions are retained. 

    Figure~\ref{fig:post_K_multimod} shows the posterior distribution of $K$, the number of distinct clusters to which the 100 data points are assigned. The histogram suggests that the values 2, 3, and 4 all receive substantial posterior mass, indicating the presence of potential uncertainty in the clustering. Figure~\ref{fig:top_partition_kde_vi_multimod} displays the partition corresponding to the $\mathcal D_{VI}$-KDE point estimate (with hyperparameter $\gamma=0.5$). This estimate essentially coincides with the true clustering once the two left-most clusters are collapsed together.\footnote{In the rest of the paper, we will refer to the partitions pictured in Figures~\ref{fig:true_partition} and \ref{fig:top_partition_kde_vi_multimod} as the ``true partition'' and the ``collapsed partition,'' respectively.} Notably, this occurs even though partitions with 3 clusters receive more posterior mass than those with 2 clusters (see Figure~\ref{fig:post_K_multimod}). This behavior may be due to many sampled partitions with 3 clusters being very close to the point estimator (e.g., differing from it by only a few cluster assignments), or because the VI metric inherently favors parsimonious clusterings by weighting larger clusters more heavily; see Equation~\eqref{eq:VI}. A clearer picture will emerge with our upcoming discussion of  uncertainty quantification.
\end{example}

\section{Credible region}\label{sec:credible_region}

While point estimation provides a useful first summary of the posterior distribution, it necessarily omits the uncertainty encoded in the whole posterior distribution. To address this, a key component of our CBI methodology is the construction of a credible posterior subset of $\Theta$ based on the principles of conformal prediction~\citep{vovk2005algorithmic, shafer2008tutorial, lei2018distribution, angelopoulos2023conformal}. Conformal prediction in its basic form allows one to construct a set-valued prediction for the next observation given the first $N$ samples from an exchangeable sequence, with a precisely defined, finite-sample, and distribution-free coverage guarantee. In our context, the crucial step is to interpret Monte Carlo samples as draws from an exchangeable law, and to recast the problem of constructing a posterior credible region as that of predicting a region where the next posterior sample would fall with high probability under the sampling scheme. We now formalize this intuition.

Recall our assumption that the available samples from $\Pi$, and in particular the calibration samples $\boldsymbol{\theta}^2=(\theta_{S+1}, \ldots,\theta_T)$ (for convenience, let $N:=T-S$), are iid. For instance, we refer to {Proposition~\ref{pro:mixing} in Appendix~\ref{sec:proofs}} for a theoretical justification of this assumption, based on thinning, when the original sampler features Markov dependence as in MCMC. Define
\begin{equation}\label{eq:conformal_pval}
    \hat p(\theta;\boldsymbol{\theta}^2) := \frac{|\{\theta'\in\boldsymbol{\theta}^2 : s(\theta;\boldsymbol{\theta}^1)\geq s(\theta';\boldsymbol{\theta}^1)\}| + 1}{N+1}.
\end{equation}
Then, for any error rate $\alpha\in(0,1)$, the corresponding credible region is
\begin{equation}\label{eq:credible_region}
    \mathcal C_{1-\alpha}(\boldsymbol{\theta}^2) := \{\theta\in\Theta : \hat p(\theta;\boldsymbol{\theta}^2)\geq \alpha\}.
\end{equation}
In words, we declare $\theta\in\Theta$ to belong to the credible region $\mathcal C_{1-\alpha}(\boldsymbol{\theta}^2)$ if its $\mathcal D$-KDE score is not too low compared to those of the calibration samples---that is, if it is not too ``unlikely'' under the posterior relative to samples known to be drawn from it. Two small technical details are worth clarifying. First, the set notation in Equations~\eqref{eq:conformal_pval} and \eqref{eq:credible_region} should be understood in the multiset sense, that is, preserving duplicate elements. Accordingly, the constructions of both $\hat p(\cdot;\boldsymbol{\theta}^2)$ and $\mathcal C_{1-\alpha}(\boldsymbol{\theta}^2)$ explicitly account for the possibility of ties in $\boldsymbol{\theta}^2$. Second, we emphasize the dependence of the credible region on the calibration set but not on the training set. This is because, from now on, the latter is treated as fixed (or, equivalently, we condition on its realization), since the coverage guarantees to be discussed do not depend on any particular scoring rule. Moreover, to avoid degenerate situations, from now on we take $N>\alpha^{-1}(1-\alpha)$. The following proposition states precisely the coverage property ensured by $\mathcal C_{1-\alpha}(\boldsymbol{\theta}^2)$.

\begin{proposition}\label{pro:conformal_guarantee}
    The credible set $\mathcal C_{1-\alpha}(\boldsymbol\theta^2)$ satisfies
    \begin{equation}\label{eq:conformal_guarantee}
        \mathbb P_{(\boldsymbol{\theta}^2,\theta)\,\overset{\textnormal{iid}}{\sim}\,\Pi}\left[\theta\in\mathcal C_{1-\alpha}(\boldsymbol\theta^2)\right] \,\equiv\, \mathbb E_{\boldsymbol{\theta}^2\,\overset{\textnormal{iid}}{\sim}\,\Pi}\left[\Pi\left(\mathcal C_{1-\alpha}(\boldsymbol\theta^2)\right)\right] \,\geq\, 1-\alpha.
    \end{equation}
    Furthermore, for any $\delta \in (0,1)$ and $\bar \alpha:=\max(\alpha, 1-\alpha)$, its posterior mass satisfies
    \begin{align}\label{eq:concentration_guarantee}
        \big|\Pi\left(\mathcal C_{1-\alpha}(\boldsymbol\theta^2)\right)  - (1-\alpha)\big| \leq \frac{\bar\alpha}{N} + \left|\frac{k_{\alpha,N}}{N} - \hat{\Pi}_{s,N}\big(s_{(k_{\alpha,N})}^{-}\big)\right| + \sqrt{\frac{1}{2N}\ln\left(\frac{2}{\delta}\right)}
    \end{align}
    with $\mathbb{P}_{\boldsymbol{\theta}^2 \overset{\textnormal{iid}}{\sim} \Pi}$-probability at least $1-\delta$, where $k_{\alpha,N}:=\lceil\alpha(N+1)-1\rceil$, $s_{(k)}$ is the $k$-th order statistic of the calibration scores, while $\hat{\Pi}_{s,N}$ is their empirical CDF. Finally,
    \begin{equation}\label{eq:pvalue_property}
        \forall \lambda\in[0,1], \quad \mathbb P_{(\boldsymbol{\theta}^2,\theta)\,\overset{\textnormal{iid}}{\sim}\,\Pi}\left[\hat p(\theta;\boldsymbol{\theta}^2)\leq \lambda\right]\leq \lambda.
    \end{equation}
\end{proposition}

Proposition~\ref{pro:conformal_guarantee} clarifies the rationale behind adopting $\mathcal C_{1-\alpha}(\boldsymbol\theta^2)$ as a credible region under the posterior $\Pi$ with coverage $(1-\alpha)\times 100\%$. In particular, Equation~\eqref{eq:conformal_guarantee} ensures that the \emph{entire procedure}, i.e., scoring the calibration samples and checking whether an independent draw from $\Pi$ falls within $\mathcal C_{1-\alpha}(\boldsymbol\theta^2)$, succeeds with probability at least $1-\alpha$. While apparently weak, since it does not condition on any specific calibration set $\boldsymbol{\theta}^2$, this marginal coverage guarantee is extremely valuable from an algorithmic perspective, as it ensures ex ante that constructing the set and checking inclusion in this manner yields the desired coverage level. Moreover, given its connection to conformal prediction, this construction has a frequentist flavor: it yields a distribution-free, finite-sample coverage guarantee under $\Pi$ based on samples from it. From a purely subjective Bayesian standpoint, however, this raises no philosophical tension, since the task is to infer properties of the posterior distribution itself, which exists and is fixed, independently of whether the Bayesian model giving rise to it is meant to learn any true, fixed data-generating process. On the other hand, a key advantage of the conformal prediction approach is that the resulting coverage guarantee is not tied to any property of the posterior distribution. In particular, it does not rely on any specific structure of $\Theta$, which in many cases of interest may be non-standard (e.g., lacking any natural vector or order structure), and remains valid regardless of the size or dimension of the posterior support, which in many of our motivating applications are extremely large or infinite.

As Equation~\eqref{eq:conformal_guarantee} shows, the coverage guarantee can also be interpreted as an expected-mass statement: the posterior mass of $\mathcal C_{1-\alpha}(\boldsymbol\theta^2)$ exceeds $1-\alpha$ in expectation, where the randomness arises from the calibration set. Thus, $\mathcal C_{1-\alpha}(\boldsymbol\theta^2)$ is a random subset of $\Theta$ whose expected posterior mass is at least $1-\alpha$. Equation~\eqref{eq:concentration_guarantee} further characterizes the behavior of this random set, showing that, with high probability, its posterior mass concentrates around $1-\alpha$ at rate $N^{-1/2}$ in the calibration size $N$. Note that, unless the posterior distribution of the scores is atomless, the second term on the right-hand side of Equation~\eqref{eq:concentration_guarantee} need not vanish as $N\to\infty$ (while it equals $1/N$ when the score distribution is continuous). This reflects the well-known fact that conformal quantile-based sets achieve coverage at least as high as the nominal level $1-\alpha$, but may not approximate it tightly from above if atoms at specific quantiles exist in the target score distribution.

This notion of coverage may at first appear to differ from the traditional Bayesian analogue, which concerns a fixed, non-random credible region $\mathcal C$ satisfying $\Pi(\mathcal C)\ge 1-\alpha$. Such a region, if computable and sufficiently small, would of course be preferable, as it would provide deterministic coverage under $\Pi$. However, the very complexity of $\Theta$ that motivates our framework typically precludes the computation of any such $\mathcal C$. Moreover, even conventional credible regions on simple parameter spaces derived from Monte Carlo samples are inherently random. Consider, for instance, a one-dimensional parameter $\theta$ whose posterior cannot be computed analytically (e.g., due to lack of conjugacy). If one resorts to Monte Carlo sampling from $\Pi$ and constructs a simple credible interval $[a,b]$ from the empirical upper and lower quantiles, then $[a,b]$ inherits randomness from the sampler, and so does its coverage $\Pi([a,b])$ under the intractable target posterior $\Pi$. This algorithmic randomness in the posterior mass is rarely acknowledged in practice, since credible sets are typically constructed to have exact guaranteed coverage \emph{under the empirical Monte Carlo approximation of $\Pi$}, which is close to the true posterior under a well-designed sampling scheme. Our proposed set $\mathcal C_{1-\alpha}(\boldsymbol{\theta}^2)$ is no different in this respect: by design it achieves deterministic $(1-\alpha)\times 100\%$ coverage under the empirical posterior based on $\boldsymbol{\theta}^2$, and it is random precisely because it depends on the sampled $\boldsymbol{\theta}^2$. When it comes to coverage under the true target posterior $\Pi$, however, a key advantage of the conformal prediction framework is that it delivers precise finite-sample guarantees, both in expectation and with high probability, as in Proposition~\ref{pro:conformal_guarantee}. These guarantees hold for any calibration size $N$ and require no assumption whatsoever on the intractable $\Pi$.

Equation~\eqref{eq:pvalue_property} shows that $\hat p(\theta;\boldsymbol{\theta}^2)$ can be interpreted as a $p$-value for testing the null hypothesis that $\boldsymbol{\theta}^2$ and $\theta$ are drawn iid from $\Pi$, since its null distribution is super-uniform. This enables formal hypothesis testing under the posterior, though of a rather different nature from traditional Bayesian procedures such as tests relying on Bayes factors. The proposed conformal test has a frequentist flavor, as it treats the posterior $\Pi$ as an unknown distribution estimated from Monte Carlo samples, and the conformal $p$-value provides a classical tool to test a nonparametric null hypothesis on the joint distribution of the calibration samples and a hypothetical new draw from $\Pi$. More generally, if one prefers not to attach a strict testing interpretation to $\hat p(\theta;\boldsymbol{\theta}^2)$, this quantity can still be viewed as a normalized measure of the typicality of $\theta$ with respect to $\Pi$: larger values correspond to parameters that rank higher, under the $\mathcal D$-KDE score, than most calibration samples. Even under this heuristic interpretation, the conformal analysis based on $\hat p(\theta;\boldsymbol{\theta}^2)$ offers a more nuanced characterization of the uncertainty encoded in $\Pi$ than a simple point estimate alone could provide.

An additional appealing feature of the procedure is that it may be applied conditionally on any subregion of $\Theta$ with positive posterior mass. This is particularly useful when one wishes to perform inference while fixing certain aspects of the complex parameter. For example, as already mentioned, in clustering applications one may wish to assess posterior uncertainty conditional on partitions with at most three interpretable clusters. In this case, a valid credible region can be constructed from the calibration scores associated with partitions having at most three clusters, with coverage guaranteed relative to the posterior conditioned on the nonzero-$\Pi$-mass event of no more than three clusters.

A few remarks are also in order regarding the iid assumption on the calibration samples $\boldsymbol{\theta}^2$. Indeed, most sampling procedures for the complex models of the kind considered here are implemented through MCMC algorithms, so some degree of dependence across samples is generally expected. A first justification for the iid assumption comes from the widely adopted practice of thinning, whereby dependence is reduced by retaining only a subset of spaced-out samples. {In Appendix~\ref{sec:proofs}}, we formally analyze this strategy and show that, under standard assumptions on the sampler, the law of a sufficiently thinned calibration sample becomes close, in total variation distance, to the iid posterior law. This {additional} result is also formulated in a way that can be linked directly to the marginal and high-probability guarantees stated in Proposition~\ref{pro:conformal_guarantee}; see {Proposition~\ref{pro:mixing}} (and the ensuing discussion) {in Appendix~\ref{sec:proofs}}, as well as \cite{barber2023conformal} for a complementary perspective in the presence of non-exchangeable samples.

More generally, it is important to note that our construction is entirely invariant to the ordering of the samples in $\boldsymbol{\theta}^2$. Consequently, randomly permuting the samples leaves the calibration scores, $p$-values, and credible region unchanged, while making the calibration scores exchangeable. Hence, although the iid assumption leads to particularly clean and interpretable marginal and high-probability coverage guarantees under the posterior law $\Pi$, the next Lemma shows that exact marginal coverage remains possible even while removing the iid assumption, provided one is willing to appropriately modify the notion of posterior coverage under consideration.

\begin{lemma}\label{lem:exchangeable_coverage}
    Let $\widetilde \Pi_{N+1}$ denote the joint distribution of $N+1$ randomly permuted draws $\vartheta_1,\ldots,\vartheta_{N+1}$ from a sampling algorithm that targets $\Pi$ marginally (i.e., such that $\vartheta_t\sim \Pi$ for all $t=1,\ldots,N+1$). Then $\mathbb P_{(\boldsymbol{\theta}^2,\theta)\,\sim\,\widetilde\Pi_{N+1}}
    \left[
        \theta\in\mathcal C_{1-\alpha}(\boldsymbol\theta^2)
    \right]
    \,\geq\,
    1-\alpha.$
\end{lemma}

That is, even if one is not willing to assume iid calibration scores, a form of coverage is nevertheless obtained \emph{under the exchangeable law} of the permuted samples drawn from any (e.g., MCMC) algorithm of choice.

On a final note on Proposition~\ref{pro:conformal_guarantee}, it is well known \citep{angelopoulos2023conformal} that the coverage guarantee in Equation~\eqref{eq:conformal_guarantee} is independent of the choice of scoring rule, as it relies solely on the exchangeability between calibration and new samples. Nevertheless, the practical utility of the resulting credible region depends critically on this choice. Arbitrary scores may yield overly large regions (since even the whole $\Theta$ itself trivially satisfies the guarantee) or exclude high concentration areas if the score misaligns with $\Pi$. Designing an informative score is therefore essential. Our distance-based KDE construction offers a principled solution: it exploits (i) the geometry of $\Theta$ induced by $\mathcal D$, which helps capture the structure of the effective posterior posterior support, and (ii) the inherent concentration of training samples in high-probability regions. In our illustrations, we verify empirically that this choice yields credible regions of reasonable size and practical utility.

\begin{remark}
    The concentration inequality in Equation~\eqref{eq:concentration_guarantee} is of independent interest, and its original proof, based on the Dvoretzky--Kiefer--Wolfowitz (DKW) inequality \citep{dvoretzky1956asymptotic, massart1990tight}, can be found {in Appendix~\ref{sec:proofs}}. It is in the spirit of existing results \citep{vovk2012conditional, sarkar2023post, lei2018distribution} that study the coverage of conformal sets as a random variable conditional on the calibration scores.
\end{remark}

\subsection{Connections to metric balls as credible regions}\label{sub:balls}

Given a discrepancy $\mathcal D$ and a point estimate $\hat\theta\in\Theta$, a natural alternative credible set is the $\mathcal D$-ball around $\hat\theta$ with radius chosen to achieve (empirical) $1-\alpha$ posterior coverage; this was popularized by \cite{wade2018clustering} in the context of random partition models. Interestingly, such credible balls can be recovered within CBI by simply choosing a different scoring rule: for $\hat\theta$ any estimator computed from $\boldsymbol{\theta}^1$ (e.g., the $\mathcal D$-KDE maximizer within $\boldsymbol{\theta}^1$), define, for each calibration draw,
\(
    \tilde s(\theta;\boldsymbol{\theta}^1) := -\mathcal D(\theta,\hat\theta).
\)
Standard conformal prediction results then yield the credible region
\(
    \{\theta\in\Theta : \mathcal D(\theta,\hat\theta)\le \hat q_{1-\alpha}\},
\)
where $\hat q_{1-\alpha}$ is the $n^{-1}\lceil(n+1)(1-\alpha)\rceil$-quantile of the set $\{\mathcal D(\theta,\hat\theta):\theta\in\boldsymbol{\theta}^2\}$. Credible balls thus emerge as a special case of CBI, now equipped with the exact, assumption-free coverage guarantees of Proposition~\ref{pro:conformal_guarantee}.

\begin{figure}
    \centering
    \includegraphics[width=0.9\linewidth]{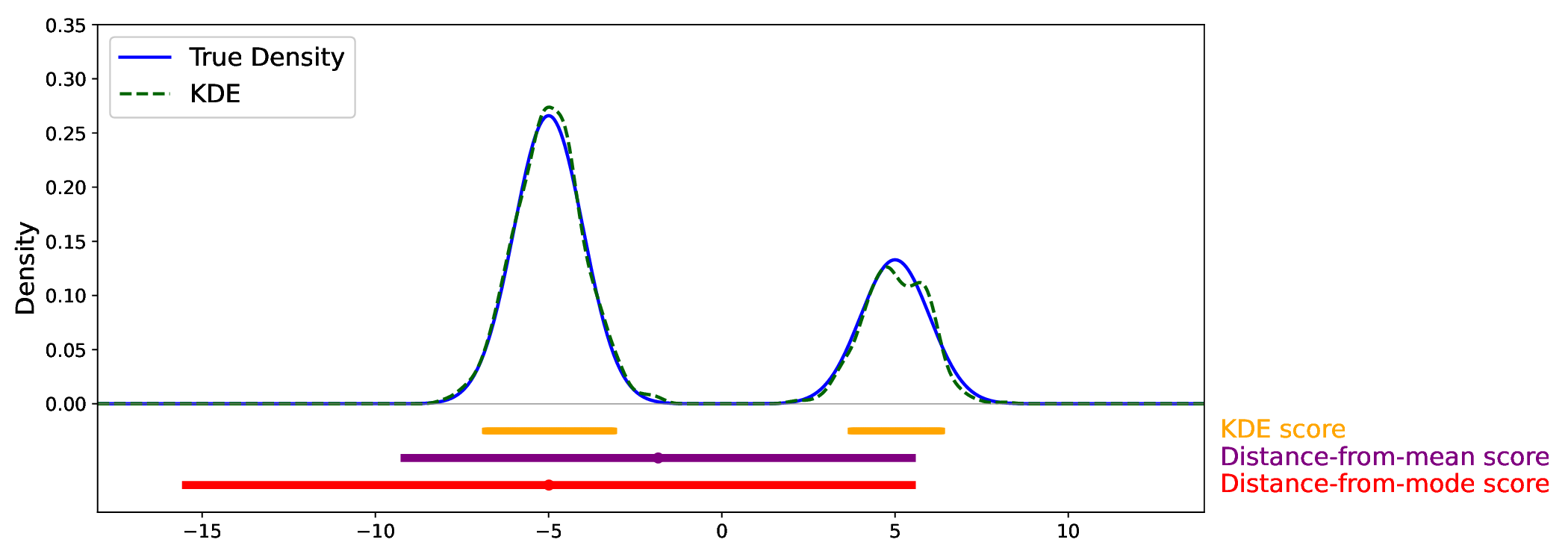}
    \caption{One-dimensional illustration comparing the KDE-based credible set with Euclidean balls centered at point estimates (all with 90\% coverage). The KDE-based set adapts to multimodality, whereas ball-type sets include large low-density regions.}
    \label{fig:interval_illustration.pdf}
\end{figure}

While this observation is conceptually useful, we argue that $\mathcal D$-KDE remains preferable in most applications, unless there is strong reason to believe the posterior is essentially unimodal around $\hat\theta$. Our KDE-based conformity score aggregates distances to all training samples, so a parameter value scores relatively high whenever it is close to \emph{any} high-density region, allowing the conformal set to track several separated such regions without including the low-density ``valleys'' between them. A ball centered at a single $\hat\theta$, by contrast, must expand uniformly in all directions until it captures the desired fraction of calibration draws; if the posterior is multimodal, this can force it through low-density regions separating $\hat\theta$ from secondary modes, and can additionally include large swaths of unrepresentative parameter space that are far from all modes. Figure~\ref{fig:interval_illustration.pdf} illustrates this with a one-dimensional continuous distribution (representing the posterior): credible sets centered at the sample mean or the KDE mode both include the low-density region between modes, with the latter also extending well beyond them, whereas the KDE-based conformal set adapts naturally to the multimodal structure. Example~\ref{ex:credible_reg} below shows that these phenomena persist in the substantially more complex space of data partitions.

\subsection{Connections to highest posterior density credible sets}

Figure~\ref{fig:interval_illustration.pdf} also suggests an interesting connection between our credible region and classical concepts in Bayesian statistics. In particular, we may rewrite
\begin{equation}\label{eq:credible_region_quantile}
\mathcal C_{1-\alpha}(\boldsymbol{\theta}^2) := \left\{\theta\in\Theta : s(\theta;\boldsymbol{\theta}^1)\geq s_{(k)} \right\},
\end{equation}
where $k:=\lceil \alpha(N+1) - 1 \rceil$ and $s_{(k)}$ denotes the $k$th order statistic of the collection $\{s(\theta;\boldsymbol{\theta}^1) : \theta\in\boldsymbol{\theta}^2\}$. Recalling the interpretation of $s(\theta;\boldsymbol{\theta}^1)$ as a posterior pseudo-density function, Equation~\eqref{eq:credible_region_quantile} may therefore be viewed as defining a highest posterior pseudo-density credible set. In practice, when the parameter space is Euclidean, as in Figure~\ref{fig:interval_illustration.pdf}, or more generally when $\Theta$ supports a meaningful notion of posterior density captured by $s(\cdot;\boldsymbol{\theta}^1)$, this interpretation may be taken literally and our method yields a classical estimated highest posterior density region whose accuracy improves as the number of training and calibration samples from $\Pi$ increases. We also refer to \cite{GUPTA2022108496} for a stimulating treatment of conformal prediction, based on a nested set construction, that further highlights these connections.

In more general settings, instead, our conformal procedure should be regarded as simply analogous to highest posterior density credible sets, and in this sense may be viewed as a generalization of them to nonstandard settings in which $\Theta$ and $\Pi$ do not support a natural notion of posterior density. This analogy also helps explain why $\mathcal C_{1-\alpha}(\boldsymbol{\theta}^2)$ is expected to be relatively efficient in terms of size compared to alternatives such as metric balls, since highest density regions (e.g., in standard Euclidean settings) minimize the volume among all regions achieving a prescribed coverage level under the corresponding distribution.

In the partition setting, it is again informative to examine the limiting regime $\gamma\to\infty$. In this case, the score $s(\theta;\boldsymbol{\theta}^1)$ converges to the posterior empirical frequency of $\theta$ in $\boldsymbol{\theta}^1$, and consequently $\mathcal C_{1-\alpha}(\boldsymbol{\theta}^2)$ approximates an empirical highest posterior mass region on the discrete space $\Theta$. This observation further highlights the advantages of using a positive but finite value of $\gamma$. Under the basic highest posterior mass construction, corresponding to $\gamma=\infty$, only partitions explicitly sampled in $\boldsymbol{\theta}^1$ may belong to the credible set. This is problematic, since highly similar partitions that are plausibly supported by the posterior may nevertheless be excluded simply because they were never sampled, a realistic possibility in large spaces that common samplers may fail to explore thoroughly. By contrast, our smoothed score allows the inclusion of nearby partitions that, while never directly observed, lie sufficiently close to a large proportion of the samples in $\boldsymbol{\theta}^1$.

\begin{example}\label{ex:credible_reg}
Consider the simulated data and MCMC output analyzed in Example~\ref{ex:point_est}. Based on the $\mathcal D_{VI}$-KDE calibration scores, we construct the credible set $\mathcal C_{0.9}(\boldsymbol{\theta}^2)$. To assess its properties, we test whether the following partitions belong to it: (i) the true data partition shown in Figure~\ref{fig:true_partition}; (ii) the same (collapsed) partition with the two leftmost clusters merged; (iii) 1,000 randomly generated partitions, each drawn independently by first sampling $K$ from the empirical posterior distribution of the number of clusters (cf. Figure~\ref{fig:post_K_multimod}), and then assigning each observation to one of $K$ clusters uniformly at random.

We find that both partitions in (i) and (ii) are contained in the set. Given the point estimate shown in Figure~\ref{fig:top_partition_kde_vi_multimod}, the inclusion of the collapsed partition is unsurprising, as it closely resembles that estimate. More interestingly, the true partition itself is also included in the set, even though point estimation alone would not have led us to expect this, since the true and collapsed partitions are quite different. This indicates the presence of posterior uncertainty, likely in the form of multimodality, which we will examine further in the next section. Moreover, none of the 1,000 randomly generated partitions fall within the credible region. This supports the view that the credible set does not achieve high coverage simply by encompassing many arbitrary partitions (e.g., those consistent with the marginal posterior on $K$ but otherwise unstructured), but rather by concentrating on representative regions of the posterior.

As a further demonstration of the reasonable size of our credible set, and as an illustration of the advantages of our $\mathcal D_{VI}$-KDE region over $\mathcal D_{VI}$-balls (centered at the $\mathcal D_{VI}$-KDE point estimate, essentially as proposed by \cite{wade2018clustering}; cf. Subsection~\ref{sub:balls}), we examine how both procedures behave on the two partitions displayed in Figure~\ref{fig:test_partitions_results_plot_vi}. The partition on the left can be reached by ``moving away'' from the collapsed partition in the ``opposite direction'' compared to the true partition (recall that both the true and collapsed partitions lie in our credible set, and in Section~\ref{sec:multimodality} below we will show that they practically coincide with the two posterior modes), as it groups into the leftmost cluster of the collapsed partition several observations that clearly belong to the rightmost one. The partition on the right of Figure~\ref{fig:test_partitions_results_plot_vi}, by contrast, lies ``between'' the two modes: it shrinks one of the true leftmost clusters of the true partition by reassigning some of its members to the other left cluster, thereby making the partition closer to the collapsed one.

In both scenarios, our $\mathcal D_{VI}$-KDE conformal $p$-value falls below the $0.1$ inclusion threshold, whereas the $p$-value associated with the credible ball remains above the threshold. This indicates the presence of low-density regions both between and away from the two modes, and while our KDE-based procedure correctly downweights these regions by focusing on the effective support of the posterior, the more crude credible ball unnecessarily includes the two unrepresentative partitions coming from such regions. This behavior is fully consistent with the one-dimensional intuition illustrated earlier in Figure~\ref{fig:interval_illustration.pdf}.

\begin{figure}
    \centering
    \includegraphics[width=0.6\linewidth]{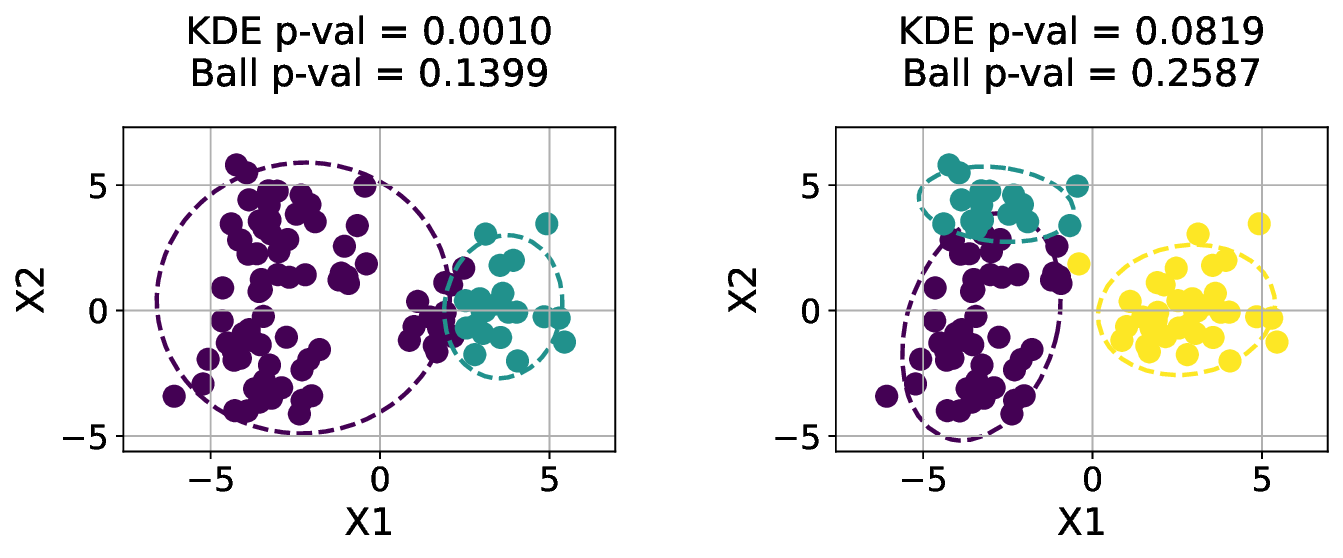}
    \caption{Two partitions tested according to the $\mathcal D_{VI}$-KDE and and $\mathcal D_{VI}$-ball conformal procedures. The partition on the left moves away from the collapsed partition ``in the opposite direction'' with respect to the true partition, while the partition on the right lies ``between'' the true and collapsed partitions.}
    \label{fig:test_partitions_results_plot_vi}
\end{figure}
\end{example}

\section{Multimodality analysis}\label{sec:multimodality}

Having defined our approach to point estimation and credible region construction, we now turn to the last piece of the CBI program, that is to address multimodality in the posterior distribution over $\Theta$. Importantly, because the parameter spaces embodied by $\Theta$ are generally non-Euclidean, we require a notion of multimodality that is operationally meaningful within these structures. In particular, we have defined a posterior pseudo-density estimate $s(\cdot;\boldsymbol{\theta}^1)$ that can be evaluated at any $\theta \in \Theta$, and we can compute a distance $\mathcal D(\theta,\theta')$ between any two parameters $\theta, \theta' \in \Theta$. These two ingredients are sufficient to define a concept of multimodality grounded in the idea of \emph{density-based clustering}. The latter provides a general framework, encompassing many algorithmic instantiations~\citep{ester1996density,kriegel2011density,rodriguez2014clustering}, based on the following principle: given a notion of density on $\Theta$ and a measure of discrepancy between its elements, one partitions the data into mutually exclusive sub-groups such that (a) each sub-group lies in a high-density region whose elements are close to each other, and (b) points from different sub-groups are far apart. A distribution is then said to be multimodal if random samples drawn from it fall into more than one sub-group, each of which identifies a mode of the distribution.

Echoing the discussion in Section~\ref{sec:point_estimation}, it is worth recalling that, also in the context of multimodality analysis, the term ``density'' should not be interpreted in a strict probabilistic sense, although such an interpretation may provide a useful analogy. Indeed, density-based clustering does not necessarily rely on a probabilistic notion of density on $\Theta$, but instead conceptualizes density in a topological or ``spatial'' sense, i.e., as a property of points lying close to many other points \citep{ester1996density}. This perspective has the advantage of avoiding the need to define a probability density on the complex space $\Theta$, requiring instead only a meaningful notion of pairwise distance between points. In standard settings where $\Theta$ is Euclidean, there are obvious connections between spatial and probabilistic notions of density which carry over to density-based clustering applications.

In our setting, the score $s(\cdot;\boldsymbol{\theta}^1)$ may be interpreted as encoding a notion of spatial density complementary to the connections with probability densities discussed in Section~\ref{sec:point_estimation}. This is evident from its dependence on the discrepancy $\mathcal D$ and is further clarified in our upcoming discussion. Consequently, our notion of posterior modes follows from the interpretation of density-based clustering as identifying internally homogeneous yet mutually heterogeneous sub-populations, each corresponding to a dense region of the parameter space that we regard as a posterior mode. To sharpen this intuition, it is again useful to consider the limiting regimes $\gamma\to0$ and $\gamma\to\infty$. In the former case, the score surface becomes uniform, with $s(\theta;\boldsymbol{\theta}^1)\to1$ for all $\theta\in\Theta$, suggesting the presence of a single diffuse mode. In the latter case, $s(\theta;\boldsymbol{\theta}^1)$ approaches the empirical frequency of $\theta$ in the training sample, so that, in discrete spaces such as the space of data partitions, one recovers the classical notion of mode as a frequent value, with multiple modes corresponding to relatively well-separated parameter values having high frequency. Neither extreme is desirable: the former artificially suppresses heterogeneity, while the latter is poorly suited to large spaces such as those considered here, where empirical frequency alone is an inadequate proxy for density because many parameter values may be sampled only once. Introducing additional spatial smoothing through the discrepancy $\mathcal D$ therefore leads to a more practically useful characterization of posterior modes.

Among the many density-based clustering algorithms, we take inspiration from the Density Peak Clustering (DPC) procedure by \citet{rodriguez2014clustering}, which is simple, computationally efficient, and well suited to our setting. In particular, since the construction of the credible region already requires the $N$ $\mathcal D$-KDE calibration scores to build the credible set, this algorithm can be applied with minimal additional computation. The only remaining step is to compute the pairwise distances $\mathcal D(\theta,\theta')$ between all distinct calibration samples $\theta, \theta'$; this has $O(N^2)$ cost, but is parallelizable across each distance computation. Then our procedure, which we term KDE-DPC, proceeds as follows:

\begin{enumerate}
    \item For each $\theta \in \boldsymbol{\theta}^2$, let 
    \(
        \delta(\theta) := \min_{\theta' \in \boldsymbol{\theta}^2 \,:\, s(\theta';\boldsymbol{\theta}^1) > s(\theta;\boldsymbol{\theta}^1)} \mathcal{D}(\theta, \theta'),
    \)
    that is, the minimum distance from $\theta$ to any other parameter with higher density.  
    For the parameter(s) with highest density (corresponding to our point estimate $\theta_\star$), set by default $\delta(\theta_\star) := \max_{\theta' \in \boldsymbol{\theta}^2} \mathcal{D}(\theta_\star, \theta').$

    \item Identify the potential cluster centers as those calibration parameters exhibiting simultaneously high density $s(\theta;\boldsymbol{\theta}^1)$ and large separation $\delta(\theta)$.

    \item Optionally, assign each calibration parameter $\theta$ to the cluster with the nearest center.
\end{enumerate}

We note that, if one were to follow \citet{rodriguez2014clustering} exactly, the density for each calibration sample $\theta$ would instead be estimated as the proportion of samples within a fixed-radius $\mathcal D$-neighborhood of $\theta$. While feasible, this would waste computation, since the KDE scores have already been computed, and it would ignore the information contributed by the training samples $\boldsymbol{\theta}^1$. Nevertheless, the overall approach is retained, as the $\mathcal D$-KDE remains distance-based, with the exponential kernel acting as a smoothed version of a strict neighborhood.

The KDE-DPC procedure exhibits several appealing features. First, as illustrated in Example~\ref{ex:multimodality}, selecting representative parameter values reduces to a simple visual task: inspecting a decision graph plotting, for each calibration sample $\theta$, $\delta(\theta)$ against $s(\theta;\boldsymbol{\theta}^1)$, with modes identified as points standing out in the north-east corner. This has the practical advantage of automatically discovering rather than pre-specifying the number of modes. Second, the posterior weight of each mode is naturally estimated as the proportion of calibration samples assigned to its cluster. Third, the method easily classifies calibration samples as inliers or outliers: north-west points are outliers (low density, far from denser samples), while those in the middle and south-east are inliers (moderate-to-high density, close to a mode), considerably enriching one's understanding of posterior uncertainty. We also note that the highest-scoring sample $\theta_\star$ is assigned an artificially large $\delta(\theta_\star)$ by construction, serving two purposes: it marks $\theta_\star$, the most representative sample, as a mode by default, and it sets a stringent threshold for any other sample to qualify as a mode, reducing spurious detections not well separated from $\theta_\star$.

\begin{example}\label{ex:multimodality}
We continue our analysis of the MCMC samples of random partitions from Examples~\ref{ex:point_est} and \ref{ex:credible_reg}. Figure~\ref{fig:density_peak_decision_graph_vi_multimod} shows the KDE-DPC decision graph, with points colored by number of clusters. Two points (highlighted in red) clearly stand out in the north-east corner, identifying the representative modes shown in Figure~\ref{fig:density_peak_centers_vi_multimod} (ordered by density): one with two clusters, obtained by merging the two leftmost clusters of the true partition (coinciding with $\theta_\star$ from Example~\ref{ex:point_est}), and another nearly identical to the true partition itself. We refer to {Appendix~\ref{sec:unimod_exp}} for a similar experiment where the posterior is instead unimodal. Coloring by number of clusters reveals that, while the posterior assigns non-negligible mass to four-cluster partitions, none emerge as representative modes, but rather all appear either as inliers near one of the two main modes or as outliers in low-density regions. This suggests that, although such configurations receive some posterior support under the fitted PY mixture model, they are likely spurious from the standpoint of a parsimonious description of posterior uncertainty.

\begin{figure}
    \centering
    \caption{Multimodality analysis results (Gaussian mixture data, analyzed with a PY Gaussian density mixture model).}
    \begin{subfigure}{0.35\linewidth}
        \centering
        \includegraphics[width=\linewidth]{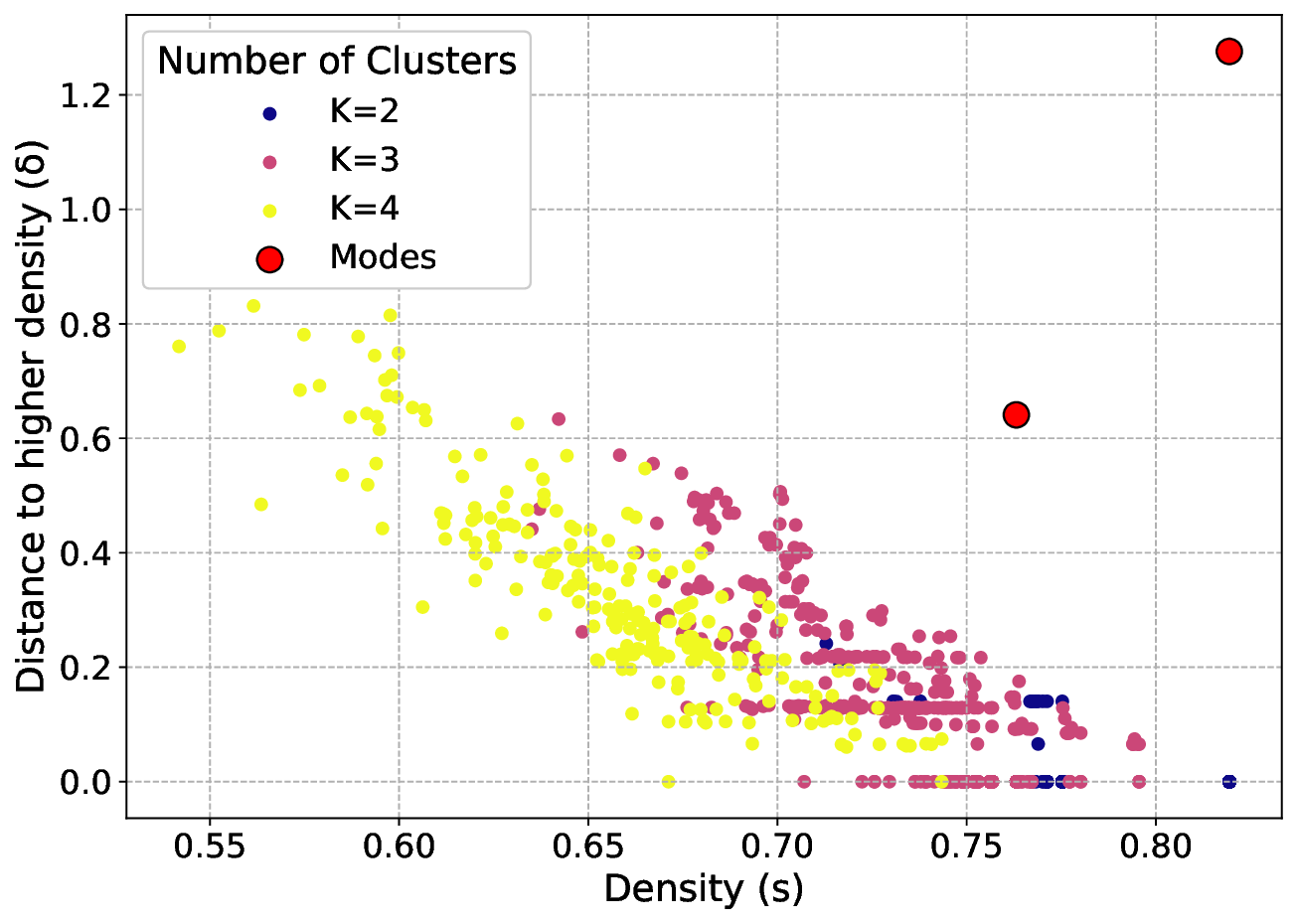}
        \caption{KDE-DPC decision graph.}
        \label{fig:density_peak_decision_graph_vi_multimod}
    \end{subfigure}
    \hfill
    \begin{subfigure}{0.6\linewidth}
        \centering
        \includegraphics[width=\linewidth]{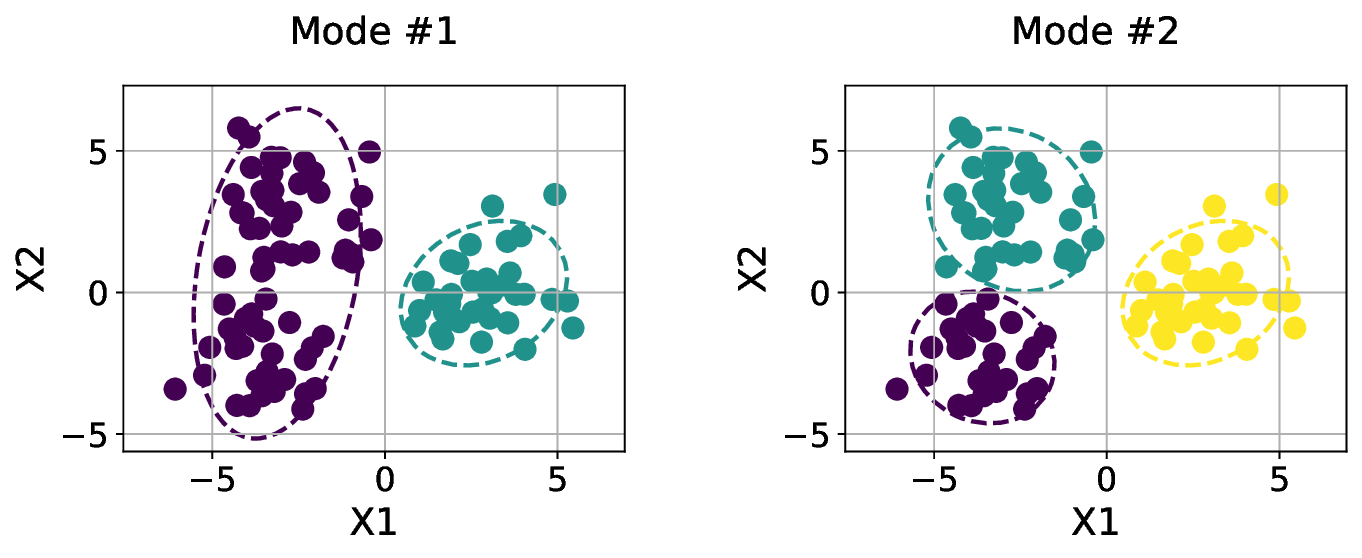}
        \caption{KDE-DPC representative partitions.}
        \label{fig:density_peak_centers_vi_multimod}
    \end{subfigure}
\end{figure}

Our KDE-DPC output is analogous to that of \citet{balocchi2025understanding}, who summarize the posterior over partitions via a Wasserstein-distance-minimizing mixture of point masses; both approaches yield posterior modes capturing separated regions of concentration. Our approach offers a few advantages. First, it needs only density scores (already computed for the credible region) and pairwise VI distances---which are both parallelizable---whereas their $k$-medoids-like algorithm is inherently sequential. Second, their method requires pre-specifying the number of modes, typically selected by re-running the algorithm across candidate values and picking the best improvement in posterior approximation, which is computationally cumbersome and possibly prone to overfitting; our decision graph instead reveals the number of modes automatically. Finally, \citet{balocchi2025understanding} propose insightful post-hoc uncertainty characterization techniques (e.g., plotting the meet between representative partitions, or each point's VI contribution) that apply verbatim to the modes discovered by our KDE-DPC; we refer to \cite{balocchi2025understanding}'s thorough discussion for further details.

\end{example}

\section{Hyperparameter choice}\label{sec:hyperparameter}
Once posterior training and calibration samples have been collected, the only hyperparameter required to run CBI is the positive constant $\gamma$ appearing in the definition of the pseudo-KDE score. In our applications, the default choice $\gamma=0.5$ consistently produces interpretable and practically useful results. Nevertheless, a broader discussion of how to choose this quantity is in order. Importantly, though, we emphasize that the coverage guarantees presented throughout the paper do not depend on the value of $\gamma$ and hold for any choice of this hyperparameter. Rather, $\gamma$ affects (i) the resulting credible set in terms of its size and composition, and (ii) the modes identified through the DPC procedure, as discussed at several points in the paper.

Broadly speaking, $\gamma$ should be viewed as a parameter interpolating continuously between two extreme regimes. with $\gamma=0$, the score $s(\theta;\boldsymbol{\theta}^1)$ is uniform over $\Theta$, leading to a credible region that coincides with the entire parameter space and to a degenerate mode detection step. At the opposite extreme, when $\gamma=\infty$, the score turns into the empirical frequency of $\theta$ within $\boldsymbol{\theta}^1$, thereby reducing the procedure to a pure frequency approach based on empirical MAP estimation and a highest posterior mass credible region. The first regime is undesirable because it artificially suppresses all structural information contained in the posterior samples. The second regime may instead be appropriate when one believes that only parameter values explicitly observed in $\boldsymbol{\theta}^1$ should be regarded as plausible under the posterior; this may be reasonable in some discrete settings, but it is clearly inappropriate for atomless posteriors. A positive but finite value of $\gamma$ interpolates between these extremes, retaining the influence of empirical frequency while simultaneously allowing nearby parameter values to borrow support from one another through the discrepancy $\mathcal D$. In this sense, the procedure induces a form of ``borrowing of typicality'' across neighboring points in the parameter space. The choice of $\gamma$ therefore reflects a tradeoff whose appropriate resolution depends on the application at hand. In our experiments, $\gamma=0.5$ provided a satisfactory balance.

As already discussed, the choice of $\gamma$ is also consequential for multimodality analysis. For instance, in the random partition setting, choosing $\gamma$ excessively large causes partitions in dense regions to receive pseudo-density scores comparable to those assigned to isolated samples in low posterior mass regions, often of order approximately $1/S$ because many partitions are visited only once. Since this regime ignores similarity information encoded by $\mathcal D$, it hinders the identification of meaningful high-density regions. Conversely, taking $\gamma$ too small flattens the score surface and artificially suggests the presence of a single diffuse mode even when distinct regions of the parameter space are separated by low-density areas. Once again, $\gamma$ interpolates between these two extremes, and should therefore be chosen so as to balance oversmoothing against undersmoothing of the posterior geometry.

In the absence of strong prior knowledge allowing one to set $\gamma$ appropriately, and if one has reason to avoid the default $0.5$, we envision the following selection approaches:
\begin{enumerate}
    \item \textbf{Scale matching.} One can set
    \(
        \gamma = c/\bar{\mathcal D},
    \)
    where $\bar{\mathcal D}$ is a summary statistic of the pairwise distances among the samples in $\boldsymbol{\theta}^1$, such as the median or mean pairwise discrepancy, and $c>0$ is a user-specified constant. This choice calibrates the exponential kernel ensuring that the score is invariant to global rescaling of the discrepancy;

    \item \textbf{Leave-one-out cross-validation.} Define
    \[
        \gamma_\star
        \in
        \arg\max_{\gamma>0}
        \sum_{t=1}^S
        \log\!\left(
            \frac{\gamma}{S-1}
            \sum_{t'\neq t}
             e^{-\gamma\,\mathcal D(\theta_t,\theta_{t'})}
        \right).
    \]
    This corresponds to the standard leave-one-out bandwidth selection criterion used in kernel density estimation, adapted to our CBI setting. Notice that, to obtain a meaningful optimization problem, the score is now pre-multiplied by $\gamma$, similarly to a proper KDE and so to discourage too small values of the hyperparameter. Hence, the objective trades off (i) each sample being assigned a high score by the remaining samples, and (ii) values of $\gamma$ not too close to the uniform degenerate case of $\gamma=0$. This optimization problem is inexpensive and can be carried out efficiently over a grid of candidate values for $\gamma$.
\end{enumerate}

\section{Real data applications}\label{sec:real_data}

We are now in a position to illustrate our CBI methodology on two real-world datasets: the classic Galaxy velocities data and ERP functional data.

\subsection{Galaxy velocities data}
 
\begin{figure}
        \centering
        \caption{Galaxy velocities dataset clustering experiment (analyzed with a PY Gaussian density mixture model).}
        \begin{subfigure}{0.32\linewidth}
            \centering
            \includegraphics[width=\linewidth]{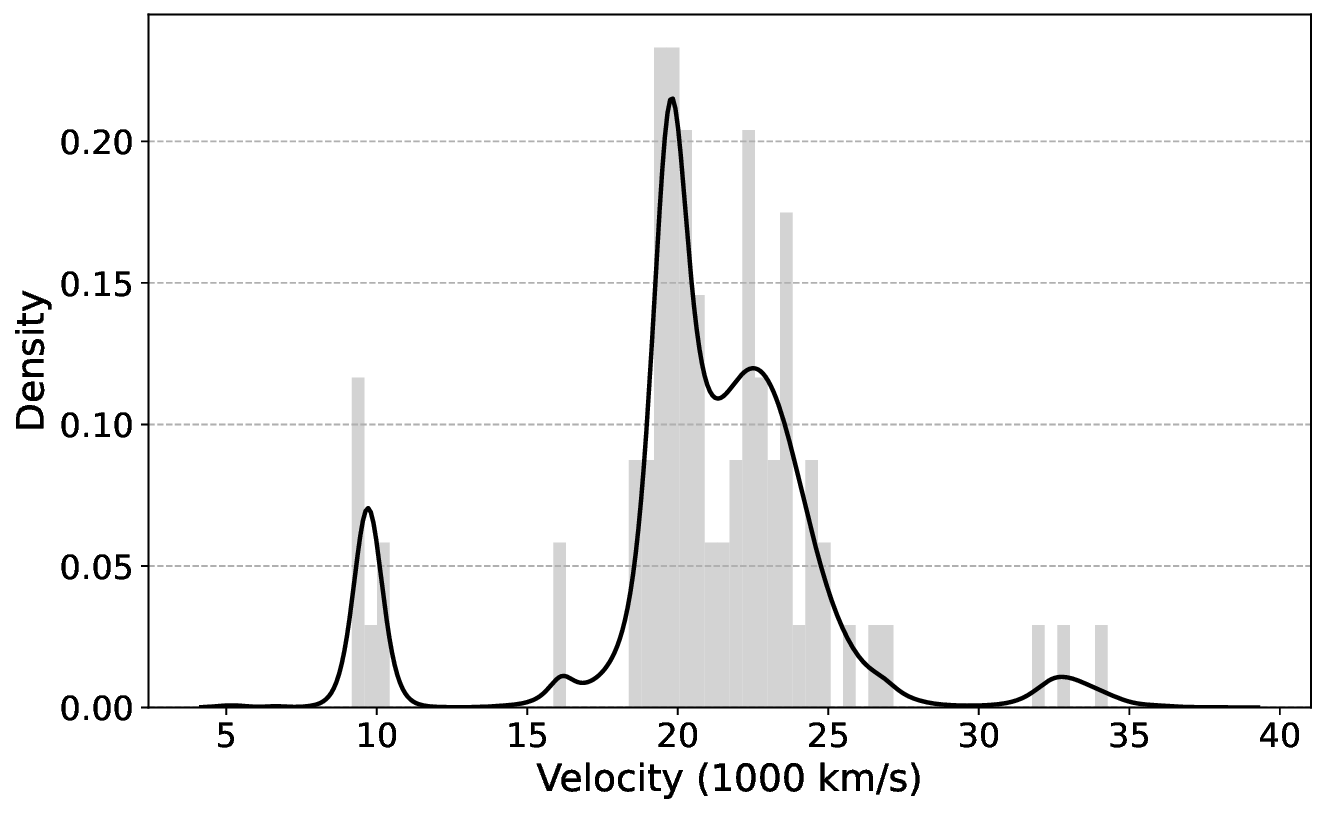}
            \caption{Histogram and posterior predictive density.}
            \label{fig:post_pred_galaxy}
        \end{subfigure}
        \hfill
        \begin{subfigure}{0.32\linewidth}
            \centering
            \includegraphics[width=\linewidth]{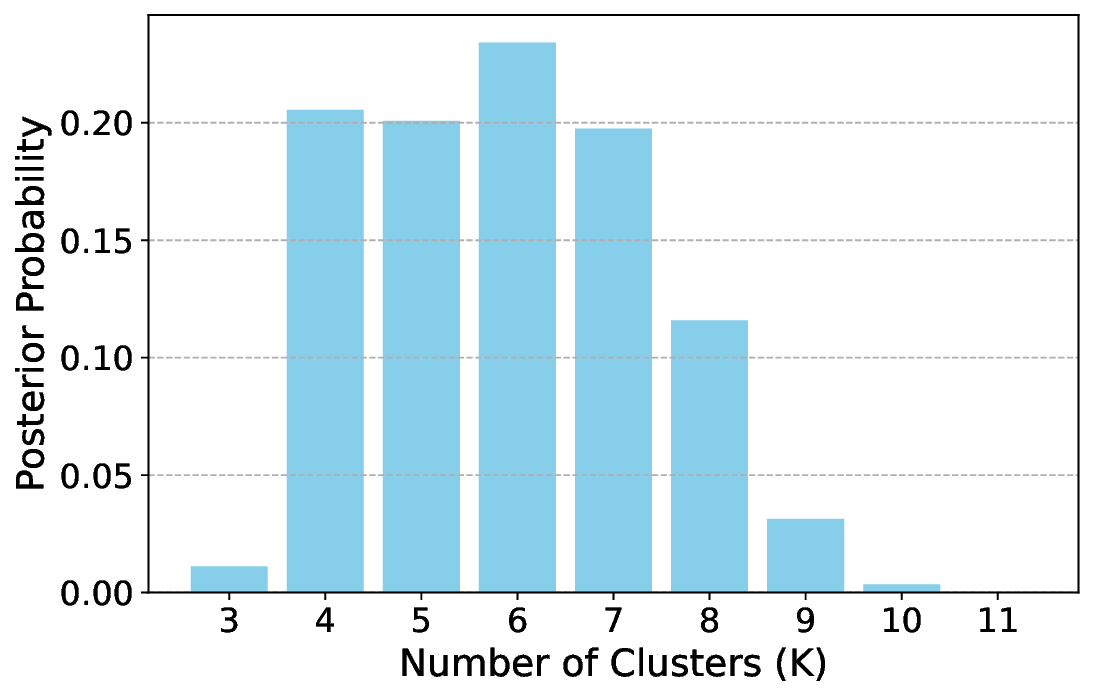}
            \caption{Posterior distribution of $K$.}
            \label{fig:posterior_k_galaxy}
        \end{subfigure}
        \hfill
        \begin{subfigure}{0.25\linewidth}
            \centering
            \includegraphics[width=\linewidth]{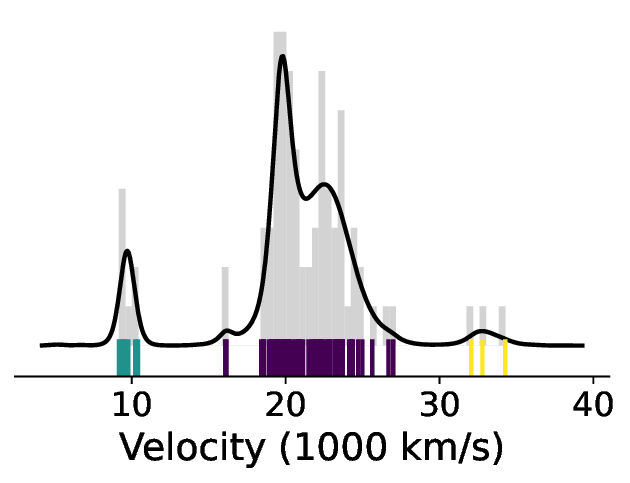}
            \caption{$\mathcal D_{VI}$-KDE point estimator.}
            \label{fig:top_density_kde_galaxy}
        \end{subfigure}
\end{figure}

The Galaxy velocities dataset is a simple benchmark for mixture-based density estimation and clustering, containing velocity measurements (in units of 1,000 km/second) for 82 galaxies observed in six well-separated conic sections of the Corona Borealis region~\citep{roeder1990density}. We analyze this dataset using a Pitman--Yor Gaussian mixture model, fitted via a marginal Gibbs sampler implemented in \texttt{pyrichlet}~\citep{selva2025pyrichlet}, using 5,000 training and 1,000 calibration MCMC samples after appropriate burn-in and thinning. Figure~\ref{fig:post_pred_galaxy} shows a histogram of the data together with the posterior predictive estimate of the density.

\begin{figure}
    \centering
    \caption{Multimodality analysis results for the Galaxy velocities data.} 
    \begin{subfigure}{0.36\linewidth}
        \centering
        \includegraphics[width=\linewidth]{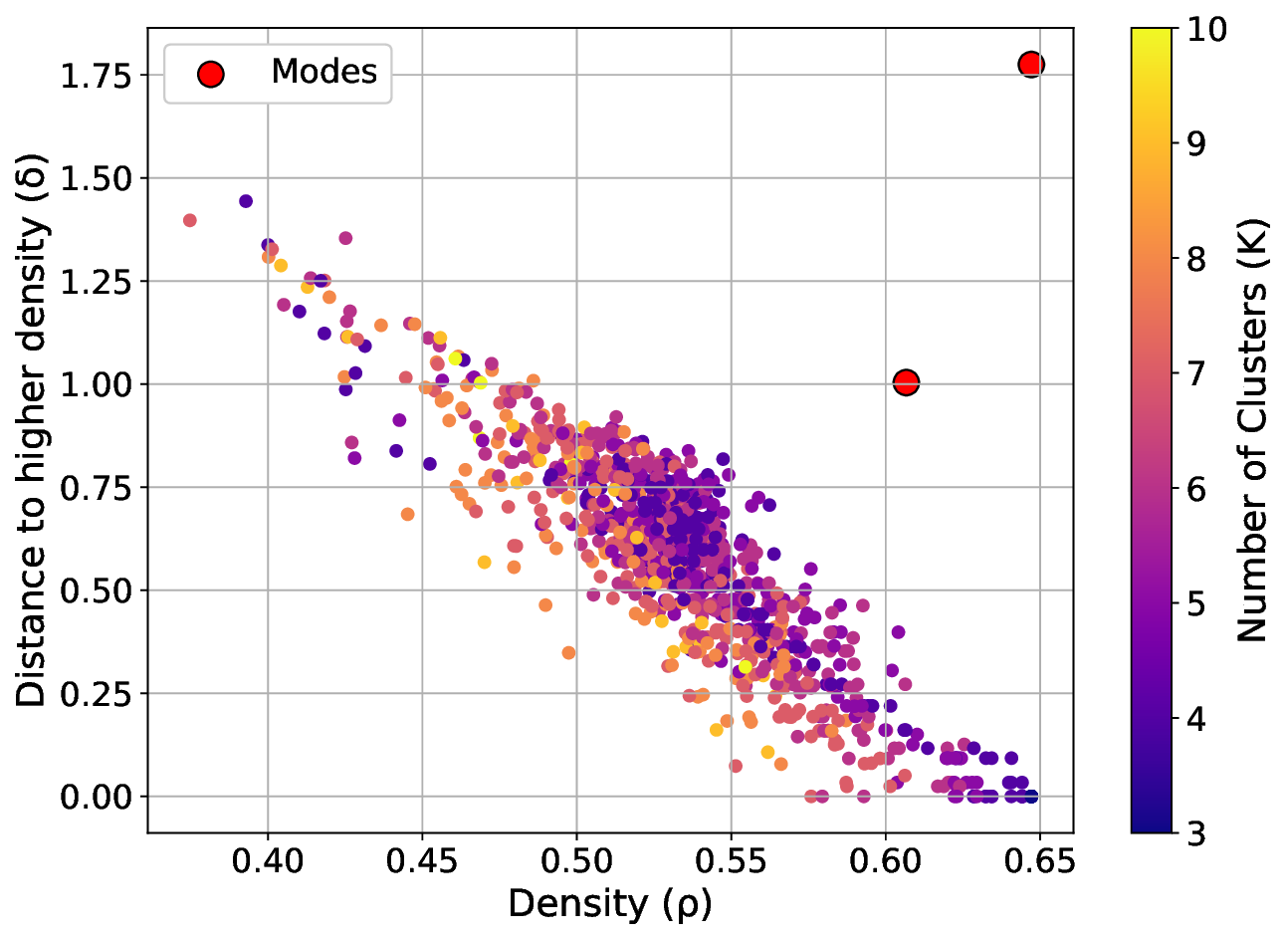}
        \caption{KDE-DPC decision graph.}
        \label{fig:dpc_decision_graph_galaxy}
    \end{subfigure}
    \hfill
    \begin{subfigure}{0.55\linewidth}
        \centering
        \includegraphics[width=\linewidth]{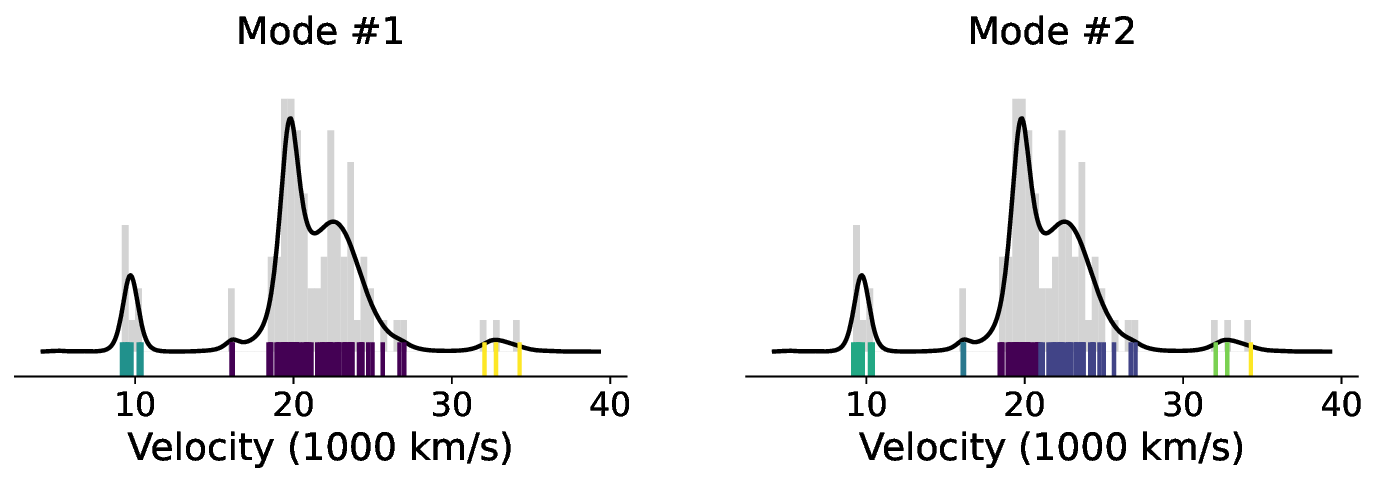}
        \caption{KDE-DPC representative partitions.}
        \label{fig:dpc_centers_galaxy}
    \end{subfigure}
\end{figure}

Figure~\ref{fig:posterior_k_galaxy} visualizes the estimated posterior distribution of the number of clusters in the data, which is spread over a relatively wide range of values, indicating substantial uncertainty. Figure~\ref{fig:top_density_kde_galaxy} displays the $\mathcal{D}_{VI}$-KDE point estimate of the data partition, which reveals three well-separated clusters. At first glance, this may appear inconsistent with the relatively small posterior mass assigned to partitions with three clusters in Figure~\ref{fig:posterior_k_galaxy}; however, our point estimation procedure seeks the partition maximizing the $\mathcal D_{VI}$-KDE score, and since the very design of $\mathcal{D}_{VI}$ favors more parsimonious partitions, this result is in fact expected and possibly desirable, particularly given that many of the sampled partitions (e.g., those with four or five clusters) differ only slightly from the selected three-cluster configuration (see also the point coloring in Figure~\ref{fig:dpc_decision_graph_galaxy}). The substantial posterior uncertainty is further reflected in the $90\%$ conformal credible set, which includes all $k$-means clusterings with $k\in\{3,4,5,6\}$; as in the previous simulation study, a size check against 1,000 randomly generated partitions confirms that none fall inside the conformal set. Figures~\ref{fig:dpc_decision_graph_galaxy} and~\ref{fig:dpc_centers_galaxy} report the results from the DPC-based multimodality analysis. Two well-separated modes are identified, corresponding to the three-cluster point estimate and a six-cluster refinement of it, both plausible given the empirical data distribution and both receiving support from the posterior.

We also use the Galaxy velocities dataset to illustrate several aspects of CBI that were discussed conceptually in the previous sections; full details on all analyses summarized below are provided {in Appendix~\ref{sec:galaxy_further}}. First, as emphasized throughout the paper, the choice of the hyperparameter $\gamma$ may affect the output of CBI. Re-running the entire pipeline over $\gamma\in\{0.05,\,0.1,\,0.5,\,1.0,\,5.0\}$ (with $\gamma=0.5$ our default) leaves the qualitative conclusions unchanged: the same two posterior modes (three and six clusters) are consistently identified, and the two data-driven selection procedures of Section~\ref{sec:hyperparameter} return values of $\gamma$ of the same order of magnitude as our default. All of these findings hint at the method's robustness to the choice of $\gamma$.
 
Moreover, across all considered values of $\gamma$, the relative difference in pseudo-KDE score between the two detected modes remains negligible, so the choice of the three-cluster configuration as the point estimate should not be overinterpreted: it simply reflects the fact that $\mathcal D_{VI}$ explicitly favors larger (and therefore fewer overall) clusters. Repeating the analysis with the Binder loss, which measures the proportion of pairwise coclustering disagreements without penalizing small clusters, instead yields a unique five-cluster mode. We emphasize that, while in a simple, well-specified setting like this there may be little need to penalize small clusters and the Binder loss may return a more intuitive point estimate, an Occam's razor effect such as the one induced by $\mathcal D_{VI}$ is often desirable, since Bayesian nonparametric clustering models are well-known to favor overly fragmented partitions unless carefully calibrated. Even in this scenario, as with the Galaxies dataset, the multimodality analysis may be then viewed as a further robustness layer to explore alternative parameter configurations supported by the posterior.
 
Finally, motivated by the well-known sensitivity of Bayesian clustering procedures to the choice of partition prior \citep{lijoi2007controlling}, we repeat the analysis under a Dirichlet process mixture prior, either calibrated to match the expected number of clusters of the baseline Pitman--Yor specification or, alternatively, misspecified toward a much larger expected number of clusters. Under the calibrated prior, results are broadly similar, with the secondary mode disappearing as the posterior concentrates more  around the primary one. Under the misspecified prior, despite substantial posterior mass on highly fragmented partitions, the point estimate still consists of only six clusters, and the $90\%$ credible region contains all $k$-means partitions with $k\in\{3,4,5,6\}$. This indicates a certain level of robustness to different and possibly misspecified underlying posteriors.

\subsection{ERP data experiment}
 
Our second application of CBI concerns functional data clustering using a specialized Bayesian random partition model. Specifically, we build on the analysis of \citet{song2025repulsive}, who employ a projection Determinantal Point Process repulsive mixture prior~\citep{xu2016bayesian,xie2020bayesian,petralia2012repulsive,beraha2025bayesian,cremaschi2025repulsion} to analyze data from the publicly available ERP CORE project~\citep{kappenman2021erp}, obtaining interpretable clusters of ERP waveforms recorded from 34 young adult participants exposed to certain visual stimuli using the first four functional principal components of the data. We refer the reader to Section~7 of \citet{song2025repulsive} for additional scientific context, and to {Appendix~\ref{sec:erp_further}} for a visualization of the raw data. The employed MCMC algorithm generates posterior samples of the random partition of the 34 subjects, which we use as our input.
 
The KDE-DPC decision graph in {Appendix~\ref{sec:erp_further} (Figure~\ref{fig:pDPP_dpc_decision_graph})} shows that the posterior distribution over partitions is unimodal, and the corresponding $\mathcal D_{VI}$-KDE point estimate, featuring four distinct clusters, coincides with the posterior mode and with the point estimate obtained by~\citet{song2025repulsive} (see their Figure~5) using the SALSO algorithm~\citep{dahl2022search}. As we report in {Appendix~\ref{sec:erp_further}}, these results are robust to different choices of $\gamma$. Moreover, when we select the latter via the two strategies proposed in Section~\ref{sec:hyperparameter}, we obtain values equal or very close to the $0.5$ default used in the rest of the experiment.

By constructing the $90\%$ credible region from the calibration scores, we test two extreme hypotheses. First, we assess whether the posterior supports full heterogeneity, i.e., the trivial partition $\theta_{34}:=(1,2,\ldots,34)$ with 34 clusters. It does not: the corresponding conformal $p$-value $\hat p(\theta_{34};\boldsymbol{\theta}^2)$ equals $0.000999$, indicating that such a configuration is highly atypical under the posterior. This corroborates the finding by~\citet{song2025repulsive} that participant subpopulations exhibit meaningful homogeneity. Next, we test whether complete homogeneity (a single cluster) can be rejected as atypical under the posterior. Interestingly, the one-cluster partition is included in the credible set, with conformal $p$-value equal to $0.140859$, indicating that while it is not highly typical, it cannot be decisively ruled out. A robustness check using 1,000 partitions generated uniformly at random confirms that none are included in the credible region, reassuring of the test power. We nevertheless emphasize that this $p$-value is valid only under the specific null hypothesis that the one-cluster partition is a posterior draw independent of the calibration samples. Even ignoring this formal interpretation, the result should be strictly understood in terms of posterior typicality \emph{under the VI metric}: the $\mathcal D_{VI}$-KDE score blends empirical posterior mass with the VI geometry, implying not that the one-cluster partition is likely (in fact, it never appears among the MCMC samples we use), but that it is not exceedingly atypical relative to posterior samples under the VI metric. More broadly, this null finding provides a good reminder that, while the KDE scoring rule yields a valid credible region with guaranteed posterior coverage by leveraging an informative metric such as $\mathcal D_{VI}$, the results of conformal testing should be interpreted strictly within their intended scope.

\section{Discussion}\label{sec:discussion}

This article has introduced CBI, a simple, robust and efficient pipeline to turn Monte Carlo samples from a Bayesian posterior distribution on a complex parameter space into interpretable insights about a point estimate from that posterior, a credible region based on conformal prediction ideas, and possible posterior multimodalities, all simply requiring a notion of discrepancy between parameters.

A few limitations are worth emphasizing. First, the usefulness of CBI critically depends on the quality of the input Monte Carlo samples: if the sampler has not converged, mixes poorly, or fails to explore disconnected posterior modes, the outputs of CBI will inherit these deficiencies. CBI therefore does not remove the need to ensure good performance of the adopted sampler. Second, although the distance-based KDE score is a simple and robust choice for CBI, some posterior distributions may not be well captured by any readily available distance combined with a simple kernel. This could be addressed by developing more flexible pseudo-density estimators, for instance leveraging deep learning architectures suited to parameter spaces such as graphs. Third, the current scope of CBI is limited to a single posterior distribution, whereas much recent work in Bayesian nonparametrics focuses on dependent priors and posteriors for modeling heterogeneous but related populations~\citep[among many others]{maceachern2000dependent,teh2006hierarchical,rodriguez2008nested,lijoi2014bayesian}. A rigorous treatment of dependent posteriors remains open and would substantially enhance applicability.

These limitations, together with other considerations, leave several research questions open. Further work is needed to study, both theoretically and empirically, the interplay between the posterior sampler, the scoring rule, and CBI itself, and how these jointly affect the size and structure of the conformal credible region. Multi-sample extensions of CBI would enable applications to dependent observation settings, such as assessing homogeneity between posteriors arising from dependent priors. Future research should also investigate CBI on parameter spaces beyond partitions, such as graphs, matrices, and regression functions; we provide preliminary illustrations in {Appendices~\ref{sec:mixmeasure_exp} and~\ref{sec:covariance_exp}} for mixing measures (with $\mathcal D$ as the Wasserstein-1 metric) and covariance matrices (with $\mathcal D$ as the operator norm). Finally, integrating alternative density-based clustering methods could offer a more refined characterization of posterior multimodality, distinguishing whether representative parameters correspond to distinct high-mass regions separated by low-mass areas, or to a single, spread-out region of concentration.

\begin{acks}[Acknowledgments]
The authors thank the Editor-in-Chief, an anonymous Associate Editor, and two anonymous referees at Bayesian Analysis for their detailed comments, which significantly improved the paper. The authors are also grateful to Mario Beraha, Michele Guindani, Peter M\"uller, Khai Nguyen, Pratik Patil, Stephen G. Walker, and to the participants of BAYSM 2026 and of the 2026 ICERM Workshop on Nonparametric Bayesian Inference - Computational Issues for their insightful comments. They especially thank Yunshan Duan, Ziyi Song, and Wenyi Wang for sharing their model outputs used for the real data applications presented in the paper.
\end{acks}

\appendix

\section{Mathematical proofs and further results}\label{sec:proofs}

{In this Appendix}, we present the proofs of {Proposition~\ref{pro:conformal_guarantee}} and {Lemma~\ref{lem:exchangeable_coverage}} in the main body of the article, and present a further result on the theoretical justification for the iid assumption used in {Proposition~\ref{pro:conformal_guarantee}} itself.

\subsection{Proof of \texorpdfstring{{Proposition~\ref{pro:conformal_guarantee}}}{Proposition 5.1}}

{Equations~\eqref{eq:conformal_guarantee} and~\eqref{eq:pvalue_property}} are classic results in conformal prediction. Hence, we omit their proof and refer the reader to classic references such as~\cite{vovk2005algorithmic,angelopoulos2023conformal}. Instead, we prove the concentration inequality in {Equation~\eqref{eq:concentration_guarantee}}.

\textbf{Step 1.}
The inclusion criterion for the credible region is equivalent to $s(\theta; \boldsymbol{\theta}^1) \ge s_{(k)}$, where $s_{(k)}$ is the $k$-th order statistic of the calibration scores and $k := \lceil \alpha(N+1) - 1 \rceil$. Let $\Pi_s(s) := \Pi(\{\theta\in\Theta : s(\theta; \boldsymbol{\theta}^1) \leq s\})$ denote the posterior CDF of the scores. The posterior mass of the random credible set is precisely
$$
\Pi(\mathcal C_{1-\alpha}(\boldsymbol\theta^2)) = 1 - \Pi(\{ \theta : s(\theta; \boldsymbol{\theta}^1) < s_{(k)}\}) = 1 - \Pi_s(s_{(k)}^{-}),
$$
where the minus sign denotes a left-hand limit. The quantity to be bounded is therefore $|\Pi(\mathcal C_{1-\alpha}(\boldsymbol\theta^2)) - (1-\alpha)| = |\alpha - \Pi_s(s_{(k)}^{-})|$.

\textbf{Step 2.}
For any $\delta \in (0,1)$, the DKW inequality~\citep{dvoretzky1956asymptotic, massart1990tight} gives
$$
\mathbb{P}_{\boldsymbol{\theta}^2\,\overset{\textnormal{iid}}{\sim}\,\Pi}\left( \sup_{s \in \mathbb{R}} |\hat{\Pi}_{s,N}(s) - \Pi_s(s)| \leq \epsilon_N(\delta) \right) \geq 1-\delta,
$$
where $\epsilon_N(\delta) = \sqrt{\frac{1}{2N}\ln(\frac{2}{\delta})}$. This implies that, on the same high-probability event, the bound also holds for the left-hand limits: $|\hat{\Pi}_{s,N}(s^{-}) - \Pi_s(s^{-})| \leq \epsilon_N(\delta)$ for all $s$.

\textbf{Step 3.}
The total deviation is bounded using the triangle inequality:
$$
|\alpha - \Pi_s(s_{(k)}^{-})| \leq \left|\alpha - \frac{k}{N}\right| + \left|\frac{k}{N} - \hat{\Pi}_{s,N}(s_{(k)}^{-})\right| + |\hat{\Pi}_{s,N}(s_{(k)}^{-}) - \Pi_s(s_{(k)}^{-})|.
$$
Each of the three terms on the right-hand side is now bounded.
\begin{enumerate}
    \item \textit{Deterministic term:} The definition of $k$ implies $\alpha(N+1)-1 \le k < \alpha(N+1)$, which gives the bound:
    $$
    |\alpha - k/N| \leq \frac{\max(\alpha, 1-\alpha)}{N}.
    $$
    
    \item \textit{Empirical jump size:} The term $|\frac{k}{N} - \hat{\Pi}_{s,N}(s_{(k)}^{-})|$ represents the proportion of samples with scores exactly equal to $s_{(k)}$. This is because $\hat{\Pi}_{s,N}(s_{(k)}^{-}) = \frac{1}{N}|\{t : s(\theta_t; \boldsymbol{\theta}^1) < s_{(k)}\}|$. If $m$ samples have a score of $s_{(k)}$, then there are $k-m$ scores strictly less than $s_{(k)}$, so $\hat{\Pi}_{s,N}(s_{(k)}^{-}) = (k-m)/N$. The term thus equals $m/N$. In the absence of ties ($m=1$), this term is $1/N$, so it vanishes deterministically if the posterior distribution of the scores is continuous.
    
    \item \textit{DKW Error:} From Step 2, with probability at least $1-\delta$, this stochastic term is bounded by:
    $$
    |\hat{\Pi}_{s,N}(s_{(k)}^{-}) - \Pi_s(s_{(k)}^{-})| \leq \epsilon_N(\delta).
    $$
\end{enumerate}

\textbf{Step 4.} Combining the bounds for the three terms yields the final inequality.

\subsection{Proof of \texorpdfstring{{Lemma~\ref{lem:exchangeable_coverage}}}{Lemma 5.2}}

The conclusions of the Lemma trivially hold: $(\boldsymbol{\theta}^2,\theta)\sim\tilde \Pi_{N+1}$ is by construction an exchangeable (finite) sequence, so the classical conformal guarantee in the statement of the Lemma follows immediately \citep[see, e.g.,][]{angelopoulos2023conformal}.

\subsection{Theoretical justification of iid assumption}

Here we provide theoretical justification for the iid assumption on the calibration scores. Recall that we collect $N := T - S$ calibration samples. For simplicity, re-index them as $\theta_0, \theta_M, \ldots, \theta_{M(N-1)}$, where $M$ acts as a thinning parameter determining which samples are included in the calibration set (that is, every $M$th element from the original sample is retained). While the iid assumption is crucial for obtaining a valid credible region under conformal prediction principles, in practice we often have access not to iid samples from $\Pi$, but to Markov-dependent samples produced by an MCMC algorithm. However, such a sampler is typically designed to be \emph{$\Pi$-mixing}, meaning that
\(
    \lim_{t\to \infty}\sup_{\theta_0\in\Theta}\Vert \mathcal L_t(\cdot\mid\theta_0) - \Pi \Vert_{TV}=0,
\)
where $\mathcal L_t(\cdot\mid\theta_0)$ denotes the distribution of $\theta_t$ conditional on the chain starting at $\theta_0$, and $\Vert\cdot\Vert_{TV}$ denotes the total variation (TV) distance between probability measures. Recall that, for two probability measures $\mu$ and $\nu$ on a sample space $\Omega$, the TV distance is defined as
\[
    \Vert \mu - \nu \Vert_{TV}
    := \frac{1}{2}\sup_{A\subseteq\Omega}|\mu(A) - \nu(A)|
    = \frac{1}{2}\sup_{h\in[-1,1]^\Omega}
        \left|\int_\Omega h\,\mathrm d\mu - \int_\Omega h\,\mathrm d\nu\right|.
\]
Under stronger assumptions, one can sometimes establish quantitative convergence, i.e., that the chain is \emph{$\Pi$-mixing at rate $\varepsilon_t$}, meaning that
\(
    \sup_{\theta_0\in\Theta}\Vert \mathcal L_t(\cdot\mid\theta_0) - \Pi \Vert_{TV} \leq \varepsilon_t,
\)
for some decreasing sequence $(\varepsilon_t)_{t\in\mathbb N}$; see \citet{levin2017markov} for a standard and comprehensive treatment.

The following result shows that, as long as the chain is mixing, the joint law of the calibration scores
\(
    (s_0, s_M, \ldots, s_{M(N-1)})
\) computed from a suitably thinned calibration sample
\(
    (\theta_0, \theta_M, \ldots, \theta_{M(N-1)}),
\)
approaches that of a posterior iid vector in TV distance (as $M\to\infty$).

\begin{proposition}\label{pro:mixing}
    Let $M,N\in\mathbb N$, let $\mu$ denote the posterior score distribution, $\mu_{N,M}$ the joint law of $(s_0, s_M, \ldots, s_{M(N-1)})$, and $\mu^{(N)}$ the $N$-fold product of $\mu$. If the Markov chain $(\theta_t)_{t\ge 0}$ is $\Pi$-mixing and started in stationarity ($\theta_0\sim\Pi$), then
    \[
        \lim_{M\to\infty}\left\Vert \mu_{N,M} - \mu^{(N)}\right\Vert_{TV} =0.
    \]
    Moreover, if the chain is $\Pi$-mixing at rate $\varepsilon_t$, then
    \begin{equation}\label{eq:TV_bound}
        \left\Vert \mu_{N,M} - \mu^{(N)}\right\Vert_{TV}\leq (N-1)\varepsilon_{M}.
    \end{equation}
\end{proposition}

In practice, Proposition~\ref{pro:mixing} ensures that, even if the sampled parameters are Markov dependent,\footnote{Notice that, in our random partition application, the MCMC algorithm operates on a higher-dimensional space than $\Theta$ (i.e., including mixture parameters in addition to the data partition). The resulting chain is then Markov on this larger space, but not necessarily when projected on $\Theta$. This does not affect Proposition~\ref{pro:mixing}, since the scores can be viewed as functions from the higher-dimensional state space to the real line. For simplicity, however, we state the result assuming the chain is defined directly on $\Theta$.} appropriate thinning allows one to treat them as effectively iid. In fact, if the chain is $\Pi$-mixing at rate $\varepsilon_t$, one can even choose the spacing $M$ to achieve a desired coverage level in light of {Proposition~\ref{pro:conformal_guarantee}}: by the definition of the TV distance in terms of bounded test functions, Proposition~\ref{pro:mixing} implies that selecting $M$ large enough ensures an expected value of $\Pi(\mathcal C_{1-\alpha}(\boldsymbol{\theta}^2))$ close to $1-\alpha$ (as $\boldsymbol{\theta}^2\mapsto\Pi(\mathcal C_{1-\alpha}(\boldsymbol{\theta}^2))$ is bounded by 1). Similarly, using the set-based definition of TV distance, one can tune $M$ to obtain an approximate high-probability coverage statement.

We conclude this discussion by noting that the linear dependence on $N$ of the bound in Equation~\eqref{eq:TV_bound} arises from a rather crude argument that treats the calibration samples as ordered and thus Markov dependent. In practice, however, we have already shown that the scores may be seen as exchangeable for the purposes of building $\mathcal C_{1-\alpha}(\boldsymbol{\theta}^2)$. From this viewpoint, a larger $N$ can even improve the TV bound, since shuffling more samples tends to reduce dependence more effectively. We do not pursue this refinement further, as it adds little methodological insight beyond the main point that, after appropriate thinning and burn-in, the calibration samples can be safely regarded as iid draws from the target posterior distribution.

\begin{proof}

We show a proof of the second, quantitative statement, which readily implies a proof of the first asymptotic property. The proof proceeds via the dual representation of the total variation distance:
\begin{equation*}
    \left\Vert \mu_{N,M} - \mu^{(N)} \right\Vert_{TV} = \frac{1}{2} \sup_{h \in [-1,1]^{\mathbb{R}^N}} \left| \mathbb{E}[h(s)] - \mathbb{E}[h(s')] \right|,
\end{equation*}
where $s = (s_0, s_M, \ldots, s_{M(N-1)})$ is the vector of scores derived from the thinned Markov chain, with $s_{Mk} = f(\theta_{Mk})$ for a measurable scoring function $f$, and $s'=(s'_0, \dots, s'_{N-1})$ is a vector of scores derived from an independent and identically distributed sequence of parameters $(\theta'_0, \dots, \theta'_{N-1})$ from $\Pi$.

For $k=0, \ldots, N-1$, define a hybrid sequence of parameters $\vec{\theta}^{(k)} = (\theta^{(k)}_0, \ldots, \theta^{(k)}_{M(N-1)})$. The law of this sequence is constructed as follows: the initial state $\theta_0$ is drawn from $\Pi$; for $j=1, \ldots, k$, the parameter $\theta_{Mj}$ is drawn from the law given by the $M$-step Markov transition kernel $\mathcal L_M(\cdot\mid\theta_{M(j-1)})$; for the remaining steps, $j=k+1, \ldots, N-1$, the parameters $\theta_{Mj}$ are drawn independently from $\Pi$. Let $E_k[ht]$ be the expectation of $h$ applied to the scores derived from $\vec{\theta}^{(k)}$. Note that $E_{N-1}[ht] = \mathbb{E}[h(s)]$ and $E_0[ht] = \mathbb{E}[h(s')]$. Then, telescoping, we get
\begin{equation*}
    \mathbb{E}[h(s)] - \mathbb{E}[h(s')] = E_{N-1}[ht] - E_0[ht] = \sum_{k=1}^{N-1} \left( E_k[ht] - E_{k-1}[ht] \right).
\end{equation*}
Consider a single term $\Delta_k = E_k[ht] - E_{k-1}[ht]$. The laws generating these two expectations differ only in the distribution of the $k$-th parameter, $\theta_{Mk}$. Let $\vec{\theta}_{<k} = (\theta_0, \dots, \theta_{M(k-1)})$ and define the function $h_k$ by taking the conditional expectation of $h$ over the subsequent independent parameters:
\begin{align*}
    &h_k(x; \vec{\theta}_{<k}) \\
    &= \mathbb{E}_{\theta'_{M(k+1)}, \dots, \theta'_{M(N-1)} \overset{\text{iid}}{\sim} \Pi} \left[ h\left(f(\theta_0), \dots, f(\theta_{M(k-1)}), f(x), f(\theta'_{M(k+1)}), \dots, f(\theta'_{M(N-1)})\right) \right].
\end{align*}
Since $h \in [-1,1]^{\mathbb{R}^N}$, it follows that $h_k: \Theta \to [-1,1]$.
The difference $\Delta_k$ can then be expressed as:
\begin{equation*}
    \Delta_k = \mathbb{E}_{\vec{\theta}_{<k}} \left[ \mathbb{E}_{\theta_{Mk} \sim \mathcal{L}_M(\cdot\mid\theta_{M(k-1)})} [h_k(\theta_{Mk}; \vec{\theta}_{<k})] - \mathbb{E}_{\theta_{Mk} \sim \Pi} [h_k(\theta_{Mk}; \vec{\theta}_{<k})] \right].
\end{equation*}
For any fixed history $\vec{\theta}_{<k}$, the inner term is a difference of expectations bounded by the TV distance between the corresponding measures:
\begin{equation*}
    \sup_{\theta_{M(k-1)}}\left| \mathbb{E}_{\mathcal{L}_M(\cdot|\theta_{M(k-1)})} [h_k] - \mathbb{E}_{\Pi} [h_k] \right| \leq 2\sup_{\theta_{M(k-1)}} \left\Vert \mathcal{L}_M(\cdot\mid\theta_{M(k-1)}) - \Pi \right\Vert_{TV} \leq 2 \varepsilon_M.
\end{equation*}
Since this bound does not depend on the history, taking the outer expectation yields $|\Delta_k| \leq 2\varepsilon_M$.
Summing the bounds for each term in the telescoping series gives:
\begin{equation*}
    \left| \mathbb{E}[h(s)] - \mathbb{E}[h(s')] \right| \leq \sum_{k=1}^{N-1} |\Delta_k| \leq \sum_{k=1}^{N-1} 2\varepsilon_M = 2(N-1)\varepsilon_M.
\end{equation*}
Substituting this into the dual formulation for the total variation distance completes the proof:
\begin{equation*}
    \left\Vert \mu_{N,M} - \mu^{(N)} \right\Vert_{TV} = \frac{1}{2} \sup_{h} \left| \mathbb{E}[h(s)] - \mathbb{E}[h(s')] \right| \leq \frac{1}{2} \cdot 2(N-1)\varepsilon_M = (N-1)\varepsilon_M.
\end{equation*}
\end{proof}

\section{Further details on main-text real-data experiments}\label{sec:real_data_further}
 
{This Appendix} collects further details, sensitivity analyses, and figures supporting the two real-data experiments presented in {Section~\ref{sec:real_data}}, namely the Galaxy velocities dataset ({Appendix~\ref{sec:galaxy_further}}) and the ERP functional data ({Appendix~\ref{sec:erp_further}}).
 
\subsection{Galaxy velocities data}\label{sec:galaxy_further}
 
We present a set of sensitivity analyses using the Galaxy velocities dataset ($n = 82$ observations) to explore how the behavior of CBI is influenced by the choice of the hyperparameter $\gamma$, the choice of partition discrepancy  $\mathcal{D}$, and the specification of the random partition prior.
 
\subsubsection{Hyperparameter selection}
To study the impact of the pseudo-KDE smoothing parameter, we re-run the entire CBI pipeline on the baseline Pitman--Yor mixture posterior MCMC samples over the grid $\gamma \in \{0.05, 0.1, 0.5, 1.0, 5.0\}$, where $\gamma = 0.5$ serves as the default choice. Using the reference modes identified at $\gamma = 0.5$ (Mode \#1 with $K = 3$ clusters and Mode \#2 with $K = 6$ clusters), we evaluate their conformal $p$-values across the grid. As summarized in Table~\ref{tab:multigamma_summary}, the qualitative conclusions remain stable across all considered values of $\gamma$. Both posterior modes are consistently captured inside the conformal credible set, with the $p$-value for Mode \#1 remaining exactly $1$ and the $p$-value for Mode \#2 remaining stable (ranging from $0.9451$ at small $\gamma$ values down to $0.9191$ at $\gamma = 5.0$). 
 
\begin{table}[ht]
\centering
\caption{Conformal $p$-values for the two reference modes across different values of $\gamma$. Both modes are accepted into the credible region across all configurations ($\alpha = 0.1$).}
\label{tab:multigamma_summary}
\begin{tabular}{cccc}
\hline
$\gamma$ & $p(\text{Mode \#1}, K=3)$ & $p(\text{Mode \#2}, K=6)$ & Number of Detected Modes \\ \hline
0.05     & 1.0000                     & 0.9451                     & 2                        \\
0.10     & 1.0000                     & 0.9451                     & 2                        \\
0.50     & 1.0000                     & 0.9451                     & 2                        \\
1.00     & 1.0000                     & 0.9411                     & 2                        \\
5.00     & 1.0000                     & 0.9191                     & 2                        \\ \hline
\end{tabular}
\end{table}
 
We also implement the two data-driven tuning strategies detailed in {Section~\ref{sec:hyperparameter}}:
\begin{itemize}
    \item \textbf{Strategy 1 (Scale Matching):} the pairwise discrepancies computed over the $5,000$ training partition samples yield a mean discrepancy of $\bar{\mathcal{D}} = 1.2761$ and a median of $\bar{\mathcal{D}} = 1.2959$. Setting $\gamma = 1/\bar{\mathcal{D}}$ produces $\gamma \approx 0.7837$ (mean-based) and $\gamma \approx 0.7717$ (median-based), which matches the order of magnitude of our default choice.
    \item \textbf{Strategy 2 (Leave-One-Out Cross-Validation):} evaluating the LOO cross-validation score over the grid selects $\gamma_\star = 1.0$ as the optimal bandwidth, a value comparable to our default choice of $\gamma = 0.5$.
\end{itemize}
 
\begin{figure}
    \centering
    \includegraphics[width=0.85\textwidth]{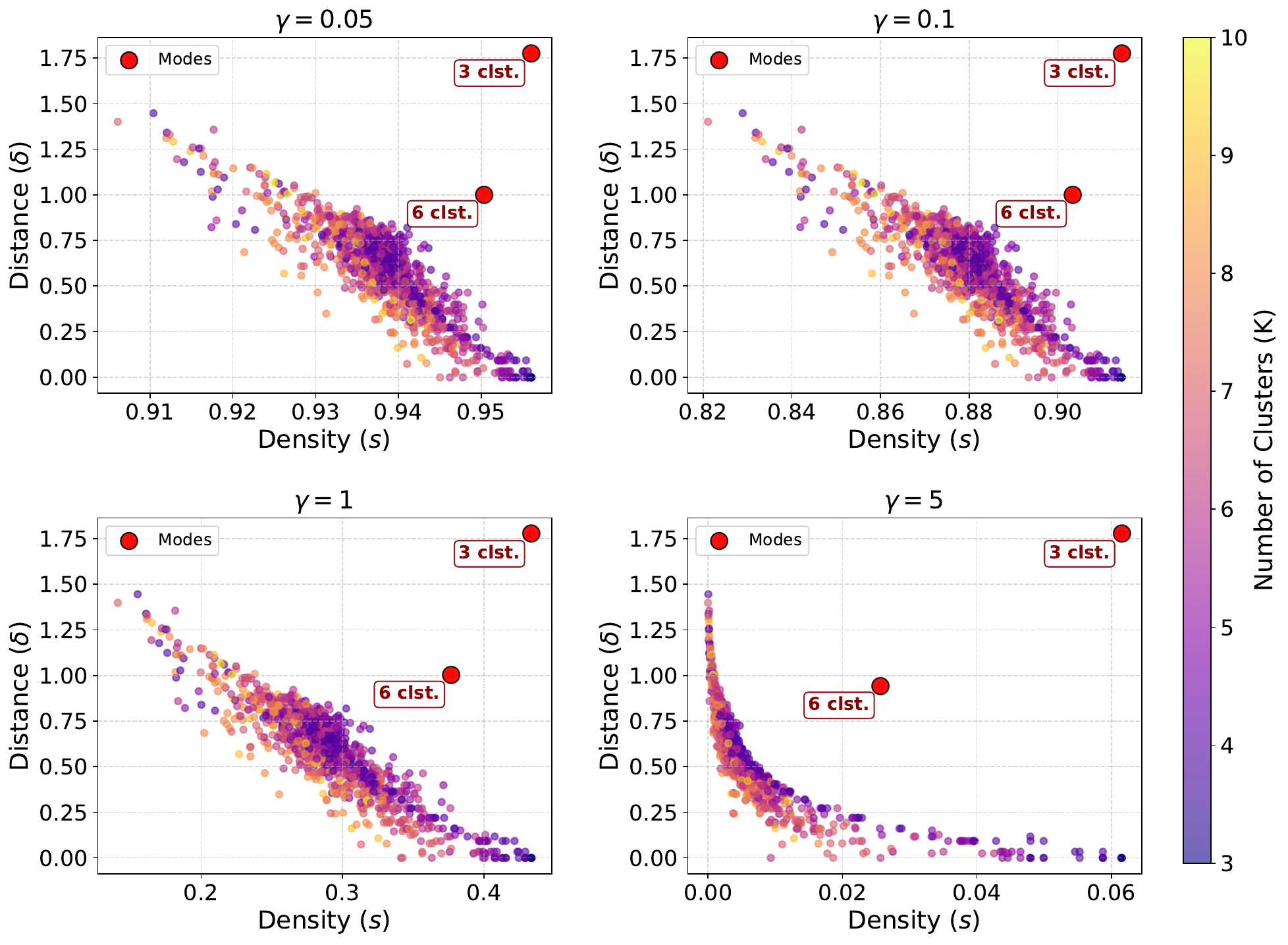}
    \caption{KDE-DPC decision graphs across $\gamma \in \{0.05, 0.1, 1.0, 5.0\}$.}
    \label{fig:dpc_multigamma}
\end{figure}

Finally, the combined DPC decision graphs across the $\gamma$ grid are shown in Figure~\ref{fig:dpc_multigamma}. The results remain qualitatively unchanged across $\gamma$ values.
 
\subsubsection{Binder loss}
\begin{figure}
    \centering
    \begin{subfigure}{0.48\textwidth}
        \centering
        \includegraphics[width=\linewidth]{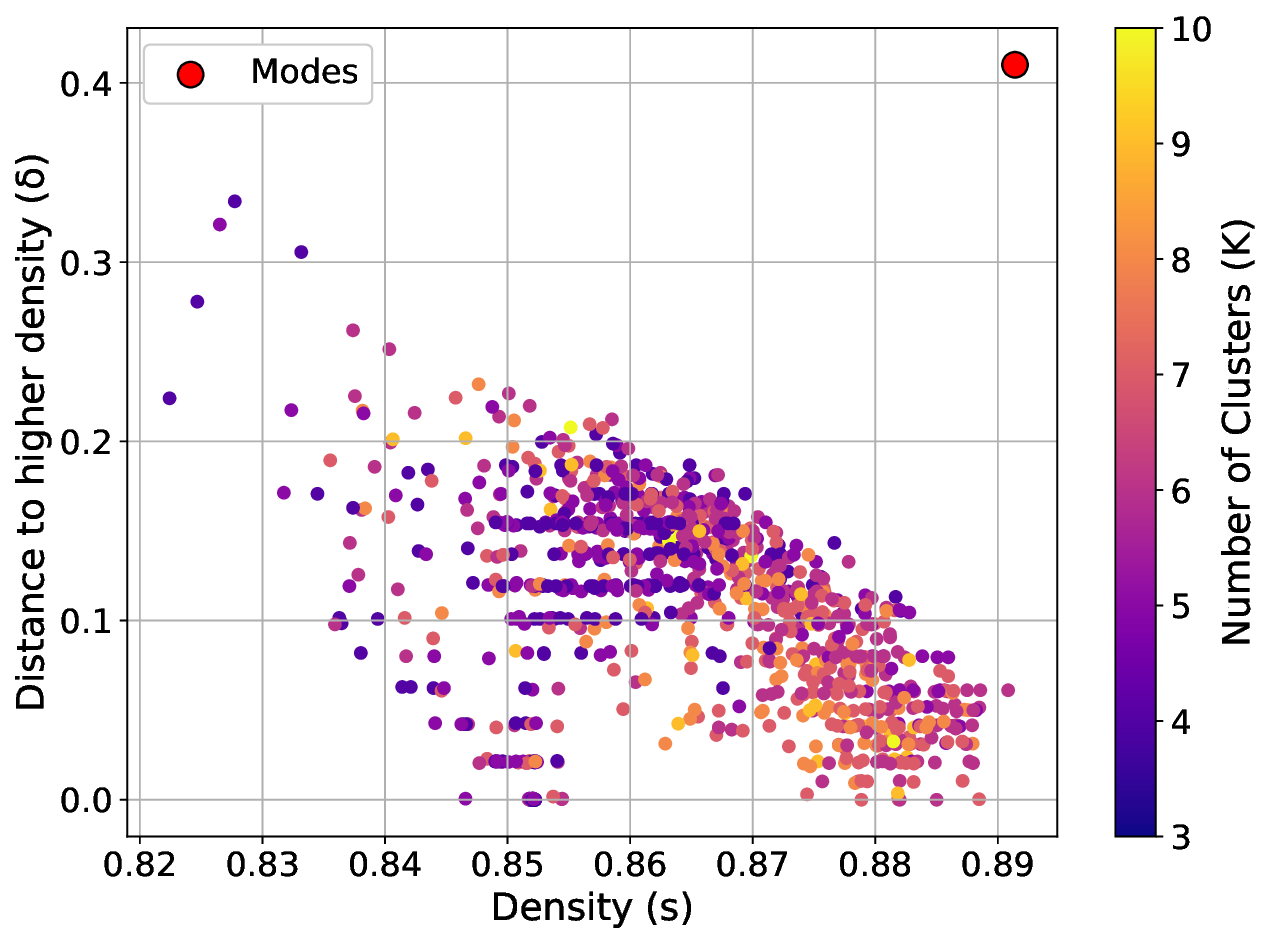}
        \caption{KDE-DPC decision graph.}
    \end{subfigure}
    \hfill
    \begin{subfigure}{0.48\textwidth}
        \centering
        \includegraphics[width=\linewidth]{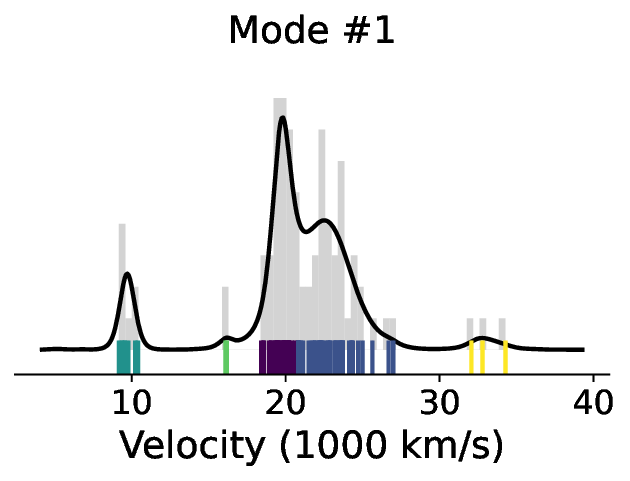}
        \caption{Identified 5-cluster mode.}
    \end{subfigure}
    \caption{Multimodality analysis under the Binder loss.}
    \label{fig:dpc_binder}
\end{figure}

To illustrate the behavior of CBI under alternative discrepancy measures, we replace $\mathcal{D}_{VI}$ for the Binder loss\begin{equation*}
    \mathcal D_{B}(\theta, \theta')
    :=
    \frac{2}{n(n-1)}
    \sum_{i < j}
    \left|
        1(c_i = c_j) - 1(c'_i = c'_j)
    \right|,
\end{equation*}
which measures the proportion of pairwise co-clustering disagreements and does not explicitly penalize small clusters. The DPC-based multimodality analysis method now identifies a single posterior mode corresponding to a 5-cluster partition; see Figure~\ref{fig:dpc_binder}. Moreover, we test standard $k$-means partitions against this pseudo-KDE to evaluate membership in the $90\%$ conformal region ($\alpha = 0.1$). The resulting $p$-values are:
\begin{itemize}
    \item $k$-means ($k=3$): $p\text{-value} = 0.0110$;
    \item $k$-means ($k=4$): $p\text{-value} = 0.7962$;
    \item $k$-means ($k=5$): $p\text{-value} = 0.3357$;
    \item $k$-means ($k=6$): $p\text{-value} = 0.4505$.
\end{itemize}
In contrast to the $\mathcal{D}_{VI}$ metric, which includes the parsimonious 3-cluster partition in the credible set and picks it as a point estimate, the Binder loss shifts the credible set toward larger cluster sizes ($k \in \{4, 5, 6\}$). The posterior is found to be unimodal and with a visually plausible five-cluster point estimate (Figure~\ref{fig:dpc_binder}). The usual size diagnostic is also satisfied, as $0$ out of $1,000$ random partitions sampled from the posterior distribution of $K$ fall inside the conformal set.
 
\subsubsection{Robustness to the choice of random partition prior}
 
Lastly, we inspect the sensitivity of CBI to the choice of the random partition prior. Specifically, we consider two alternative scenarios: (i) a Dirichlet process mixture prior calibrated so that the expected number of clusters among the $82$ observations matches that of the baseline Pitman--Yor specification, and (ii) a Dirichlet process mixture prior with a deliberately misspecified prior expectation corresponding to approximately $16$ clusters. 
 
For a random partition model governed by a Pitman-Yor or Dirichlet process prior over a sample size of $n$ observations, the expected number of unique clusters $\mathbb{E}[K_n]$ can be calculated analytically. Under a Dirichlet process (DP) prior with concentration parameter $a$ and discount parameter $d = 0$, the expected value satisfies:
\begin{equation}\label{eq:expected_k_dp}
    \mathbb{E}[K_n] = \sum_{i=0}^{n-1} \frac{a}{a + i}.
\end{equation}
More generally, under a Pitman-Yor (PY) process prior with concentration parameter $a$ and discount parameter $0 < d < 1$, the expected number of unique clusters is given by:
\begin{equation}\label{eq:expected_k_py}
    \mathbb{E}[K_n] = \frac{a}{d} \left[ \frac{\Gamma(a + d + n)\Gamma(a)}{\Gamma(a + n)\Gamma(a + d)} - 1 \right];
\end{equation}
see \cite{pitman2006combinatorial}, Equation (3.13). For the baseline specification adopted in {Section~\ref{sec:real_data}} ($a = 2.0$, $d = 0.05$, and $n = 82$), Equation \eqref{eq:expected_k_py} yields $\mathbb{E}[K_n] \approx 8.8229$. To investigate prior sensitivity, we solve for a matching concentration parameter under a DP prior ($d = 0$) targeting this same prior expectation. A grid-search procedure reveals that $a = 2.3058$ matches this setting closely, achieving $\mathbb{E}[K_n] \approx 8.8205$. Instead, to evaluate the behavior of CBI under prior misspecification, we consider a DP prior with concentration parameter $a = 6.0$, which implies a much larger prior expectation of $\mathbb{E}[K_n] \approx 16.5932$.
 
\begin{figure}
    \centering
    \begin{subfigure}{0.48\textwidth}
        \centering
        \includegraphics[width=\linewidth]{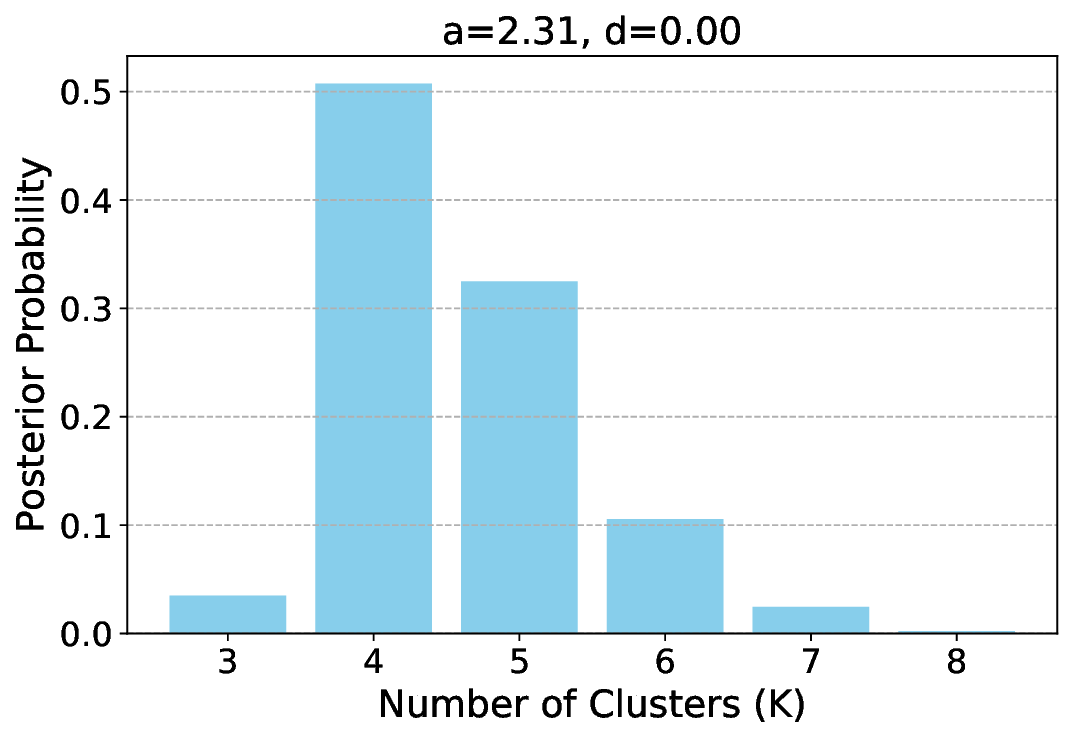}
        \caption{Posterior of $K_n$ (calibrated DP prior).}
        \label{fig:9a}
    \end{subfigure}
    \hfill
    \begin{subfigure}{0.48\textwidth}
        \centering
        \includegraphics[width=\linewidth]{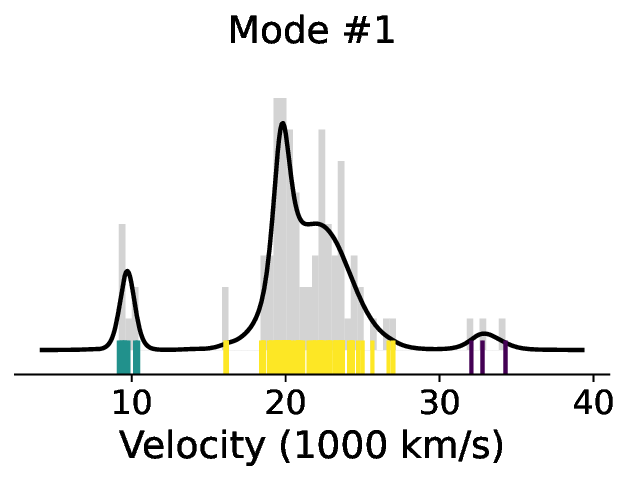}
        \caption{Posterior mode (calibrated DP prior).}
        \label{fig:9b}
    \end{subfigure}
    
    \vspace{0.3cm}
    
    \begin{subfigure}{0.48\textwidth}
        \centering
        \includegraphics[width=\linewidth]{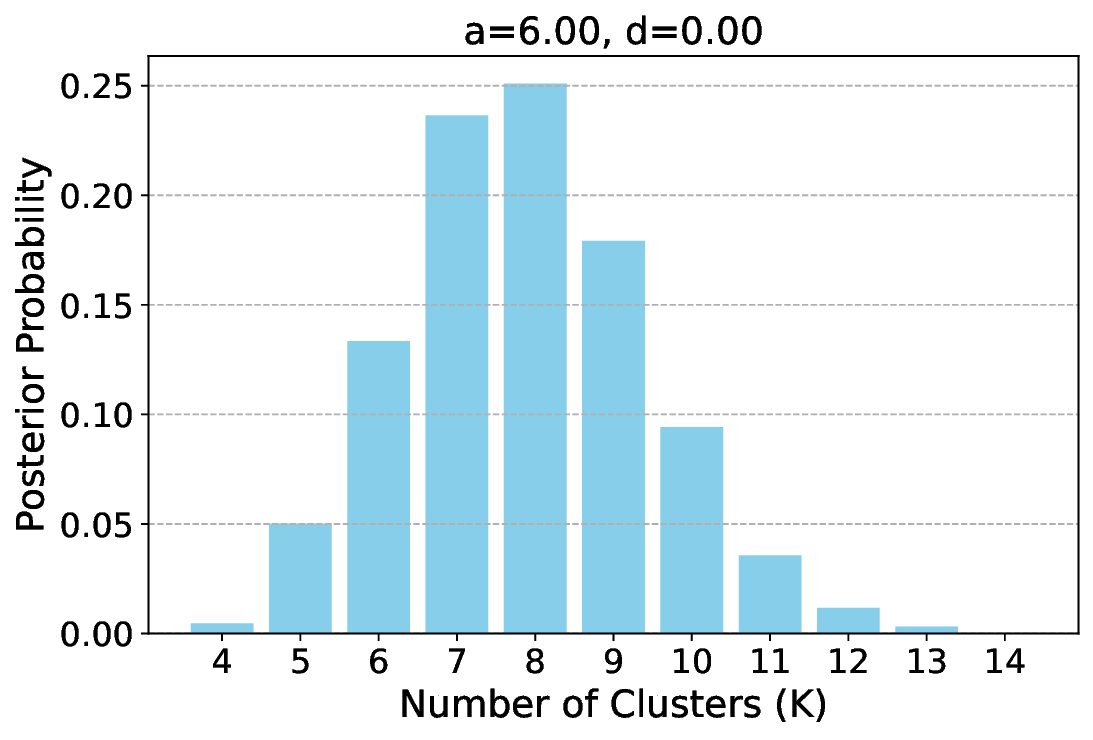}
        \caption{Posterior of $K_n$ (misspecified DP prior).}
        \label{fig:9c}
    \end{subfigure}
    \hfill
    \begin{subfigure}{0.48\textwidth}
        \centering
        \includegraphics[width=\linewidth]{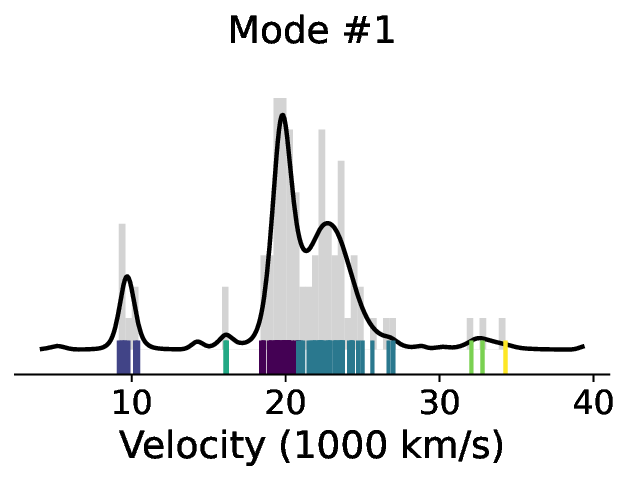}
        \caption{Posterior mode (misspecified DP prior).}
        \label{fig:9d}
    \end{subfigure}
    
    \caption{Posterior distribution of the number of clusters $K_n$ alongside their corresponding posterior modes across the two alternative Dirichlet Process prior specifications on the Galaxy velocities dataset.}
    \label{fig:robustness_plots}
\end{figure}

In scenario (i), the results are similar to those obtained under the baseline Pitman--Yor prior, except that the posterior becomes more concentrated around partitions with three or four clusters and the clear secondary mode involving larger numbers of clusters disappears. This is consistent with the stronger concentration properties of the Dirichlet process prior relative to more flexible alternatives, which leads to a less heterogeneous posterior distribution. Consequently, only a single mode is detected (Figures~\ref{fig:9a} and \ref{fig:9b}), and the resulting conformal region narrows, accepting only the $k$-means configuration with $k=4$ ($p\text{-value} = 0.7752$) while rejecting $k=3$ ($p = 0.0639$), $k=5$ ($p = 0.0340$), and $k=6$ ($p = 0.0160$).
 
In the misspecified scenario (ii), by contrast, the usefulness of $\mathcal D_{VI}$ becomes particularly evident. Although the raw posterior sample space places substantial mass on highly fragmented partitions due to prior misspecification, $\mathcal{D}_{VI}$ exerts an Occam's razor effect. DPC identifies a single mode corresponding to a 6-cluster partition (Figures~\ref{fig:9c} and \ref{fig:9d}), and the $90\%$ conformal region expands to absorb the prior-induced uncertainty, accepting all $k$-means candidate partitions for $k \in \{3, 4, 5, 6\}$ with high $p$-values ($0.2767, 0.9510, 0.6923,$ and $0.8352$, respectively).
 
Across both prior settings, the proportion of $1,000$ randomly generated partitions that fall within the conformal set remains exactly $0$, validating the size calibration of CBI even under prior misspecification.
 
\subsection{ERP data}\label{sec:erp_further}
 
This subsection collects the figures and additional discussion accompanying the ERP data experiment of {Section~\ref{sec:real_data}}, in which CBI is applied to the posterior samples of partitions of 34 participants obtained by \citet{song2025repulsive} using a projection Determinantal Point Process (pDPP) repulsive mixture prior fit to the ERP CORE data~\citep{kappenman2021erp}.
 
\begin{figure}[t!]
    \centering
    \caption{CBI results for ERP data analyzed using the pDPP mixture model of~\citet{song2025repulsive}.}
 
    \begin{subfigure}{0.4\linewidth}
        \centering
        \includegraphics[width=\linewidth]{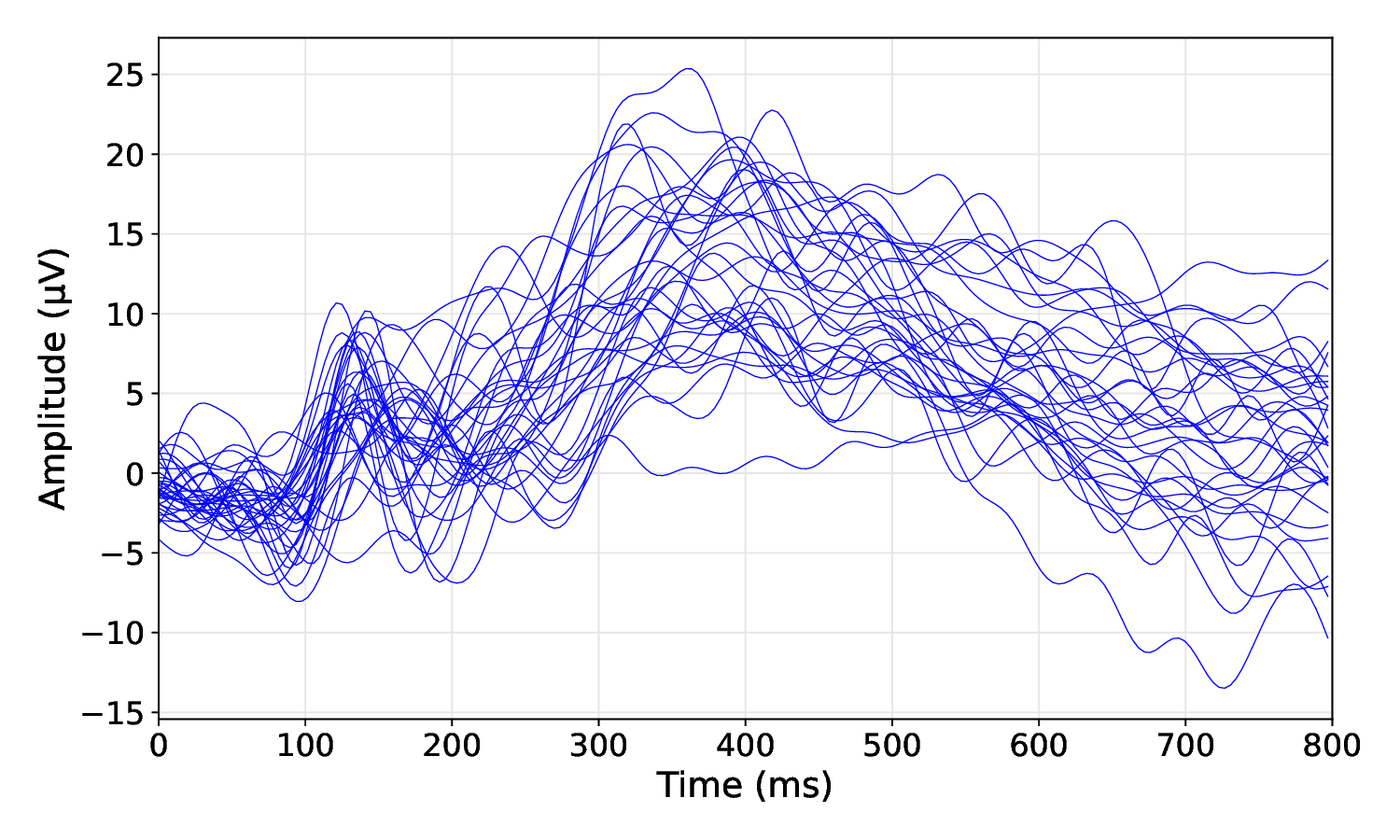}
        \caption{Raw ERP data.}
        \label{fig:pDPP_raw_erp_plot}
    \end{subfigure}
    \hfill
    \begin{subfigure}{0.35\linewidth}
        \centering
        \includegraphics[width=\linewidth]{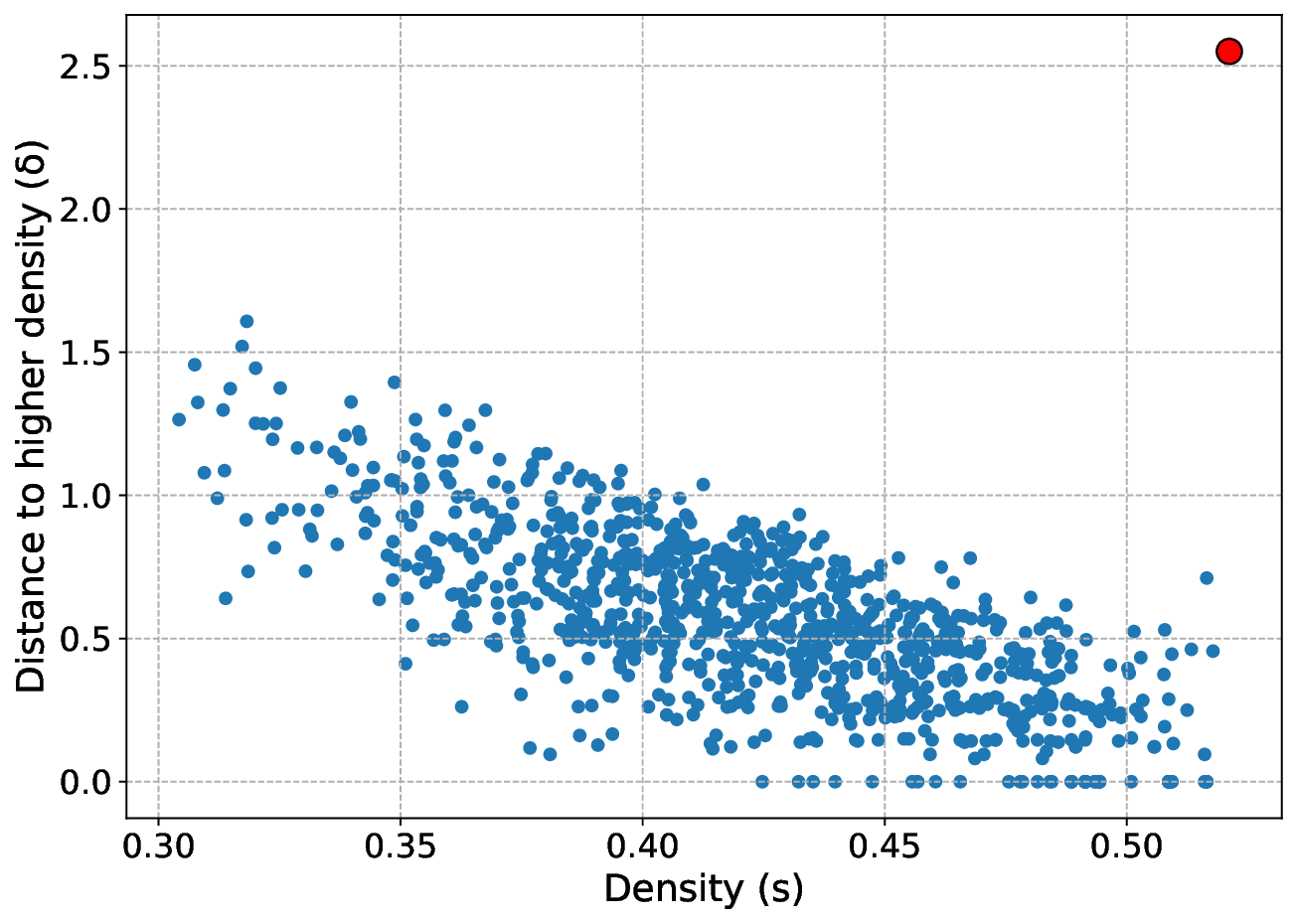}
        \caption{KDE-DPC decision graph.}
        \label{fig:pDPP_dpc_decision_graph}
    \end{subfigure}
    \vfill    
    \begin{subfigure}{0.6\linewidth}
        \centering
        \includegraphics[width=\linewidth]{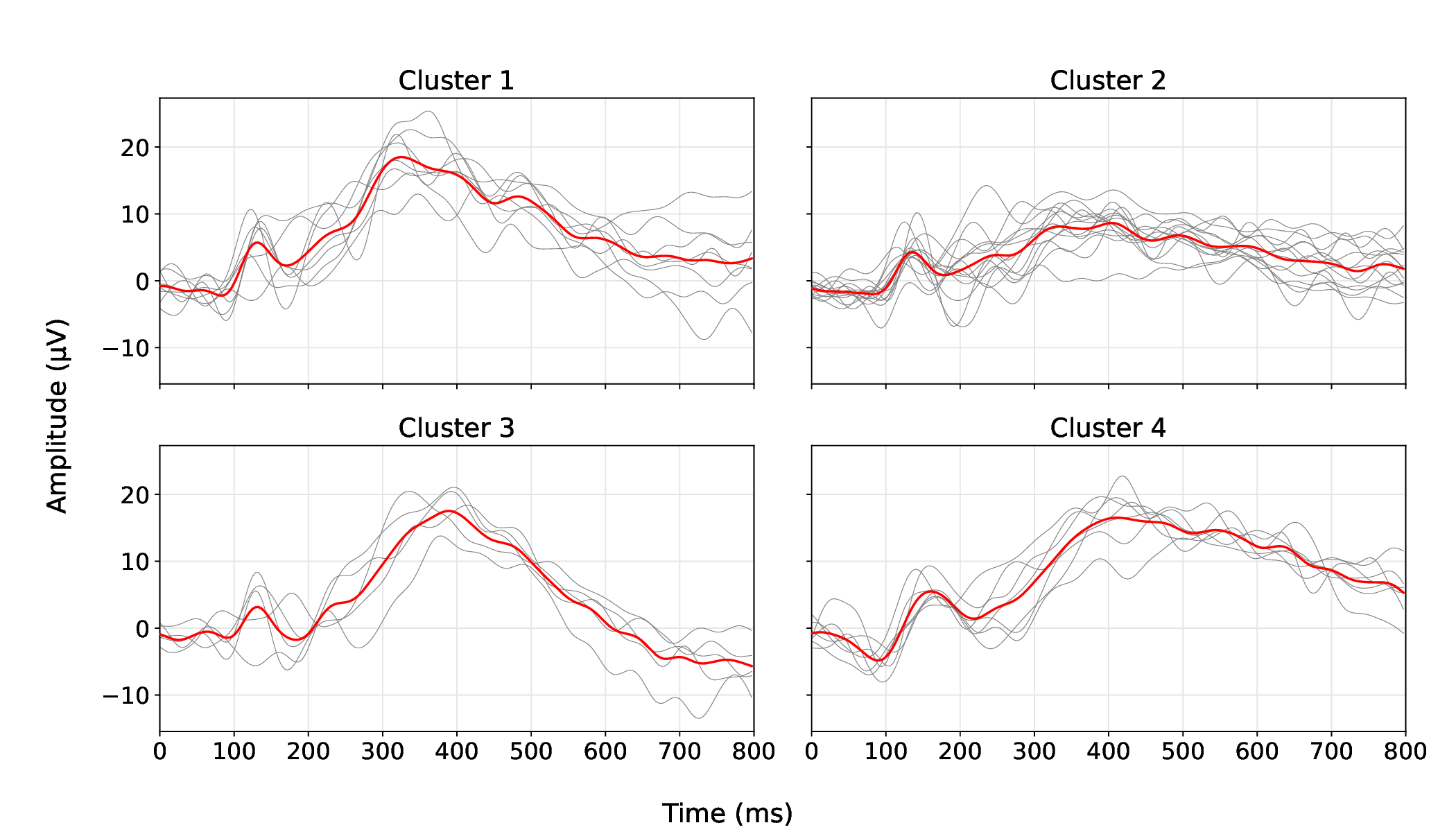}
        \caption{$\mathcal D_{VI}$-KDE point estimate. Note: cluster members are plotted in gray, while cluster-specific means are in red.}
        \label{fig:pDPP_kde_point_estimate}
    \end{subfigure}
    
\end{figure}

Figure~\ref{fig:pDPP_raw_erp_plot} visualizes the analyzed data in terms of waveform amplitudes measured up to 800 milliseconds after stimulus onset. As shown in Figure~\ref{fig:pDPP_dpc_decision_graph}, the KDE-DPC decision graph reveals that the posterior distribution over partitions is unimodal, concentrating around a single representative partition. The corresponding $\mathcal D_{VI}$-KDE point estimate, shown in Figure~\ref{fig:pDPP_kde_point_estimate}, features four distinct clusters and coincides with the posterior mode. Interestingly, this estimate matches the point estimate obtained by \citet{song2025repulsive} (see their Figure~5) using the SALSO algorithm~\citep{dahl2022search}, which corroborates the original point estimate found in that study.
 
\begin{figure}
    \centering
    \includegraphics[width=0.9\linewidth]{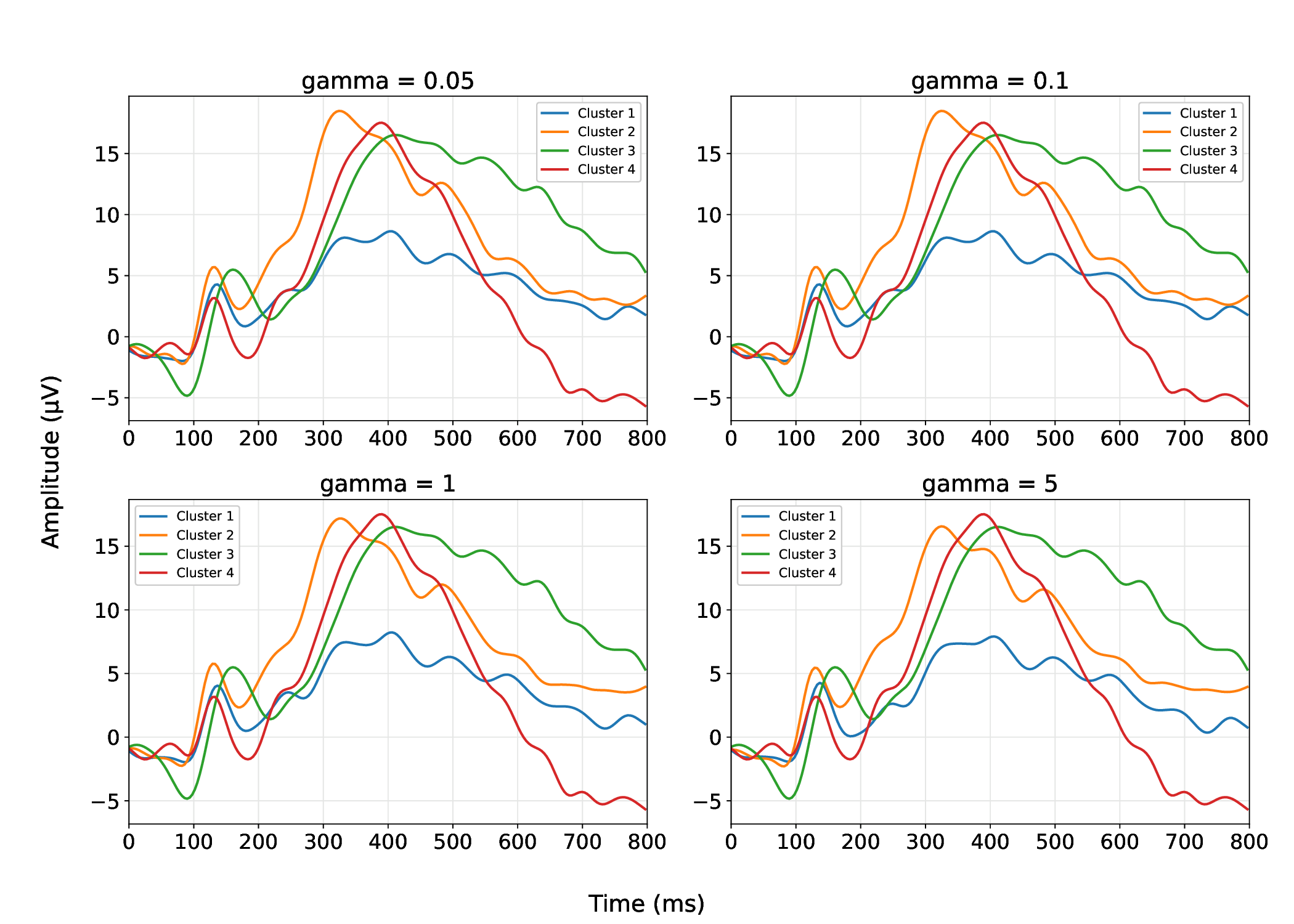}
    \caption{Point estimates of the random partition for the ERP dataset (cluster-specific mean functions, colored by cluster) for different values of the hyperparameter $\gamma$.}
    \label{fig:pDPP_point_estimates_multigamma}
\end{figure}

To assess the sensitivity of these results to the value of $\gamma$, we apply the two proposed hyperparameter selection methods, sweeping $\gamma \in \{0.05, 0.1, 0.5, 1, 5\}$ for the leave-one-out criterion. Both scale-matching strategies point to a bandwidth close to the value used throughout the main analysis: matching to the mean pairwise $\mathcal D_{VI}$ discrepancy among the training partitions gives $\gamma \approx 0.566$, and matching to the median gives $\gamma \approx 0.559$. This is corroborated by leave-one-out cross-validation over the grid, which selects $\gamma_\star = 0.5$, i.e., our default choice. Consistent with this, Figure~\ref{fig:pDPP_point_estimates_multigamma} shows that the KDE point estimate is essentially unchanged across the grid: the cluster-specific mean functions recovered under $\gamma \in \{0.05, 0.1, 1, 5.0\}$ are all nearly indistinguishable from the $\gamma = 0.5$.

\begin{figure}
    \centering
    \includegraphics[width=0.9\linewidth]{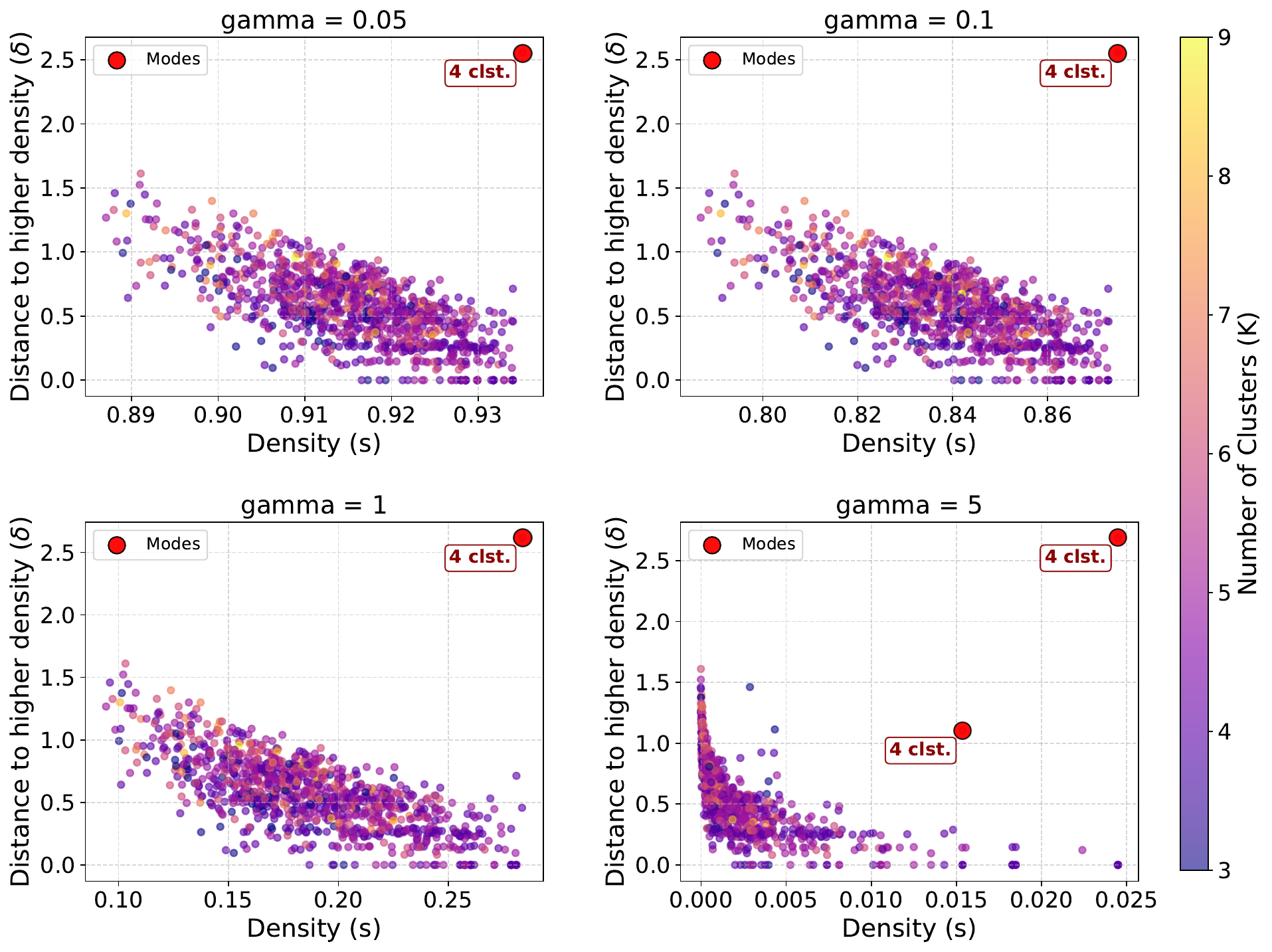}
    \caption{KDE-DPC decision graph for the ERP dataset, for different values of the hyperparameter $\gamma$.}
    \label{fig:pDPP_dpc_multigamma}
\end{figure}

\begin{figure}
    \centering
    \includegraphics[width=0.6\linewidth]{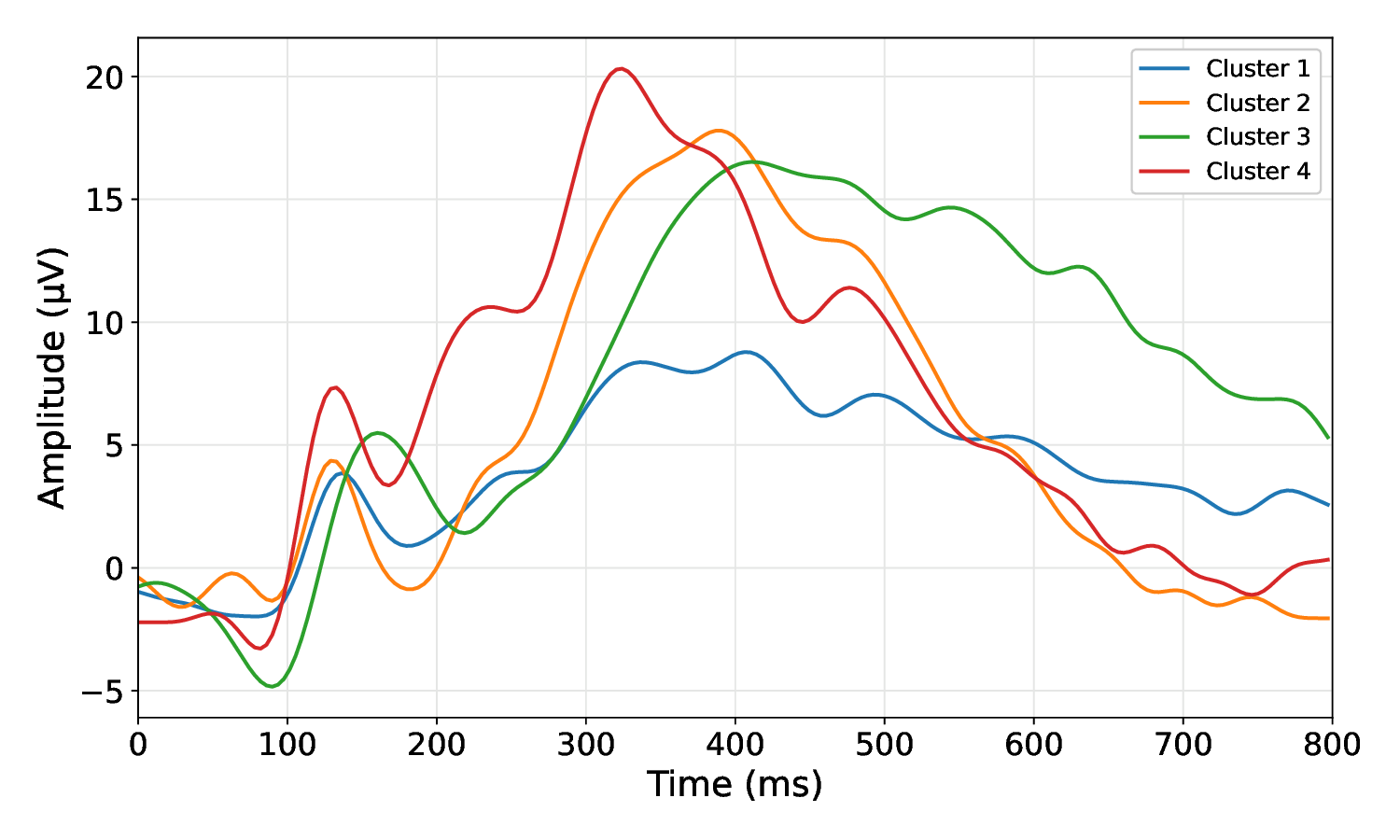}
    \caption{Secondary mode in the ERP dataset experiment for $\gamma=5.0$.}
    \label{fig:pDPP_gamma5_secondary_mode}
\end{figure}

Interestingly, when $\gamma = 5.0$, the KDE-DPC decision graph in Figure~\ref{fig:pDPP_dpc_multigamma} reveals a secondary mode in addition to the primary one, in contrast to the unimodal decision graphs obtained for all other values of $\gamma$. However, as shown in Figure~\ref{fig:pDPP_gamma5_secondary_mode}, this secondary mode is itself very similar to the primary one, i.e., to the $\gamma = 5.0$ point estimate discussed above. This apparent bimodality is likely an artifact of setting $\gamma$ too high. Indeed, as $\gamma$ grows, the resulting scores converge to raw empirical frequencies, which are near zero for most calibration samples (see, for instance, the high concentration of points near the vertical axis of the decision graph for $\gamma = 5.0$). As a result, a partition that occurs moderately often in the calibration sample and is not too different from the primary mode, while being essentially indistinguishable in density from many other partitions under smaller values of $\gamma$, can stand out as having a comparatively high score under higher values of $\gamma$, and consequently be identified, quite uninformatively, as a separate mode. This provides further evidence that the proposed selection methods, together with the default value we adopt, are practically useful for obtaining interpretable output.

\section{Additional experiment 1: unimodal posterior distribution over partitions}\label{sec:unimod_exp}

\begin{figure}
        \centering
        \caption{Gaussian mixture simulated data, analyzed using a PY(0.01, 0.01) Gaussian mixture prior.} 
        \begin{subfigure}{0.3\textwidth}
            \centering
            \includegraphics[width=\linewidth]{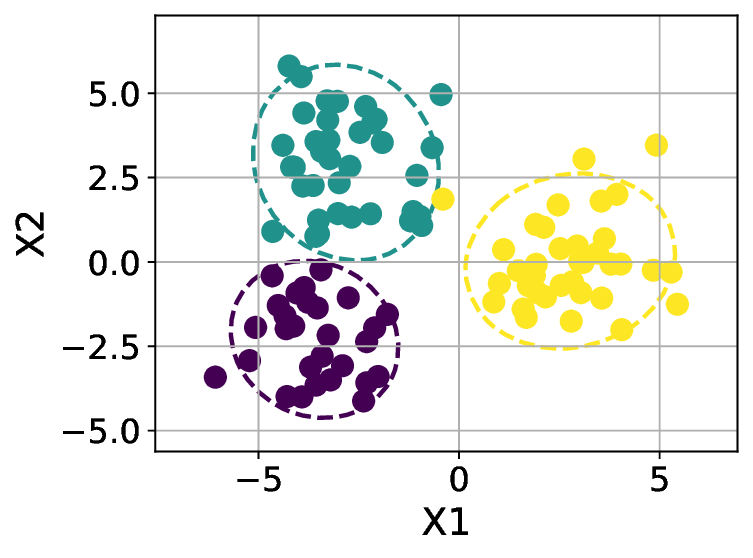}
            \caption{True data clustering.}
            \label{fig:true_partition_unimod}
        \end{subfigure}
        \hfill
        \begin{subfigure}{0.38\textwidth}
            \centering
            \includegraphics[width=\linewidth]{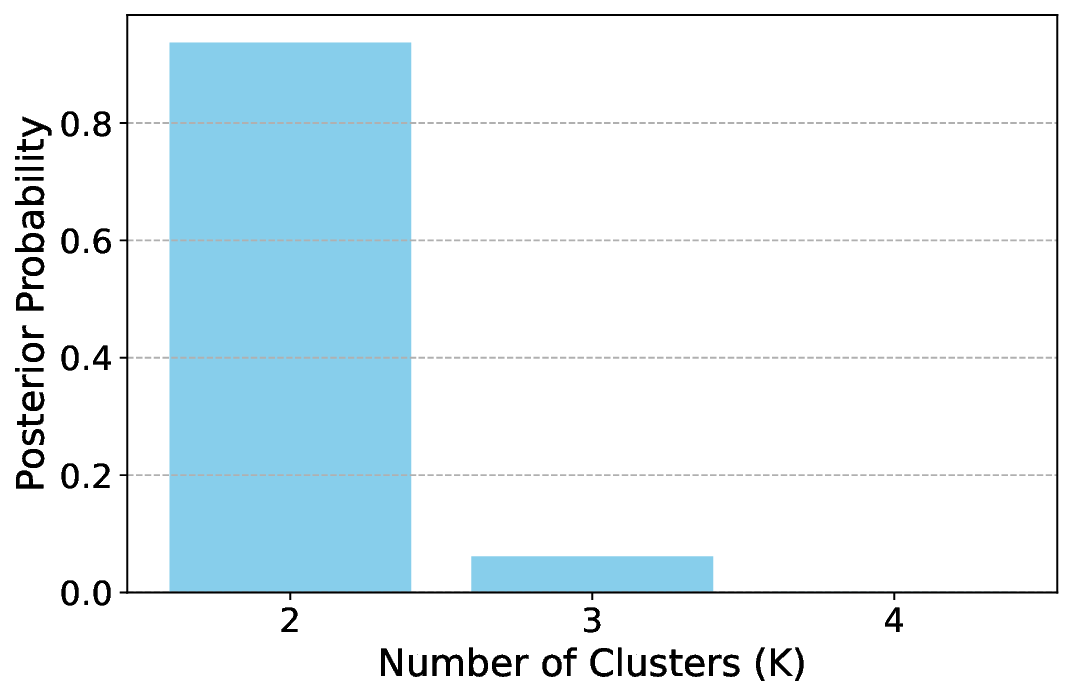}
            \caption{Posterior distribution of $K$.}
            \label{fig:post_K_unimod}
        \end{subfigure}
        \hfill
        \begin{subfigure}{0.3\textwidth}
            \centering
            \includegraphics[width=\linewidth]{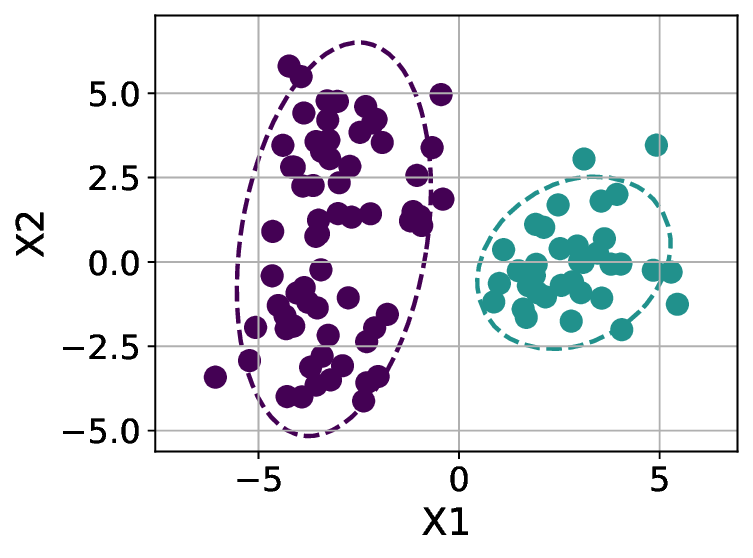}
            \caption{$\mathcal D_{VI}$-KDE point estimator.}
            \label{fig:top_partition_kde_vi_unimod}
        \end{subfigure}
    \end{figure}

We run a Bayesian clustering experiment analogous to the one used to illustrate CBI in {Examples~\ref{ex:point_est}--\ref{ex:multimodality}}, using the same simulated two-dimensional dataset (Figure~\ref{fig:true_partition_unimod}). The only difference is that the Pitman-Yor (PY) Gaussian mixture prior is now specified with a smaller concentration parameter, set to $0.01$. As is well known~\citep{deblasi2013gibbs}, lowering this parameter increases prior mass on partitions with fewer clusters. This shift is reflected in the posterior over the number of clusters $K$, which becomes more concentrated on small values. Consistent with this, the $\mathcal{D}_{VI}$-KDE point estimate remains the same two-cluster configuration identified in the example with higher concentration parameter (Figures~\ref{fig:post_K_unimod} and~\ref{fig:top_partition_kde_vi_unimod}).

\begin{figure}
        \centering
        \includegraphics[width=0.45\textwidth]{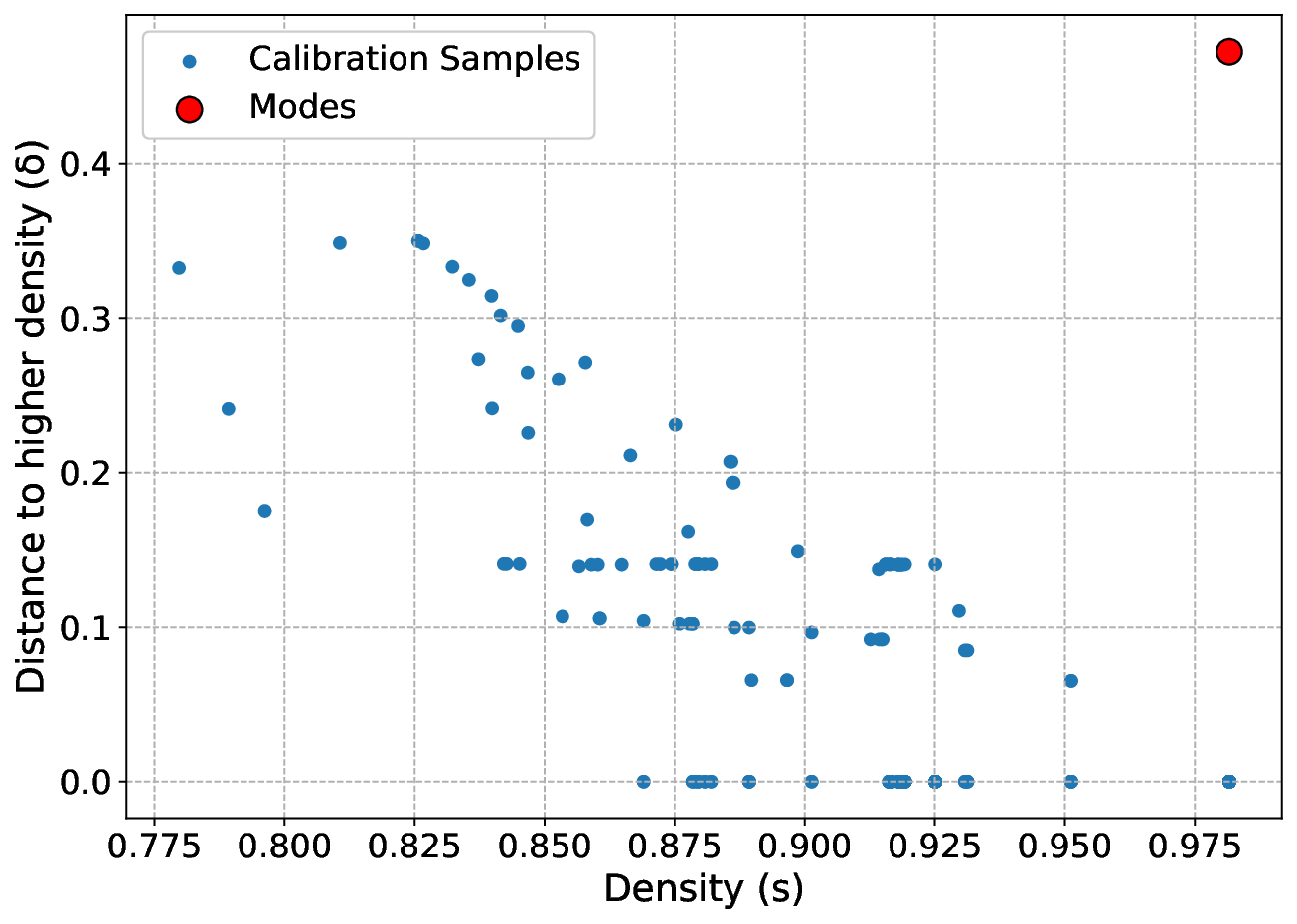}
        \caption{KDE-DPC decision graph (unimodal PY Gaussian mixture posterior on simulated data).}
        \label{fig:density_peak_decision_graph_vi_unimod}
\end{figure}

When constructing the conformal credible region with 90\% coverage, we find that it includes the “true partition’’ when the two leftmost clusters are merged, but no longer the original three-cluster partition. The credible set nevertheless passes the size diagnostic based on 1,000 randomly generated partitions whose number of clusters follows the posterior distribution of $K$: none of these random partitions fall inside the credible region. Finally, the multimodality analysis in Figure~\ref{fig:density_peak_decision_graph_vi_unimod} confirms that, under this posterior—corresponding to a prior favoring fewer clusters—the true three-cluster configuration ceases to be a posterior mode, leaving the point estimate as the only identified representative.

\section{Additional experiment 2: Colorectal cancer spatial transcriptomics data (dependent posterior distributions over partitions)}\label{sec:ST_application}

In a recent study, \citet{duan2025spatially} analyze spatial transcriptomics data on colorectal cancer~\citep{lee2020lineage}, with the aim of discovering spatially aligned subpopulations of tumor, immune, and stromal cells. Spatial alignment refers to the spatial co-localization of clusters across cell types, where spatial coordinates correspond to the two-dimensional positions of cells within a medical image. To capture this structure, the authors propose a spatially-aligned random partition model that combines a PY Gaussian mixture model for gene expression features with a prior dependence structure that induces spatial co-localization in the latent clustering across cell types. Their MCMC output consists of sampled partitions for each cell type: cells of different types are clustered separately, but the partitions are modeled as dependent through prior-induced spatial alignment. The results of their analysis indeed reveal subpopulations (clusters) of cells from different types that are spatially co-localized.

\begin{figure}
    \centering
    \includegraphics[width=0.9\linewidth]{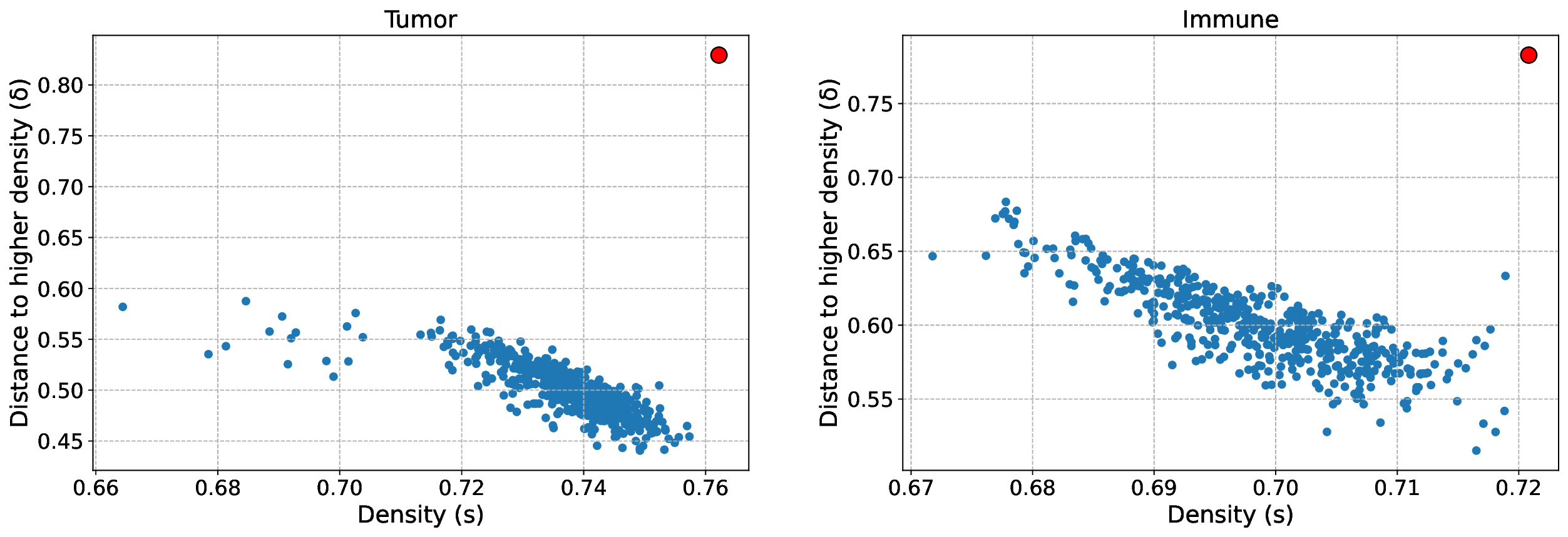}
    \caption{KDE-DPC decision graphs for posterior distributions over partitions of tumor and immune cells.}
    \label{fig:Tumor_Immune_dpc_decision_graph_combined}
\end{figure}

We apply our CBI pipeline to these MCMC samples using the VI metric between partitions. For simplicity, we restrict attention to tumor and immune cells only. Figure~\ref{fig:Tumor_Immune_dpc_decision_graph_combined} shows the KDE-DPC decision graphs for the marginal posterior distributions over partitions of both cell types. In both cases, the posterior appears unimodal, indicating that the only meaningful partitions summarizing these posteriors are the point estimates presented next.

\begin{figure}
    \centering
    
    \begin{subfigure}{0.6\linewidth}
        \centering
        \includegraphics[width=\linewidth]{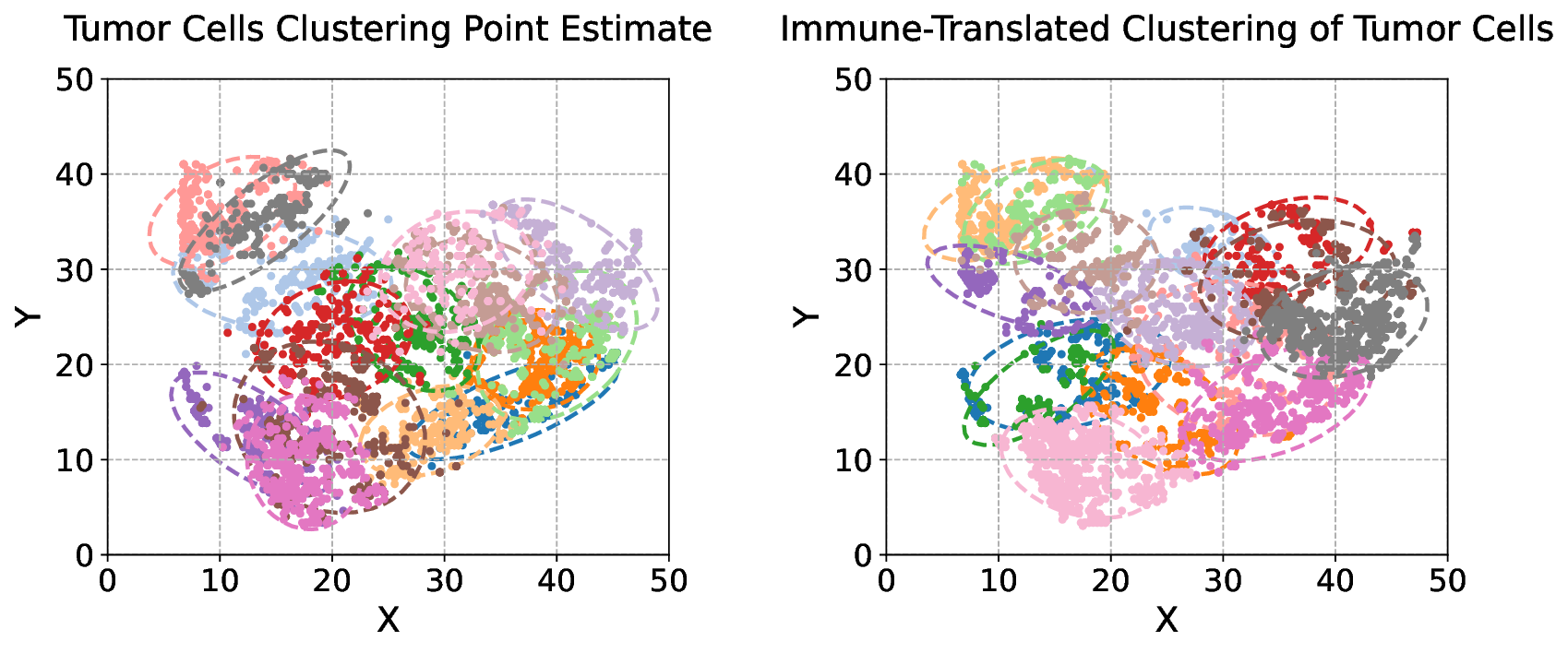}
    \end{subfigure}
    
    \vspace{0.1em}

    \begin{subfigure}{0.6\linewidth}
        \centering
        \includegraphics[width=\linewidth]{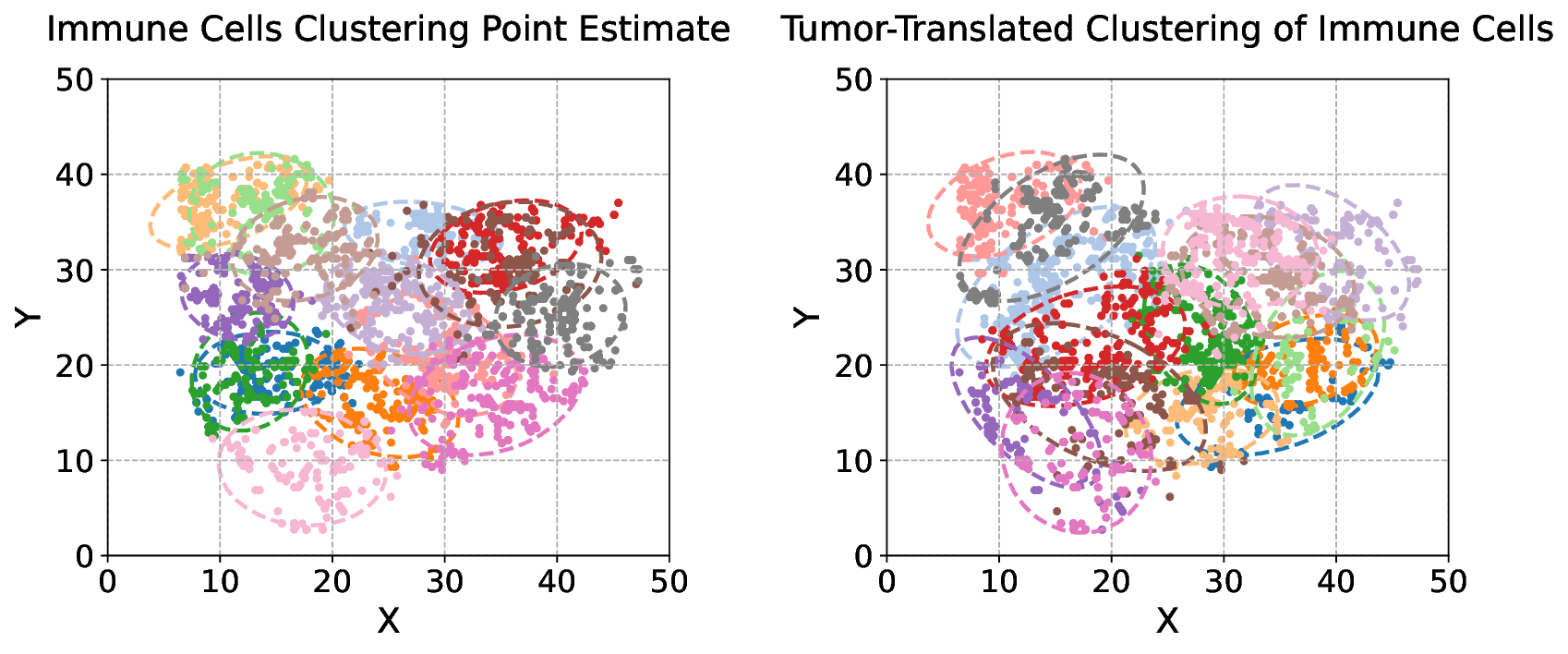}
    \end{subfigure}
    \caption{Left: $\mathcal{D}_{VI}$-KDE point estimates for tumor and immune cell partitions. 
    Right: partitions obtained by spatially translating the $\mathcal{D}_{VI}$-KDE point estimates across cell types.}
    \label{fig:ST_clusterings_combined}
\end{figure}

The left-hand side plots in Figure~\ref{fig:ST_clusterings_combined} show the $\mathcal{D}_{VI}$-KDE point estimates for tumor and immune cell partitions. Recall that the model of \citet{duan2025spatially} performs density estimation and clustering on high-dimensional gene expression features while using spatial coordinates to induce alignment. For visualization purposes, we display partitions only in the spatial domain, since spatial alignment is the focus of the analysis. Both point estimates reveal spatially coherent clusters, in line with the findings of~\citet{duan2025spatially}.

Finally, we use the conformal credible regions (constructed separately for both marginal posteriors) to test a hypothesis of \emph{global spatial alignment} between the partitions of different cell types. Specifically, given the point estimate for type A cells (either tumor or immune), we spatially translate it to type B cells by matching each type B cell to its closest (spatially) type A cell, and assigning that type B cell to the cluster of the matched type A cell. The right-hand side plots in Figure~\ref{fig:ST_clusterings_combined} illustrate these translated partitions. We then check whether each translated partition of type B cells lies within the 90\% conformal credible region for the posterior distribution over partitions of type B cells. We emphasize that this constitutes a test of global alignment, as it requires the entire translated partition to lie within the credible region. This is distinct from the goal of \citet{duan2025spatially}, who adopt a more local notion of alignment seeking to identify subsets of clusters that are spatially co-located, rather than assuming full alignment across all clusters. As a result of our analysis, both spatially translated partitions fall outside the corresponding credible regions, supporting the hypothesis that only subsets of tumor and immune cell clusters are spatially co-located. This is consistent with the findings of \citet{duan2025spatially}.

\section{Additional experiment 3: posterior distribution over mixing measures}\label{sec:mixmeasure_exp}

In this experiment, we apply CBI to infer the mixing measure in a Gaussian mixture model. We use again the Galaxy velocities dataset and analyze it as in {Section~\ref{sec:real_data}} under the same PY Gaussian mixture model with unchanged prior hyperparameters. The model for the observations $X_1, \ldots, X_{n}$ is
\begin{equation*}
    X_i \mid \tilde P \overset{\textnormal{iid}}{\sim} \int_{\mathbb R \times \mathbb R_+} N(\cdot \mid \mu, \sigma^2)\, \tilde P(\mathrm d\mu, \mathrm d\sigma^2),
\end{equation*}
where $N(\cdot \mid \mu, \sigma^2)$ denotes a Gaussian density with mean $\mu$ and variance $\sigma^2$, and $\tilde P$ is a random probability measure on the space of means and variances endowed with a PY process prior~\citep{perman1992size, pitman1997two}. Both the prior and posterior distributions of $\tilde P$ place their mass on discrete measures with countably many atoms. When appropriate conditional sampling methods are employed~\citep{papaspiliopoulos2008retrospective, walker2007sampling, kalli2011slice}, posterior samples of $\tilde P$ can be obtained directly.

Since the Gibbs sampler used in the clustering experiment of {Section~\ref{sec:real_data}} marginalizes out $\tilde P$ via latent cluster assignments, we refit the model using the conditional sampler implemented in \texttt{numpyro}~\citep{phan2019composable, bingham2019pyro}, which employs the NUTS algorithm~\citep{hoffman2014nuts}. We apply the CBI pipeline to 5,000 training and 1,000 calibration samples from the posterior distribution of $\tilde P$, truncated to 10 atoms. Distances between mixing measures are computed using the Wasserstein-1 distance~\citep{villani2008optimal, nguyen2013convergence}, implemented in the \texttt{pot} library~\citep{flamary2021pot, flamary2024pot}.

\begin{figure}
    \centering
    \caption{CBI applied to posterior samples of mixing measures (Galaxy velocities dataset analyzed with a PY Gaussian mixture model).}
    \begin{subfigure}{0.35\textwidth}
        \centering
        \includegraphics[width=\linewidth]{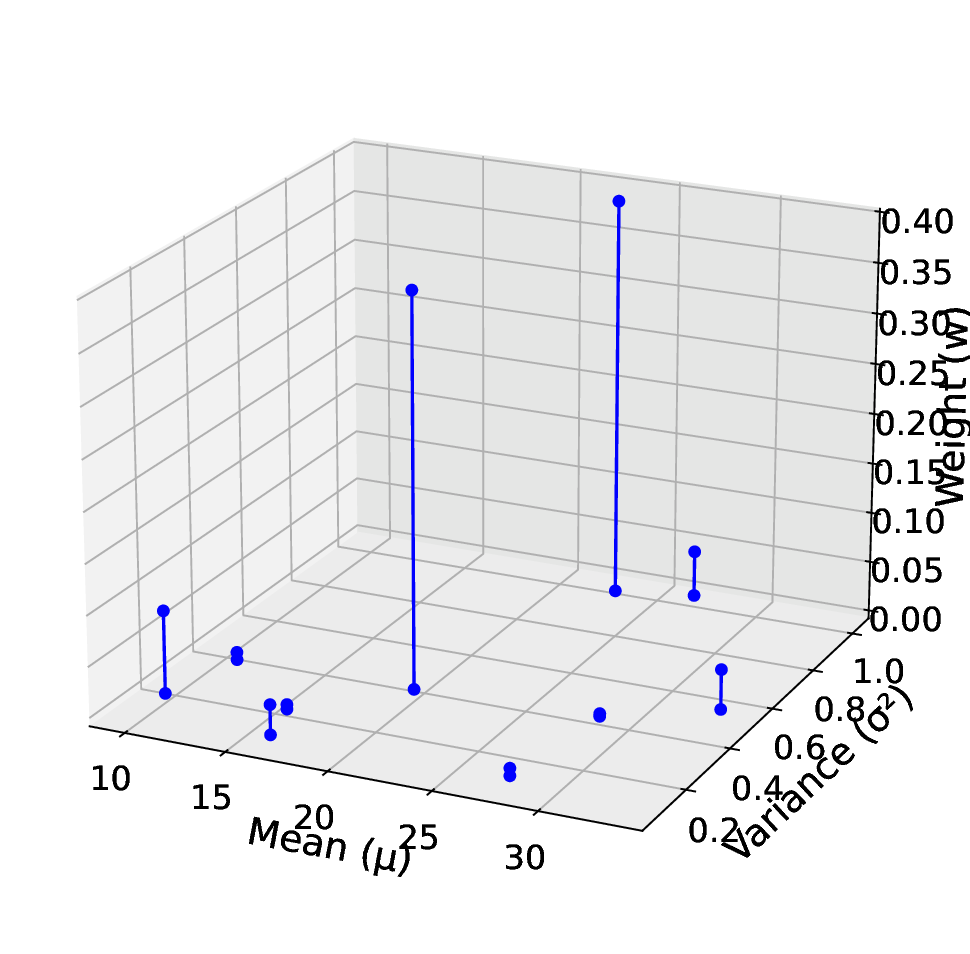}
        \caption{Wasserstein-KDE point estimate.}
        \label{fig:mode_mixing_measure_3d}
    \end{subfigure}
    \hspace{1cm}
    \begin{subfigure}{0.35\textwidth}
        \centering
        \includegraphics[width=\linewidth]{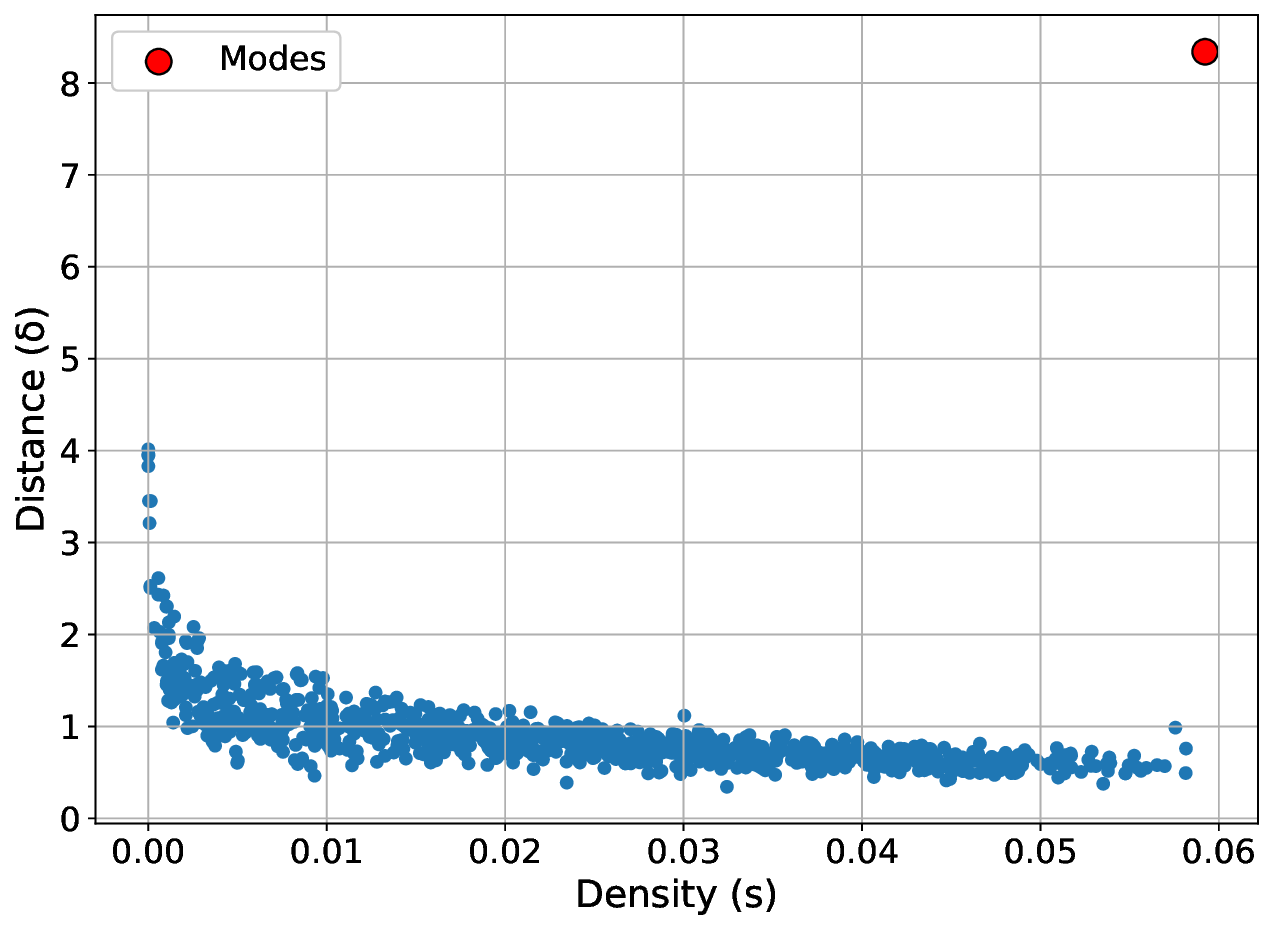}
        \caption{KDE-DPC decision graph.}
        \label{fig:dpc_decision_graph_mixing_measure}
    \end{subfigure}    
\end{figure}

Figures~\ref{fig:mode_mixing_measure_3d} and~\ref{fig:dpc_decision_graph_mixing_measure} show that the posterior concentrates around a single mixing measure, which is in line with the empirical distribution of the observed velocities. The 90\% conformal credible region contains the empirical $k$-means mixing measures with $k=4,5,6$, but excludes that with $k=3$.\footnote{The region also passes a basic size check performed by verifying whether 1,000 mixing measures randomly generated according to the prior are included in the region---none of them is.} Compared with the analysis in {Section~\ref{sec:real_data}}, where a three-cluster configuration was the highest-KDE point estimate under the VI distance on partitions, this result suggests a subtle distinction: while three clusters best summarize the posterior over partitions, more than three nonzero components are required for the corresponding mixing measure, and hence the implied density, to be plausible under the posterior. This provides a refined understanding of the posterior structure captured by the fitted Bayesian model.

\section{Additional experiment 4: posterior distribution over covariance matrices}\label{sec:covariance_exp}

Our final additional experiment applies CBI to covariance estimation. We consider the four-dimensional Iris dataset~\citep{anderson1936species}, restricted to the \emph{Setosa} species. After standardizing the observations $X_1, \ldots, X_n$, we fit a simple Gaussian model of the form
\begin{equation*}
    X_i \mid \Sigma \overset{\textnormal{iid}}{\sim} N(\cdot\mid 0, \Sigma),
\end{equation*}
with an LKJ prior on the correlation matrix \citep{lewandowski2009generating}. We draw 5{,}000 training and 1{,}000 calibration posterior samples for $\Sigma$ using \texttt{numpyro}, and apply CBI by measuring distances between covariance matrices according to the operator norm
\[
\Vert \Sigma \Vert_{\mathrm{op}} := \max_{v : \Vert v \Vert = 1} \Vert \Sigma v \Vert.
\]

\begin{figure}
    \centering
    \caption{CBI applied to posterior samples of covariance matrices (Iris dataset analyzed with a Gaussian covariance model).}
    \begin{subfigure}{0.4\textwidth}
        \centering
        \includegraphics[width=\linewidth]{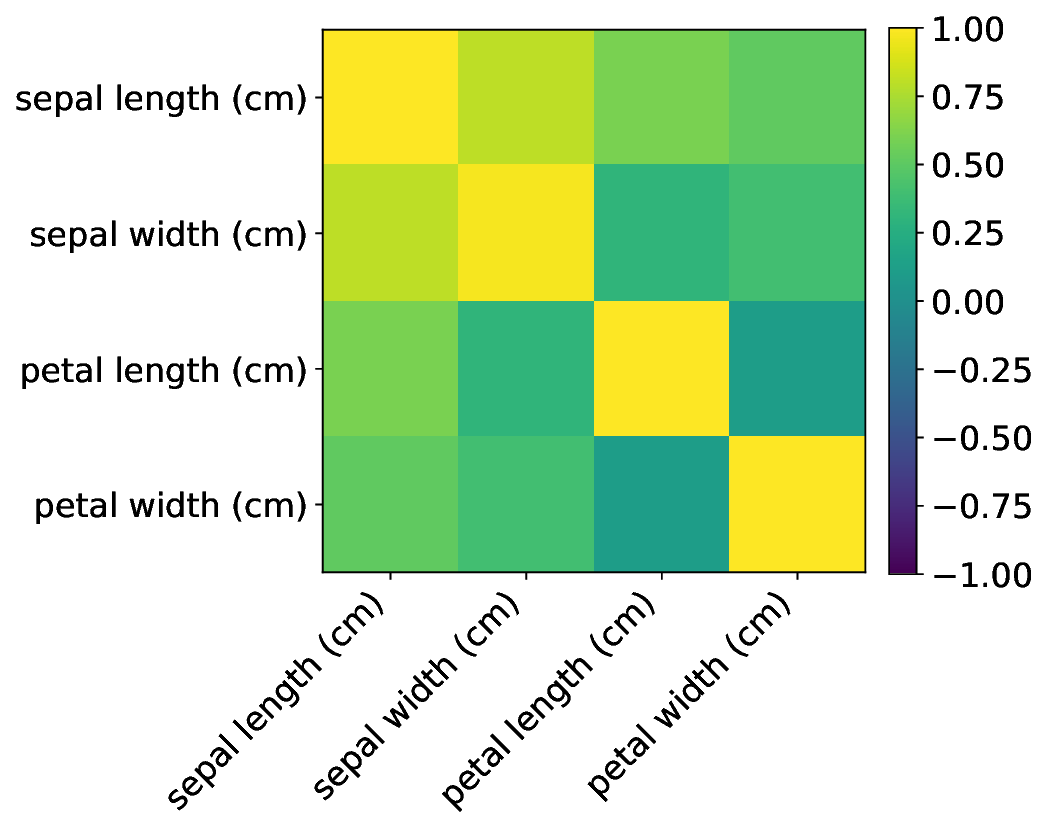}
        \caption{$\Vert\cdot\Vert_{\mathrm{op}}$-KDE point estimate.}
        \label{fig:posterior_mode_setosa}
    \end{subfigure}
    \hfill
    \begin{subfigure}{0.35\textwidth}
        \centering
        \includegraphics[width=\linewidth]{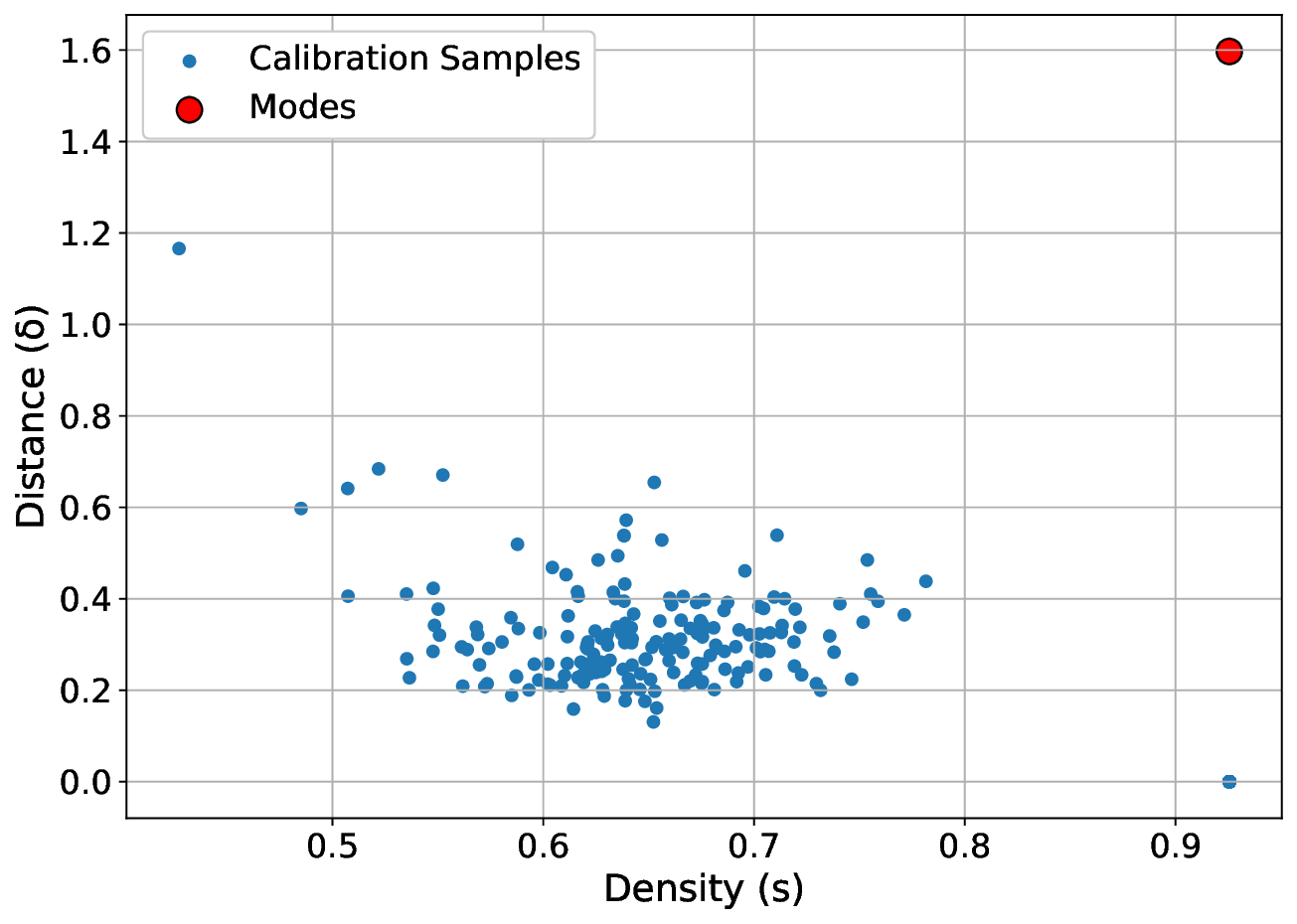}
        \caption{KDE-DPC decision graph.}
        \label{fig:dpc_decision_plot_setosa}
    \end{subfigure}
\end{figure}

Figures~\ref{fig:posterior_mode_setosa} and~\ref{fig:dpc_decision_plot_setosa} show that the posterior distribution is concentrated around a single, well-defined mode corresponding to the $\Vert\cdot\Vert_{\mathrm{op}}$-KDE point estimate. When constructing the 90\% conformal credible region, we find that it contains the empirical covariance matrix, but, encouragingly, none of 1,000 randomly generated covariance matrices drawn from the prior.

\bibliographystyle{ba}
\bibliography{main}
\begin{thebibliography}{75}
\newcommand{\enquote}[1]{``#1''}
\expandafter\ifx\csname natexlab\endcsname\relax\def\natexlab#1{#1}\fi
\expandafter\ifx\csname url\endcsname\relax
  \def\url#1{{\tt #1}}\fi
\expandafter\ifx\csname urlprefix\endcsname\relax\def\urlprefix{URL }\fi
\ifx\endbibitem\undefined \let\endbibitem\relax\fi

\bibitem[{Anderson(1936)}]{anderson1936species}
Anderson, E. (1936).
\newblock \enquote{The species problem in {I}ris.}
\newblock {\em Annals of the Missouri Botanical Garden\/}, 23(3): 457--509.
\endbibitem

\bibitem[{Angelopoulos and Bates(2023)}]{angelopoulos2023conformal}
Angelopoulos, A.~N. and Bates, S. (2023).
\newblock \enquote{Conformal {P}rediction: {A} {G}entle {I}ntroduction.}
\newblock {\em Foundations and Trends{\textregistered} in Machine Learning\/}, 16(4): 494--591.
\endbibitem

\bibitem[{Balocchi and Wade(2025)}]{balocchi2025understanding}
Balocchi, C. and Wade, S. (2025).
\newblock \enquote{Understanding uncertainty in {B}ayesian cluster analysis.}
\newblock {\em arXiv preprint arXiv:2506.16295\/}.
\endbibitem

\bibitem[{Barber et~al.(2023)Barber, Candes, Ramdas, and Tibshirani}]{barber2023conformal}
Barber, R.~F., Candes, E.~J., Ramdas, A., and Tibshirani, R.~J. (2023).
\newblock \enquote{Conformal prediction beyond exchangeability.}
\newblock {\em The Annals of Statistics\/}, 51(2): 816--845.
\endbibitem

\bibitem[{Bariletto and Walker(2026)}]{bariletto2026scalable}
Bariletto, N. and Walker, S.~G. (2026).
\newblock \enquote{{Scalable Posterior Uncertainty for Flexible Density-Based Clustering}.}
\newblock {\em arXiv preprint arXiv:2603.03188\/}.
\endbibitem

\bibitem[{Barrios et~al.(2013)Barrios, Lijoi, Nieto-Barajas, and Pr{\"u}nster}]{barrios2013modeling}
Barrios, E., Lijoi, A., Nieto-Barajas, L.~E., and Pr{\"u}nster, I. (2013).
\newblock \enquote{{Modeling with Normalized Random Measure Mixture Models}.}
\newblock {\em Statistical Science\/}, 28(3): 313--334.
\endbibitem

\bibitem[{Beraha et~al.(2025)Beraha, Argiento, Camerlenghi, and Guglielmi}]{beraha2025bayesian}
Beraha, M., Argiento, R., Camerlenghi, F., and Guglielmi, A. (2025).
\newblock \enquote{Bayesian mixture models with repulsive and attractive atoms.}
\newblock {\em Journal of the Royal Statistical Society Series B: Statistical Methodology\/}, qkaf027.
\endbibitem

\bibitem[{Binder(1978)}]{binder1978bayesian}
Binder, D.~A. (1978).
\newblock \enquote{Bayesian cluster analysis.}
\newblock {\em Biometrika\/}, 65(1): 31--38.
\endbibitem

\bibitem[{Bingham et~al.(2019)Bingham, Chen, Jankowiak, Obermeyer, Pradhan, Karaletsos, Singh, Szerlip, Horsfall, and Goodman}]{bingham2019pyro}
Bingham, E., Chen, J.~P., Jankowiak, M., Obermeyer, F., Pradhan, N., Karaletsos, T., Singh, R., Szerlip, P.~A., Horsfall, P., and Goodman, N.~D. (2019).
\newblock \enquote{{Pyro: Deep Universal Probabilistic Programming}.}
\newblock {\em Journal of Machine Learning Research\/}, 20: 28:1--28:6.
\endbibitem

\bibitem[{Bolfarine et~al.(2025)Bolfarine, Lopes, and Carvalho}]{bolfarine2025lower}
Bolfarine, H., Lopes, H.~F., and Carvalho, C.~M. (2025).
\newblock \enquote{{Lower-dimensional posterior density and cluster summaries for overparameterized Bayesian models}.}
\newblock {\em arXiv preprint arXiv:2506.09850\/}.
\endbibitem

\bibitem[{Buch et~al.(2024)Buch, Dewaskar, and Dunson}]{buch2024bayesian}
Buch, D., Dewaskar, M., and Dunson, D.~B. (2024).
\newblock \enquote{Bayesian level-set clustering.}
\newblock {\em arXiv preprint arXiv:2403.04912\/}.
\endbibitem

\bibitem[{Cremaschi et~al.(2025)Cremaschi, Wertz, and De~Iorio}]{cremaschi2025repulsion}
Cremaschi, A., Wertz, T.~M., and De~Iorio, M. (2025).
\newblock \enquote{Repulsion, chaos, and equilibrium in mixture models.}
\newblock {\em Journal of the Royal Statistical Society Series B: Statistical Methodology\/}, 87(2): 389--432.
\endbibitem

\bibitem[{Dahl(2006)}]{dahl2006model}
Dahl, D.~B. (2006).
\newblock \enquote{{Model-based clustering for expression data via a Dirichlet process mixture model}.}
\newblock In Do, K.~A., M{\"u}ller, P., and Vannucci, M. (eds.), {\em Bayesian Inference for Gene Expression and Proteomic\/}, 201--218. Cambridge University Press.
\endbibitem

\bibitem[{Dahl et~al.(2022)Dahl, Johnson, and M{\"u}ller}]{dahl2022search}
Dahl, D.~B., Johnson, D.~J., and M{\"u}ller, P. (2022).
\newblock \enquote{Search algorithms and loss functions for Bayesian clustering.}
\newblock {\em Journal of Computational and Graphical Statistics\/}, 31(4): 1189--1201.
\endbibitem

\bibitem[{De~Blasi et~al.(2013)De~Blasi, Favaro, Lijoi, Mena, Pr{\"u}nster, and Ruggiero}]{deblasi2013gibbs}
De~Blasi, P., Favaro, S., Lijoi, A., Mena, R.~H., Pr{\"u}nster, I., and Ruggiero, M. (2013).
\newblock \enquote{{Are Gibbs-type priors the most natural generalization of the Dirichlet process?}}
\newblock {\em IEEE transactions on pattern analysis and machine intelligence\/}, 37(2): 212--229.
\endbibitem

\bibitem[{Duan et~al.(2025)Duan, Guo, Yan, Wang, and Mueller}]{duan2025spatially}
Duan, Y., Guo, S., Yan, H., Wang, W., and Mueller, P. (2025).
\newblock \enquote{Spatially aligned random partition models on spatially resolved transcriptomics data.}
\newblock {\em bioRxiv\/}, 2025--04.
\endbibitem

\bibitem[{Dvoretzky et~al.(1956)Dvoretzky, Kiefer, and Wolfowitz}]{dvoretzky1956asymptotic}
Dvoretzky, A., Kiefer, J., and Wolfowitz, J. (1956).
\newblock \enquote{Asymptotic minimax character of the sample distribution function and of the classical multinomial estimator.}
\newblock {\em The Annals of Mathematical Statistics\/}, 642--669.
\endbibitem

\bibitem[{Escobar and West(1995)}]{escobar1995bayesian}
Escobar, M.~D. and West, M. (1995).
\newblock \enquote{Bayesian density estimation and inference using mixtures.}
\newblock {\em Journal of the American Statistical Association\/}, 90(430): 577--588.
\endbibitem

\bibitem[{Escobar and West(1998)}]{escobar1998computing}
--- (1998).
\newblock \enquote{Computing nonparametric hierarchical models.}
\newblock In {\em Practical nonparametric and semiparametric Bayesian statistics\/}, 1--22. Springer.
\endbibitem

\bibitem[{Ester et~al.(1996)Ester, Kriegel, Sander, Xu et~al.}]{ester1996density}
Ester, M., Kriegel, H.-P., Sander, J., Xu, X., et~al. (1996).
\newblock \enquote{A density-based algorithm for discovering clusters in large spatial databases with noise.}
\newblock {\em Knowledge Discovery and Data Mining\/}, 96(34): 226--231.
\endbibitem

\bibitem[{Ferraty and Vieu(2006)}]{ferraty2006nonparametric}
Ferraty, F. and Vieu, P. (2006).
\newblock {\em {Nonparametric Functional Data Analysis: Theory and Practice}\/}.
\newblock Springer.
\endbibitem

\bibitem[{Flamary et~al.(2021)Flamary, Courty, Gramfort, Alaya, Boisbunon, Chambon, Chapel, Corenflos, Fatras, Fournier, Gautheron, Gayraud, Janati, Rakotomamonjy, Redko, Rolet, Schutz, Seguy, Sutherland, Tavenard, Tong, and Vayer}]{flamary2021pot}
Flamary, R., Courty, N., Gramfort, A., Alaya, M.~Z., Boisbunon, A., Chambon, S., Chapel, L., Corenflos, A., Fatras, K., Fournier, N., Gautheron, L., Gayraud, N.~T., Janati, H., Rakotomamonjy, A., Redko, I., Rolet, A., Schutz, A., Seguy, V., Sutherland, D.~J., Tavenard, R., Tong, A., and Vayer, T. (2021).
\newblock \enquote{{POT: Python Optimal Transport}.}
\newblock {\em Journal of Machine Learning Research\/}, 22(78): 1--8.
\endbibitem

\bibitem[{Flamary et~al.(2024)Flamary, Vincent-Cuaz, Courty, Gramfort, Kachaiev, Quang~Tran, David, Bonet, Cassereau, Gnassounou, Tanguy, Delon, Collas, Mazelet, Chapel, Kerdoncuff, Yu, Feickert, Krzakala, Liu, and Fernandes~Montesuma}]{flamary2024pot}
Flamary, R., Vincent-Cuaz, C., Courty, N., Gramfort, A., Kachaiev, O., Quang~Tran, H., David, L., Bonet, C., Cassereau, N., Gnassounou, T., Tanguy, E., Delon, J., Collas, A., Mazelet, S., Chapel, L., Kerdoncuff, T., Yu, X., Feickert, M., Krzakala, P., Liu, T., and Fernandes~Montesuma, E. (2024).
\newblock \enquote{{POT Python Optimal Transport (version 0.9.5)}.}
\endbibitem

\bibitem[{Fritsch and Ickstadt(2009)}]{fritsch2009clustering}
Fritsch, A. and Ickstadt, K. (2009).
\newblock \enquote{Improved criteria for clustering based on the posterior similarity matrix.}
\newblock {\em Bayesian Analysis\/}, 4(2): 367--392.
\endbibitem

\bibitem[{Fruhwirth-Schnatter et~al.(2019)Fruhwirth-Schnatter, Celeux, and Robert}]{fruhwirth2019handbook}
Fruhwirth-Schnatter, S., Celeux, G., and Robert, C.~P. (2019).
\newblock {\em Handbook of mixture analysis\/}.
\newblock CRC press.
\endbibitem

\bibitem[{Grazian(2023)}]{grazian2023review}
Grazian, C. (2023).
\newblock \enquote{A review on Bayesian model-based clustering.}
\newblock {\em arXiv preprint arXiv:2303.17182\/}.
\endbibitem

\bibitem[{Gupta et~al.(2022)Gupta, Kuchibhotla, and Ramdas}]{GUPTA2022108496}
Gupta, C., Kuchibhotla, A.~K., and Ramdas, A. (2022).
\newblock \enquote{Nested conformal prediction and quantile out-of-bag ensemble methods.}
\newblock {\em Pattern Recognition\/}, 127: 108496.
\endbibitem

\bibitem[{Hoffman et~al.(2014)Hoffman, Gelman et~al.}]{hoffman2014nuts}
Hoffman, M.~D., Gelman, A., et~al. (2014).
\newblock \enquote{The No-U-Turn sampler: adaptively setting path lengths in {Hamiltonian Monte Carlo}.}
\newblock {\em Journal of Machine Learning Research\/}, 15(1): 1593--1623.
\endbibitem

\bibitem[{Kalli et~al.(2011)Kalli, Griffin, and Walker}]{kalli2011slice}
Kalli, M., Griffin, J.~E., and Walker, S.~G. (2011).
\newblock \enquote{Slice sampling mixture models.}
\newblock {\em Statistics and Computing\/}, 21(1): 93--105.
\endbibitem

\bibitem[{Kappenman et~al.(2021)Kappenman, Farrens, Zhang, Stewart, and Luck}]{kappenman2021erp}
Kappenman, E.~S., Farrens, J.~L., Zhang, W., Stewart, A.~X., and Luck, S.~J. (2021).
\newblock \enquote{{ERP CORE: An open resource for human event-related potential research}.}
\newblock {\em NeuroImage\/}, 225: 117465.
\endbibitem

\bibitem[{Kriegel et~al.(2011)Kriegel, Kr{\"o}ger, Sander, and Zimek}]{kriegel2011density}
Kriegel, H.-P., Kr{\"o}ger, P., Sander, J., and Zimek, A. (2011).
\newblock \enquote{Density-based clustering.}
\newblock {\em Wiley Interdisciplinary Reviews: Data Mining and Knowledge Discovery\/}, 1(3): 231--240.
\endbibitem

\bibitem[{Lau and Green(2007)}]{lau2007clustering}
Lau, J.~W. and Green, P.~J. (2007).
\newblock \enquote{Bayesian model-based clustering procedures.}
\newblock {\em Journal of Computational and Graphical Statistics\/}, 16(3): 526--558.
\endbibitem

\bibitem[{Lee et~al.(2020)Lee, Hong, Etlioglu, Cho, Pomella, Van~den Bosch, Vanhecke, Verbandt, Hong, Min et~al.}]{lee2020lineage}
Lee, H.-O., Hong, Y., Etlioglu, H.~E., Cho, Y.~B., Pomella, V., Van~den Bosch, B., Vanhecke, J., Verbandt, S., Hong, H., Min, J.-W., et~al. (2020).
\newblock \enquote{Lineage-dependent gene expression programs influence the immune landscape of colorectal cancer.}
\newblock {\em Nature genetics\/}, 52(6): 594--603.
\endbibitem

\bibitem[{Lehmann and Romano(2005)}]{lehmann2005testing}
Lehmann, E.~L. and Romano, J.~P. (2005).
\newblock {\em Testing statistical hypotheses\/}.
\newblock Springer.
\endbibitem

\bibitem[{Lei et~al.(2018)Lei, G’Sell, Rinaldo, Tibshirani, and Wasserman}]{lei2018distribution}
Lei, J., G’Sell, M., Rinaldo, A., Tibshirani, R.~J., and Wasserman, L. (2018).
\newblock \enquote{Distribution-free predictive inference for regression.}
\newblock {\em Journal of the American Statistical Association\/}, 113(523): 1094--1111.
\endbibitem

\bibitem[{Levin and Peres(2017)}]{levin2017markov}
Levin, D.~A. and Peres, Y. (2017).
\newblock {\em Markov chains and mixing times\/}, volume 107.
\newblock American Mathematical Society.
\endbibitem

\bibitem[{Lewandowski et~al.(2009)Lewandowski, Kurowicka, and Joe}]{lewandowski2009generating}
Lewandowski, D., Kurowicka, D., and Joe, H. (2009).
\newblock \enquote{Generating random correlation matrices based on vines and extended onion method.}
\newblock {\em Journal of Multivariate Analysis\/}, 100(9): 1989--2001.
\endbibitem

\bibitem[{Lijoi et~al.(2005)Lijoi, Mena, and Pr{\"u}nster}]{lijoi2005hierarchical}
Lijoi, A., Mena, R.~H., and Pr{\"u}nster, I. (2005).
\newblock \enquote{{Hierarchical mixture modeling with normalized inverse-Gaussian priors}.}
\newblock {\em Journal of the American Statistical Association\/}, 100(472): 1278--1291.
\endbibitem

\bibitem[{Lijoi et~al.(2007)Lijoi, Mena, and Pr{\"u}nster}]{lijoi2007controlling}
--- (2007).
\newblock \enquote{Controlling the reinforcement in {B}ayesian non-parametric mixture models.}
\newblock {\em Journal of the Royal Statistical Society Series B: Statistical Methodology\/}, 69(4): 715--740.
\endbibitem

\bibitem[{Lijoi et~al.(2014)Lijoi, Nipoti, and Pr{\"u}nster}]{lijoi2014bayesian}
Lijoi, A., Nipoti, B., and Pr{\"u}nster, I. (2014).
\newblock \enquote{Bayesian inference with dependent normalized completely random measures.}
\newblock {\em Bernoulli\/}, 20(3): 1260--1291.
\endbibitem

\bibitem[{Lijoi et~al.(2020)Lijoi, Pr{\"u}nster, and Rigon}]{lijoi2020pitman}
Lijoi, A., Pr{\"u}nster, I., and Rigon, T. (2020).
\newblock \enquote{{The Pitman--Yor multinomial process for mixture modelling}.}
\newblock {\em Biometrika\/}, 107(4): 891--906.
\endbibitem

\bibitem[{Lo(1984)}]{lo1984class}
Lo, A.~Y. (1984).
\newblock \enquote{On a class of Bayesian nonparametric estimates: I. Density estimates.}
\newblock {\em The Annals of Statistics\/}, 12(1): 351--357.
\endbibitem

\bibitem[{MacEachern(2000)}]{maceachern2000dependent}
MacEachern, S.~N. (2000).
\newblock \enquote{{Dependent Dirichlet Processes}.}
\newblock {\em Unpublished manuscript, Department of Statistics, The Ohio State University\/}.
\endbibitem

\bibitem[{MacEachern and M{\"u}ller(1998)}]{maceachern1998estimating}
MacEachern, S.~N. and M{\"u}ller, P. (1998).
\newblock \enquote{Estimating mixture of Dirichlet process models.}
\newblock {\em Journal of Computational and Graphical Statistics\/}, 7(2): 223--238.
\endbibitem

\bibitem[{Massart(1990)}]{massart1990tight}
Massart, P. (1990).
\newblock \enquote{The tight constant in the {Dvoretzky-Kiefer-Wolfowitz} inequality.}
\newblock {\em The Annals of Probability\/}, 1269--1283.
\endbibitem

\bibitem[{Meil{\u{a}}(2007)}]{meila2007comparing}
Meil{\u{a}}, M. (2007).
\newblock \enquote{Comparing clusterings---an information based distance.}
\newblock {\em Journal of Multivariate Analysis\/}, 98(5): 873--895.
\endbibitem

\bibitem[{Nguyen and Mueller(2026)}]{nguyen2024summarizing}
Nguyen, K. and Mueller, P. (2026).
\newblock \enquote{Summarizing Nonparametric Bayesian Mixture Posteriors—Sliced Optimal Transport Metrics for Gaussian Mixtures.}
\newblock {\em Journal of Computational and Graphical Statistics\/}, 1--12.
\endbibitem

\bibitem[{Nguyen(2013)}]{nguyen2013convergence}
Nguyen, X. (2013).
\newblock \enquote{{Convergence of latent mixing measures in finite and infinite mixture models}.}
\newblock {\em The Annals of Statistics\/}, 41(1): 370 -- 400.
\endbibitem

\bibitem[{Papaspiliopoulos and Roberts(2008)}]{papaspiliopoulos2008retrospective}
Papaspiliopoulos, O. and Roberts, G.~O. (2008).
\newblock \enquote{Retrospective Markov chain Monte Carlo methods for Dirichlet process hierarchical models.}
\newblock {\em Biometrika\/}, 169--186.
\endbibitem

\bibitem[{Perman et~al.(1992)Perman, Pitman, and Yor}]{perman1992size}
Perman, M., Pitman, J., and Yor, M. (1992).
\newblock \enquote{{Size-biased sampling of Poisson point processes and excursions}.}
\newblock {\em Probability Theory and Related Fields\/}, 92(1): 21--39.
\endbibitem

\bibitem[{Petralia et~al.(2012)Petralia, Rao, and Dunson}]{petralia2012repulsive}
Petralia, F., Rao, V., and Dunson, D. (2012).
\newblock \enquote{Repulsive mixtures.}
\newblock {\em Advances in neural information processing systems\/}, 25.
\endbibitem

\bibitem[{Phan et~al.(2019)Phan, Pradhan, and Jankowiak}]{phan2019composable}
Phan, D., Pradhan, N., and Jankowiak, M. (2019).
\newblock \enquote{{Composable Effects for Flexible and Accelerated Probabilistic Programming in NumPyro}.}
\newblock {\em arXiv preprint arXiv:1912.11554\/}.
\endbibitem

\bibitem[{Pitman(2006)}]{pitman2006combinatorial}
Pitman, J. (2006).
\newblock {\em {Combinatorial stochastic processes: Ecole d'et{\'e} de probabilit{\'e}s de Saint-Flour XXXII-2002}\/}.
\newblock Springer Science \& Business Media.
\endbibitem

\bibitem[{Pitman and Yor(1997)}]{pitman1997two}
Pitman, J. and Yor, M. (1997).
\newblock \enquote{{The two-parameter Poisson-Dirichlet distribution derived from a stable subordinator}.}
\newblock {\em The Annals of Probability\/}, 855--900.
\endbibitem

\bibitem[{Quintana and Iglesias(2003)}]{quintana2003clustering}
Quintana, F.~A. and Iglesias, P.~L. (2003).
\newblock \enquote{Bayesian clustering and product partition models.}
\newblock {\em Journal of the Royal Statistical Society: Series B (Statistical Methodology)\/}, 65(2): 557--574.
\endbibitem

\bibitem[{Rastelli and Friel(2018)}]{rastelli2018optimal}
Rastelli, R. and Friel, N. (2018).
\newblock \enquote{Optimal {B}ayesian estimators for latent variable cluster models.}
\newblock {\em Statistics and Computing\/}, 28(6): 1169--1186.
\endbibitem

\bibitem[{Rigon et~al.(2023)Rigon, Herring, and Dunson}]{rigon2023generalized}
Rigon, T., Herring, A.~H., and Dunson, D.~B. (2023).
\newblock \enquote{A generalized {B}ayes framework for probabilistic clustering.}
\newblock {\em Biometrika\/}, 110(3): 559--578.
\endbibitem

\bibitem[{Rodriguez et~al.(2008)Rodriguez, Dunson, and Gelfand}]{rodriguez2008nested}
Rodriguez, A., Dunson, D.~B., and Gelfand, A.~E. (2008).
\newblock \enquote{The nested {D}irichlet process.}
\newblock {\em Journal of the American statistical Association\/}, 103(483): 1131--1154.
\endbibitem

\bibitem[{Rodriguez and Laio(2014)}]{rodriguez2014clustering}
Rodriguez, A. and Laio, A. (2014).
\newblock \enquote{Clustering by fast search and find of density peaks.}
\newblock {\em Science\/}, 344(6191): 1492--1496.
\endbibitem

\bibitem[{Roeder(1990)}]{roeder1990density}
Roeder, K. (1990).
\newblock \enquote{Density estimation with confidence sets exemplified by superclusters and voids in the galaxies.}
\newblock {\em Journal of the American Statistical Association\/}, 85(411): 617--624.
\endbibitem

\bibitem[{Sarkar and Kuchibhotla(2023)}]{sarkar2023post}
Sarkar, S. and Kuchibhotla, A.~K. (2023).
\newblock \enquote{Post-selection inference for conformal prediction: Trading off coverage for precision.}
\newblock {\em arXiv preprint arXiv:2304.06158\/}.
\endbibitem

\bibitem[{Selva et~al.(2025)Selva, Fuentes-Garc{\'\i}a, and Gil-Leyva}]{selva2025pyrichlet}
Selva, F., Fuentes-Garc{\'\i}a, R., and Gil-Leyva, M.~F. (2025).
\newblock \enquote{{pyrichlet: A Python Package for Density Estimation and Clustering Using Gaussian Mixture Models}.}
\newblock {\em Journal of Statistical Software\/}, 112: 1--39.
\endbibitem

\bibitem[{Shafer and Vovk(2008)}]{shafer2008tutorial}
Shafer, G. and Vovk, V. (2008).
\newblock \enquote{A tutorial on conformal prediction.}
\newblock {\em Journal of Machine Learning Research\/}, 9(3).
\endbibitem

\bibitem[{Song et~al.(2025)Song, Camerlenghi, Shen, Guindani, and Beraha}]{song2025repulsive}
Song, Z., Camerlenghi, F., Shen, W., Guindani, M., and Beraha, M. (2025).
\newblock \enquote{{Repulsive Mixture Model with Projection Determinantal Point Process}.}
\newblock {\em arXiv preprint arXiv:2510.08838\/}.
\endbibitem

\bibitem[{Teh et~al.(2006)Teh, Jordan, Beal, and Blei}]{teh2006hierarchical}
Teh, Y.~W., Jordan, M.~I., Beal, M.~J., and Blei, D.~M. (2006).
\newblock \enquote{{Hierarchical Dirichlet Processes}.}
\newblock {\em Journal of the American Statistical Association\/}, 101(476): 1566--1581.
\endbibitem

\bibitem[{Villani et~al.(2008)}]{villani2008optimal}
Villani, C. et~al. (2008).
\newblock {\em Optimal Transport: Old and New\/}, volume 338.
\newblock Springer.
\endbibitem

\bibitem[{Vinh et~al.(2010)Vinh, Epps, and Bailey}]{vinh2010information}
Vinh, N.~X., Epps, J., and Bailey, J. (2010).
\newblock \enquote{Information Theoretic Measures for Clusterings Comparison: Variants, Properties, Normalization and Correction for Chance.}
\newblock {\em Journal of Machine Learning Research\/}, 11: 2837--2854.
\endbibitem

\bibitem[{Vovk(2012)}]{vovk2012conditional}
Vovk, V. (2012).
\newblock \enquote{Conditional validity of inductive conformal predictors.}
\newblock In {\em Asian Conference on Machine Learning\/}, 475--490. PMLR.
\endbibitem

\bibitem[{Vovk et~al.(2005)Vovk, Gammerman, and Shafer}]{vovk2005algorithmic}
Vovk, V., Gammerman, A., and Shafer, G. (2005).
\newblock {\em Algorithmic Learning in a Random World\/}.
\newblock Springer.
\endbibitem

\bibitem[{Wade(2023)}]{wade2023bayesian}
Wade, S. (2023).
\newblock \enquote{Bayesian cluster analysis.}
\newblock {\em Philosophical Transactions of the Royal Society A\/}, 381(2247): 20220149.
\endbibitem

\bibitem[{Wade and Ghahramani(2018)}]{wade2018clustering}
Wade, S. and Ghahramani, Z. (2018).
\newblock \enquote{{Bayesian Cluster Analysis: Point Estimation and Credible Balls (with Discussion)}.}
\newblock {\em Bayesian Analysis\/}, 13(2): 559 -- 626.
\endbibitem

\bibitem[{Walker(2007)}]{walker2007sampling}
Walker, S.~G. (2007).
\newblock \enquote{{Sampling the Dirichlet mixture model with slices}.}
\newblock {\em Communications in Statistics---Simulation and Computation{\textregistered}\/}, 36(1): 45--54.
\endbibitem

\bibitem[{Woody et~al.(2021)Woody, Carvalho, and Murray}]{woody2021model}
Woody, S., Carvalho, C.~M., and Murray, J.~S. (2021).
\newblock \enquote{Model interpretation through lower-dimensional posterior summarization.}
\newblock {\em Journal of Computational and Graphical Statistics\/}, 30(1): 144--161.
\endbibitem

\bibitem[{Xie and Xu(2020)}]{xie2020bayesian}
Xie, F. and Xu, Y. (2020).
\newblock \enquote{{Bayesian repulsive Gaussian mixture model}.}
\newblock {\em Journal of the American Statistical Association\/}, 115(529): 187--203.
\endbibitem

\bibitem[{Xu et~al.(2016)Xu, M{\"u}ller, and Telesca}]{xu2016bayesian}
Xu, Y., M{\"u}ller, P., and Telesca, D. (2016).
\newblock \enquote{Bayesian inference for latent biologic structure with determinantal point processes (DPP).}
\newblock {\em Biometrics\/}, 72(3): 955--964.
\endbibitem

\end{thebibliography}
\end{document}